\documentclass[10pt]{article}

\usepackage{latexsym}
\usepackage{amsmath}
\usepackage{amsfonts}
\usepackage{amsbsy}
\usepackage{amssymb}
\usepackage{epsfig}
\usepackage{mathrsfs}
\usepackage{epsfig}
\usepackage{psfrag}
\usepackage{bbm}
\usepackage{graphicx,color}
\usepackage{hyperref}
\usepackage{tikz}
\usetikzlibrary{patterns}

\numberwithin{equation}{section}

\def\be{\begin{equation}}
\def\ee{\end{equation}}
\def\ba{\begin{eqnarray}}
\def\ea{\end{eqnarray}}

\newcommand\nn{\nonumber}
\newcommand\q{\quad}
\newcommand\bd{{\bf d}}

\title{Quantum group spin nets:\\ refinement limit and relation to spin foams}
\author{Bianca Dittrich, Mercedes Martin-Benito, Sebastian Steinhaus\\[5pt]
\small   Perimeter Institute for Theoretical Physics,\\
 \small 31 Caroline Street North, Waterloo, Ontario, Canada N2L 2Y5
}

 \date{}

\begin{document}

\maketitle

\begin{abstract}
So far spin foam models are hardly understood beyond a few of their basic building blocks. To make progress on this question, we define analogue spin foam models, so called spin nets, for quantum groups $\text{SU}(2)_k$ and examine their effective continuum dynamics via tensor network renormalization. In the refinement limit of this coarse graining procedure, we find a vast non-trivial fixed point structure beyond the degenerate and the $BF$ phase. In comparison to previous work, we use fixed point intertwiners, inspired by Reisenberger's construction principle \cite{reisenberger} and the recent work \cite{bw1}, as the initial parametrization. In this new parametrization  fine tuning  is not required in order to flow to these new fixed points. Encouragingly, each fixed point has an associated extended phase, which allows for the study of phase transitions in the future. Finally we also present an interpretation of spin nets in terms of melonic spin foams. The coarse graining flow of spin nets can thus be interpreted as describing the effective coupling between two spin foam vertices or space time atoms.
\end{abstract}

\section{Introduction}

The aim of spin foam quantization is to provide a non--pertubative and background--independent quantization of gravity via an auxiliary discretization. A key outstanding question for these models is to show that they reproduce familiar low energy physics, in particular a phase describing a smooth manifold, in a limit in which the scale of the discretization is much smaller than other relevant length scales in the system \cite{alexreview}.

In this work we will provide first conjectures on the possible large scale behaviour of spin foams. These are based on the investigation of the coarse graining or renormalization flow of spin net models, which, as we will show in this work, can be interpreted as a coarse graining flow of (melonic) spin foams and as describing the effective coupling between two spin foam vertices.

To lay out the context of this work, explicit investigations of the coarse graining and renormalization flow of spin foam models \cite{alexreview,cbook,eprl}, have been hindered by the overwhelming complexity of the models, apart from an incomplete understanding of possible infinities \cite{riello} and a number of conceptual questions \cite{fotini}, for instance the meaning of scale in coarse graining background independent systems. 
Regarding these questions a conceptual understanding was reached in \cite{bd12b,refining}, that allows to interpret the coarse graining flow as the construction of the continuum limit of the theory via the concept of dynamical cylindrical consistency and embedding or refinement maps.  To tackle the concrete implementation of this coarse graining flow, after establishing the importance of coarse graining for the recovery of diffeomorphism symmetries in discrete systems and therefore the recovery of a  correct continuum dynamics  \cite{bahrdittrich}, a program was started in \cite{sffinite,eckert}, which considers the corresponding question in simplified systems, that keep however key dynamical principles, in particular the imposition of so--called simplicity constraints \cite{alexreview,cbook,eprl}, of spin foams. 

The simplifications considered can be summarized into three types: (a) the replacement of the structure group $SU(2)\times SU(2)$ of the full gravitational models by a finite group \cite{sffinite,bahretal12}, (b) a dimensional reduction of spin foams to so--called spin nets \cite{sffinite,eckert}, and (c) the consideration of a regular lattice \cite{eckert,he}.

In this work we lift the simplification (a), regarding (b) we show  that spin net coarse graining can be interpreted as a coarse graining of (melonic) spin foams and confirm again \cite{bahrdittrich,he} that (c) does lead to triangulation invariant models, at least for the non--critical fixed points. This latter fact is deeply related to the restoration of a notion of diffeomorphism symmetry \cite{bahrdittrich} at the fixed points. In fact the recovery of a sensible continuum limit and the regaining of diffeomorphism symmetry might even be only possible if certain regularity conditions on the coarse graining process are satisfied \cite{refining,fat}. 

For these reasons we will be able to present for the first time\footnote{An exception is \cite{bahretal12,unpublished} which considered a sequence of $4-1$ and $5-1$ Pachner moves for spin foam models based on finite groups. However, as mentioned, the question arises whether such a flow, based on coarse graining moves \cite{refining} only, allows to make definite conclusions about the continuum limit. In contrast the tensor network algorithm \cite{levin,guwen} considered here includes, in a sense, refining, coarse graining, and entangling moves \cite{refining}.} the coarse graining flow in a non--trivial spin foam system and investigate the fixed points it leads to. This allows us to draw first conclusions about the possible phases that might describe the continuum limit of spin foams. 

To be more precise, we introduce here spin net models with structure group $SU(2)_k$, the quantum deformation (at the root of unity) of the group $SU(2)$. This makes a numerical investigation of the models possible, as in the quantum group only finitely many representations appear. On the technical level this requires the introduction of the Haar projector for the quantum group which we construct in this work. Models based on the quantum deformation of the rotation group describe gravitational systems with a cosmological constant \cite{noui}. For the 4D systems one would need $SU(2)_k \times SU(2)_k$, however also an effective description in terms of $SU(2)_k$ alone might be possible \cite{jeff}. Thus we lift the simplification, considered in \cite{eckert,bdetal13}, of the structure group almost completely. Moreover considering the behaviour of the models for growing level $k$ allows to make conjectures on the limit $k\rightarrow \infty$, which gives back the classical group $SU(2)$.

As in previous work \cite{eckert,bdetal13} we will employ tensor network techniques \cite{levin,guwen} to derive a coarse graining flow equation, which is then investigated numerically. This flow equation describes the change of intertwiner degrees of freedom under coarse graining, as exemplified for the first time in \cite{bdetal13}. As is outlined in \cite{bd12b,refining} these tensor network techniques are particularly suited for the construction of the continuum limit for spin nets and spin foams. In \cite{eckert} a tensor network representation of spin foams appeared which might allow to derive similar coarse graining flow equations directly for spin foams. Also \cite{refining} points out, that the geometric building blocks of spin foams, the simplices, allow to be interpreted as coarse graining, refining or entangling moves, in dependence of how these are glued together.  Such moves are the basic ingredients of tensor network renormalization. This might lead to an even more geometric way to derive such coarse graining flow equations.

Another innovation that we present in this work is a change in the parametrization for the (phase) space of spin net and spin foam models. This new parametrization takes the lesson from \cite{bdetal13} seriously, that the relevant degrees of freedom are given by intertwiners. We thus introduce a new parametrization, based on Reisenberger's construction principle \cite{reisenberger}, that leads us to employ so--called {\it fixed point intertwiners} \cite{bw1,bw2}, which encode the simplicity constraints. Reisenberger's principle ensures that the models so obtained are physically reasonable, i.e. that the simplicity constraints they encode appear to be the same in (physically equivalent) different recoupling schemes.

We will find that this new parametrization is highly advantageous: in contrast to the parametrization introduced in \cite{bahretal12} and employed in \cite{bdetal13}, models flow generically to non--trivial fixed points  (in particular different from the degenerate fixed point, in which only the spin representation $j=0$ appears). We conjecture that this behaviour is due to the projector property that is satisfied for models based on fixed point intertwiners. Modern spin foams are rather based on a parametrization like that employed in \cite{bahretal12}, and the reconstruction of the projector property  is rather complicated \cite{warsaw}. The findings here therefore suggest to reconsider Reisenberger's construction principle, in particular since it allows the construction of vertices of arbitrary valence \cite{reisenberger}, as was also later realized for the EPRL model \cite{eprl} in \cite{warsaw0}. Similar to the arguments on the uniqueness of the Barrett--Crane model \cite{bc} in \cite{reisenberger}, it is even possible to find a parametrization of all physically sensible models, by finding all fixed point intertwiners for a given structure group \cite{bw1,bw2}. 

~\\

The outline of the paper is the following: The next section \ref{sec:reisenberger} introduces the basic ingredients of spin foam and spin net models that we need for the work here. It will also explain the parametrization for the models based on Reisenberger's construction principle, for which we need the fixed point intertwiners. 
In section \ref{quantum-group} we introduce the essential notions and tools related with the quantum group  $\text{SU}(2)_k$ that  will be needed to define  spin net models based on this group. Using these tools we introduce the general structure of our models in section \ref{structurespinnets}. In section \ref{coarse} we adapt to the quantum group case the symmetry preserving coarse graining algorithm introduced in  \cite{bdetal13}. This will lead to a flow equation describing the behaviour of intertwiner degrees of freedom under coarse graining. We will give
 a (3D) geometrical interpretation of the flow equation which might lead to a more intuitive understanding of the coarse graining flow. Section \ref{results} contains the explicit definition of the weights for the initial models. We then investigate the behaviour under coarse graining of these models and analyze and describe the results obtained. We comment on the interpretation of spin net coarse graining in the context of spin foam coarse graining in section \ref{interpretation}. We discuss our main findings in Section \ref{discussion}. Finally appendix \ref{app} includes  technical (diagrammatic) calculations that might help  the understanding of some equations in the main text, in particular those appearing in sections \ref{quantum-group} and \ref{coarse}.

\section{Reisenberger's principle and intertwiner models} \label{sec:reisenberger}

Spin foam models are generalization of lattice gauge theories and as such are defined on 2--complexes, i.e. generalizations of graphs, in which in addition to vertices and edges, also faces are specified.  There is an important difference though: whereas for lattice gauge theories the main dynamical ingredient are weights on the faces, these weights are  not as essential for the dynamics of spin foams (they are crucial for the divergence properties of the model \cite{riello}, but are not supposed to include i.e. a coupling constant).  Rather for the dynamics of spin foams the choice of an intertwiner  for the edges is crucial -- equivalent to choosing a certain implementation of the so called simplicity constraints. 

To be more specific, consider lattice gauge theory based on a gauge (or structure) group $G$, which we assume to be either a compact semi--simple Lie group or a finite group.\footnote{For a basic introduction into spin foams as statistical systems see \cite{sffinite}.} The variables are given as group elements associated to oriented edges of a graph. The dynamics is encoded into face weights $\omega_f$, which are class functions on the group. These face weights are evaluated on the face holonomies $h_f$, i.e. the (oriented) product of group elements $g_e$ associated to the edges $e$ around a face $h_f=\prod_{e\subset f} g_e$. Let us assume that all edges agree in their orientation with the orientation induced from the face.

The partition function is then given as
\ba\label{i1}
Z=\int\,\prod_{e} \bd g_e \, \prod_f \omega_f(h_f) \q .
\ea
Here $\bd g$ denotes the (normalized) Haar measure on $G$.
We can change to a dual description by applying a group Fourier transform. Due to the invariance of the face weight under group conjugation we can expand the face weights as a linear combination of characters
\ba\label{i2}
\omega_f(h) =\sum_\rho \tilde \omega_f(\rho) \chi_\rho(h) \quad ,
\ea
where $\chi_\rho= {\rho_{m}}^m$ and ${\rho(\cdot)_{m}}^n$ denotes a matrix element of a irreducible unitary representation $\rho$ (irreps) of the group $G$. In (\ref{i2}) we sum over a complete set of (equivalence classes) of irreps. 

Expanding $h_f$ into the product of group elements and using the representation property $\rho(g_1g_2)=\rho(g_1)\cdot \rho(g_2)$ one finds that the partition function is now given as\ba\label{i3}
Z=\sum_{\rho_f} \prod_f \tilde \omega_f(\rho_f)   \,\prod_e ({\mathcal P^e_{Haar}})_{m_{f_1} m_{f_2} \cdots m_{f_{N_e}}}^{n_{f_1} n_{f_2} \cdots n_{f_{N_e}}} \q .
\ea
Here $\mathcal P^e_{Haar}$ is the so--called Haar projector, which absorbs the integrations over the group elements $g$ in (\ref{i1}). For an edge whose $N_e$ adjacent faces agree in orientation with the one of the edge  the Haar projector is given as
\ba\label{ghaar}
({\mathcal P^e_{Haar}})_{m_{f_1} m_{f_2} \cdots m_{f_{N_e}}}^{n_{f_1} n_{f_2} \cdots n_{f_{N_e}}}  &=& \int \bd g \,\,  \prod_{f \supset e} {\rho_f(g_e)_{m_f}}^{n_f}  \q .
\ea

The group integration or averaging enforces gauge invariance. Indeed the Haar projector is an intertwining map between the group invariant (singlet) subspace of the tensor product of representations
\ba
\mathcal P_{Haar}: \text{Inv}(V_{\rho_1} \cdots V_{\rho_{N_e}} )\rightarrow \text{Inv}(V_{\rho_1} \cdots V_{\rho_{N_e}} )
\ea
or, if seen as a map on $V_{\rho_1} \cdots V_{\rho_{N_e}}$ it acts as a projector onto $ \text{Inv}(V_{\rho_1} \cdots V_{\rho_{N_e}} )$. 

Choosing any orthonormal basis\footnote{Often the elements of $\text{Inv}(V_{\rho_1} \cdots V_{\rho_{N_e}} )$ themselves are referred to as intertwiners.}  $\{|\iota_d\rangle\}_d$ in $\text{Inv}(V_{\rho_1} \cdots V_{\rho_{N_e}} )$ we can write the Haar intertwiner as
\ba\label{shaar}
\mathcal P_{Haar} \,=\, \sum_{d} |\iota_d \rangle \,\langle \iota_d | \q ,
\ea
which makes the projection property $\mathcal P\cdot \mathcal P=\mathcal P$ of $\mathcal P_{Haar}$ obvious.

As mentioned the important dynamical ingredient in spin foams are intertwiners, whereas the face weights are rather chosen due to kinematical considerations, for instance as delta functions on the group in the group representation.  For non--trivial spin foam models, that is models which cannot be rewritten into standard lattice gauge theories, one replaces $\mathcal P_{Haar}$ by a smaller projector $\mathcal P$, selecting a smaller subspace of  $\text{Inv}(V_{\rho_1} \cdots V_{\rho_{N_e}} )$.

 With an appropriate choice of orthonormal basis of intertwiners we can write
\ba\label{i6}
\mathcal P^{(N)}\,=\, \sum_{d \in {\cal S}_N }  |\iota_d \rangle \,\langle \iota_d | \q ,
\ea
where ${\cal S}_N$ denotes a subset of intertwiner labels specified by the choice of simplicity constraints. Again the form (\ref{i6}) makes the projector property $\mathcal P\cdot \mathcal P=\mathcal P$ explicit. In the context of spin foams this projector property ensures a notion of (weak) independence of the underlying two--complex, namely invariance under edge subdivisions \cite{bahretal12,warsaw,bojowaldperez}.

On the other hand the form (\ref{i6}) makes it non--obvious how to relate the sets ${\cal S}_N$ for different valencies $N$ (i.e. number of faces sharing the edge $e$) of the edges. To solve this problem, Reisenberger proposed the following principle in \cite{reisenberger}: 

~\\
{\bf Reisenberger's construction:} We consider multiplicity free representation categories, where in the reduction of the tensor product of any two irreps each irrep can only appear once. In this case, by choosing a recoupling scheme,  intertwiners can be labelled by sets of representation labels: an intertwiner for valency $N$ is characterized by the $N$ representation labels plus $(N-3)$ further representation labels. Note that the choice of recoupling scheme can be encoded in a three--valent graph with $N$ outer edges, and that the representation labels are attached to the edges of this graph.  We will denote with ${\cal S}_1$ the set of allowed representations implied by the simplicity constraints. 

In Reisenberger's construction this set ${\cal S}_1$ then determines all the other sets ${\cal S}_N$ via the following principle: For every possible choice of recoupling scheme, i.e. three--valent graph, only representations in ${\cal S}_1$ should appear.
~\\

 Reisenberger applies this to the Barret-Crane model \cite{bc} and shows that choosing ${\cal S}_1$ accordingly the intertwiners $\mathcal{P}^{(N)}$ are uniquely determined.
It actually turns out that if we take the intertwiner as a model for a partition function, as introduced in \cite{bw1} (with boundary data given by the $n$ representation and summing over all the intertwiner labels) the Barrett-Crane model leads even to a triangulation invariant model.\footnote{Indeed it can be seen as a re--interpretaion of an analogue $BF$ spin net model \cite{sffinite,eckert}, which is also topological. Related, the Barrett-Crane intertwiner is given as a square of the $BF$ intertwiner \cite{sffinite}.}

In \cite{bw1} so--called intertwiner models were introduced, which are 2D statistical models determining an intertwining map between representation spaces associated to the  boundary components of the 2D bulk. The models are defined on three--valent graphs. Dualizing these graphs we obtain a triangulation (under certain regularity assumptions). It turns out that intertwiner models whose partition function are invariant under the change of (2D) triangulation, are also -- at least in the planar case -- a realization of Reisenberger's conditions. In particular each triangulation invariant model is characterized by a set ${\cal S}_1$ of representations that are allowed to appear as intertwiner labels and as boundary data.

Let us explain this statement in more detail, restricting to four--valent intertwiners, i.e.\  $N=4$.  Fixing the 4 outer representations $\rho_1,\ldots,\rho_4$,  a choice of intertwiner basis,  labeled by $\rho_5$, can then be represented graphically by the following three graphs $\Gamma$:
\ba
A=
\begin{tikzpicture}[baseline,scale=0.75]
\draw (-0.5,-0.75) -- (0,-0.25) -- (0,0.25) -- (-0.5,0.75)
      (0.5,-0.75) -- (0,-0.25)
      (0,0.25) -- (0.5,0.75)
      (-0.5,-1) node {$\rho_3$}
      (0.5,-1) node {$\rho_4$}
      (0.5,1.25) node {$\rho_1$}
      (-0.5,1.25) node {$\rho_2$}
      (0.3,0) node {$\rho_5$};
\end{tikzpicture}\q;\q
B=
\begin{tikzpicture}[baseline,scale=0.75]
\draw (-1,0.5) -- (-0.5,0) -- (0,0) -- (0.5,0.5)
      (-0.5,0) -- (-1,-0.5)
      (0,0) -- (0.5,-0.5)
      (-1,-0.75) node {$\rho_3$}
      (-1,0.75) node {$\rho_2$}
      (0.5,0.75) node {$\rho_1$}
      (0.5,-0.75) node {$\rho_4$}
      (-0.25,0.25) node {$\rho_5$};
\end{tikzpicture}\q;\q
C=\begin{tikzpicture}[baseline,scale=0.75]
\draw  (-0.75,0.75) -- (-0.25,0.25) -- (0.25,0.25) -- (0.75,0.75)
	(0.25,0.25) -- (-0.5,-0.75)
	 (-0.25,0.25) -- (0,0)
	 (0.1,-0.1) -- (0.7,-0.75)
      (-0.5,-1) node {$\rho_3$}
      (-0.75,1) node {$\rho_2$}
      (0.7,1) node {$\rho_1$}
      (0.7,-1) node {$\rho_4$}
      (0,0.5) node {$\rho_5$};
\end{tikzpicture}\q .
\ea

A specific intertwiner is then specified by the basis coefficients $a_{\Gamma}(\rho_1,\ldots,\rho_4;\rho_5)$. Here one has to make one choice of basis $\Gamma=A,B$ or $C$. Once the intertwiner coefficients are fixed in one basis, say $A$,  it can of course be re--expressed in the two other bases $B$ and $C$. Reisenberger's principle demands that $a_A$ has to vanish if any of the arguments $\rho_1,\ldots, \rho_4$ is not in ${\cal S}_1$ or if $\rho_5$ is not in ${\cal S}_1$. 
Additionally the coefficients $a_A$ have to be chosen so that also $a_B$ and $a_C$ vanish as soon as $\rho_5$ is not in  ${\cal S}_1$. 

The triangulation invariant intertwiners constructed in \cite{bw1,bw2} are characterized by a certain set ${\cal S}_1$ of allowed representations. Additionally the basis coefficients are invariant under changes which keep the graph planar, i.e. we even have $a_A=a_B$.  As for a graph with crossing (note that for a quantum group one has to specify whether this is a over or under--crossing), only a certain subset of the fixed point intertwiners satisfies $a_A=a_B=a_C$. However, using the braiding operation for quantum groups, see for instance \cite{bw1}, one can find for the change of basis from $C$ to $B$
\begin{align}
\sum_{\rho_5} a_C(\rho_1,\rho_2,\rho_3,\rho_4;\rho_5)
\begin{tikzpicture}[baseline,scale=0.75]
\draw  (-0.75,0.5) -- (-0.25,0) -- (0.25,0) -- (0.75,0.5)
	(0.25,0) -- (-0.5,-1)
	 (-0.25,0) -- (0,-0.25)
	 (0.1,-0.35) -- (0.7,-1)
      (-0.5,-1.25) node {$\rho_3$}
      (-0.75,0.75) node {$\rho_2$}
      (0.7,0.75) node {$\rho_1$}
      (0.7,-1.25) node {$\rho_4$}
      (0,0.25) node {$\rho_5$};
\end{tikzpicture} 
=
\sum_{\rho_5} a_C(\rho_1,\rho_2,\rho_3,\rho_4;\rho_5)&
\begin{tikzpicture}[baseline,scale=0.75]
\draw 	(0,0.25) -- (0, 0.75) -- (0.5, 1.25)
(0,0.75) -- (-0.5, 1.25)
(-0.25,0) -- (0,0.25) -- (0.25,0)
	(0.25,0) -- (-0.5,-1)
	 (-0.25,0) -- (0,-0.25)
	 (0.1,-0.35) -- (0.7,-1)
      (-0.5,-1.25) node {$\rho_3$}
      (-0.75,1.25) node {$\rho_2$}
      (0.75,1.25) node {$\rho_1$}
      (0.7,-1.25) node {$\rho_4$}
      (-0.25,0.5) node {$\rho_5$};
\end{tikzpicture} \nonumber\\
=
\sum_{\rho_5} a_C(\rho_1,\rho_2,\rho_3,\rho_4;\rho_5)&\,
\begin{tikzpicture}[baseline,scale=0.75]
\draw (-0.5,-1) -- (0,-0.5) -- (0,0.5) -- (-0.5,1)
      (0.5,-1) -- (0,-0.5)
      (0,0.5) -- (0.5,1)
      (-0.5,-1.25) node {$\rho_3$}
      (0.5,-1.25) node {$\rho_4$}
      (0.5,1.25) node {$\rho_1$}
      (-0.5,1.25) node {$\rho_2$}
      (0.25,0) node {$\rho_5$};
\end{tikzpicture} b(\rho_3,\rho_4,\rho_5) \q ,
\end{align}
where in the first step we used the triangulation invariance, i.e. invariance under a planar change of graph, and in the second step we resolved the braiding, described by non--vanishing coefficients  $b(\rho_3,\rho_4,\rho_5)$. For a certain set of triangulation invariant models these coefficients $b$ can be replaced by $1$ in the sum over $\rho_5$. For most models (for instance the family described in section \ref{maximalJ}) this is however not the case. Nevertheless Reisenberger's principle still holds: if $a_C$ vanishes for a  certain  representation $\rho_5$ this will also hold for the $A$ basis coefficient, given by $a_C\times b$.  Thus Reisenberger's principle is satisfied at least for four--valent intertwiners, relevant for simplicial lattices in 4D. For 3D spin foams (also on non--simplicial lattices) one could even restrict to planar graph changes only, as the intertwiners are attached to edges and so the expansion might be argued to take place in a plane orthogonal to this edge.

There exists a rich structure of these triangulation invariant models  \cite{bw1,bw2}. These models define in particular fixed points for the renormalization flow in the space of intertwiner models. This suggest to indeed use these particular fixed point models for defining intertwiners for spin foams. We will call such an intertwiner coming from a  fixed point model a {\it fixed point intertwiner}. Such a fixed point intertwiner, characterized by a set ${\cal S}_1$ of irreps, leads to a family of  intertwiners $|\iota^N_{{\cal S}_1}\rangle$, labelled by the valency $N$ of the edge, so that we can define a projector just as
\ba\label{i7}
\mathcal P_{{\cal S}_1}^{(N)} \,=\, |\iota^N_{{\cal S}_1}\rangle \, \langle \iota^N_{{\cal S}_1}| \q .
\ea
Assuming normalization the projector property is obviously satisfied. 

Note that a spin foam model defined via such a fixed point intertwiner does in general {\it not} define a fixed point of the renormalization flow in the space of spin foam models. The same holds for spin nets, that we will define as lower dimensional analogues of spin foams further below.

Alternatively we can also consider linear combinations of different fixed point intertwiners
\ba\label{i8}
\mathcal P_{\alpha_i^N}^{(N)} \,=\, \sum_i  \alpha_i^N |\iota^N_{{\cal S}^i_1}\rangle \, \langle \iota^N_{{\cal S}^i_1}| \q .
\ea
These also satisfy Reisenberger's principle (at least for $N=4$) , as now only representations in $\cup_i {\cal S}^i_1$ appear in any possible recoupling scheme. However as the  $|\iota^N_{{\cal S}^i_1}\rangle$ are not necessarily orthogonal to each other\footnote{Of course the intertwiners can in principle be orthogonalized with the Gram Schmidt procedure.}, the projector property will be only satisfied for specific choices of the coefficients $\alpha_i^N$. Nevertheless it is interesting to consider (\ref{i8}) for general coefficients, as it allows to obtain phase diagrams and in this way to determine phase transitions. The latter might allow to define a continuum limit theory with propagating degrees of freedom.

The main result of this paper is the following: Using these fixed point intertwiners to define (2D) spin net models, these spin nets flow in general to `non--trivial' triangulation invariant models without the necessity of fine tuning. 

Here we mean with `trivial' fixed points, those in which no simplicity constraints occur, and which can be expressed as standard lattice gauge theories. E.g. for the so--called degenerate fixed point only the representation $j=0$ is excited (appearing as high temperature fixed point in the  statistical physics interpretation). Another example are fixed points which can be understood as analogues of $BF$ theories. In these cases the models can all be described by the Haar projector and a choice of edge weights (the analogues of spin foam face weights for the spin nets).

Furthermore we have to explain the notion of fine tuning, as this depends highly on the parametrization that one chooses for the initial models, which are then subjected to the coarse graining flow. To compare the parametrization (\ref{i8}) introduced here and another parametrization used in \cite{bdetal13}   let us explain the latter. It is based on so--called $E$--functions, introduced in \cite{bahretal12}. These $E$--functions encode the simplicity constraints. This parametrization covers all the standard models for spin foams \cite{alexreview,eprl,bc}. In this case the Haar projector is replaced by operators of the form
\ba\label{i9}
\mathcal P^{(N)}\,=\,  \mathcal P_{Haar} \cdot E^{\rho_1}\otimes \cdots  \otimes E^{\rho_N} \cdot \mathcal P_{Haar}
\ea
where $E^{\rho_I}$ denotes a (non--intertwining) map: $ V_{\rho_I} \rightarrow V_{\rho_I}$ on a representation space associated to one of the faces adjacent to the edge in question. This gives another possibility to define consistently spin foam models for vertices or edges with arbitrary valency \cite{warsaw0}. However, for instance for the EPRL model \cite{eprl} the projector property needs to be implemented by an additional construction \cite{warsaw}, in which the factorizing structure of the operators (\ref{i9}) over the representation spaces $V_{\rho_I}$ is a priori changed. 

Indeed the form (\ref{i9})  can be expanded in an intertwiner basis again
\ba\label{i10}
\mathcal P^{(n)}\,=\,  \sum_{d,d'}  |\iota_d\rangle \langle \iota_d |\,  \mathcal P_{Haar} \cdot E^{\rho_1}\otimes \cdots  \otimes E^{\rho_n} \cdot \mathcal P_{Haar}  \,|\iota_{d'} \rangle \langle \iota_{d'} |
\ea
where now in general non--diagonal terms proportional to $|\iota_d\rangle \langle \iota_{d'} |$ appear. The question whether this is a projector turns then into the question whether a diagonaliztion leads to eigenvalues $0$ and $1$ only.

\cite{bdetal13} used the $E$--function parametrization for models based on the structure group $S_3$. (The fusion categories of the permutation group $S_3$ and $SO(3)_{k=4}$ are actually  gauge equivalent as fusion categories\footnote{We thank Oliver Buerschaper for pointing this out.}, hence we can compare the coarse graining results of \cite{bdetal13} with the examples considered here.) A phase space diagram based on this parametrization leads to the following picture: Basically all models given by this parametrization flow either to the high temperature phase or to a `trivial' $S_3$ or ${\mathbb Z}_2$ ordered phase, analogous to either $BF$ theory on $S_3$ or $BF$ theory on the subgroup ${\mathbb Z}_2$ for spin foams. Only by fine tuning to a phase transition, in the case of \cite{bdetal13} between high temperature and $S_3$ ordered phase, one might flow to a non--trivial fixed point. It was this instance of a non--trivial fixed point, which motivated the investigations in \cite{bw1}, which revealed the possibility of a very rich structure of non--trivial fixed points in spin net models. However a posteriori it was rather a case of good fortune to have seen this fixed point at all, as it appeared (a) only for a tiny subset of initial models in the parametrization used, and (b) only for a certain truncation, as higher truncations turned on couplings that would lead to a flow away from this fixed point. 

The picture that we will find for the parametrization (\ref{i8})  introduced in this work will be very different. Rather models flow generically to non--trivial fixed points. Thus with the parametrization of the phase diagram here, the fixed points define a phase, i.e. an extended region in our phase diagram\footnote{A parametrization independent definition of a stable phase would require stability with respect to all possible perturbations of the fixed point (or all possible perturbations respecting a given symmetry \cite{wenclass,schuch}), that is loosely said an extended region in the infinite dimensional phase space encoding all possible models. For practical reasons we will  use  the parametrization dependent definition of  `phases' here and investigate stability properties of the phases in future work.}. We will work with quantum groups $\text{SO}(3)_k$ which makes a systematic investigation for different choices of $k$ possible (bounded by the available computer power of course). This allows us to conjecture that the essential picture will not change in the limit $k\rightarrow \infty$ and therefore holds for general (quantum) groups.

These results give new weight to the point suggested in \cite{warsaw,bojowaldperez}, which is to build models and select measure factors such that  invariance under certain subdivisions of edges and faces is guaranteed. For edge subdivisions this leads to the projector property for the `edge operators'  $\mathcal P$. The investigations here show that models which implement this property (or are in a certain sense near such models) flow rather to non--trivial fixed points. Whereas for models of the form (\ref{i9}) the implementation of the projector property is rather involved \cite{warsaw} and as \cite{bdetal13} shows for the structure group $S_3$, most of these models flow to trivial fixed points.

~\\

Finally let us explain spin net models.
 Very simply whereas for spin foams the $\mathcal P^{(N)}$ are based on edges, for spin nets the $\mathcal P^{(N)}$ are based on vertices and define the vertex weights. The edges are labelled by the representation labels $\rho$, as well as two magnetic indices, labelling a basis in $V_\rho$ and $V_{\rho^*}$ (the dual representation space) respectively. Thus every edge carries a Hilbert space
\ba\label{intro4}
{\cal H}_e\,=\ \oplus_\rho V_{\rho} \otimes V_{\rho^*} \quad ,
\ea
where the sum is over all irreducible representations. The contraction of indices of vertex weights ${\mathcal P}^{(N)}$ and the sum over representation labels $j$  corresponds then to integrating out all degrees of freedom associated to the edges.

Thus the partition function for a spin net is defined as 
\ba\label{intro5}
Z=\sum_{\rho_e,m_e,n_e}  \prod_v \mathcal P_v^{(N_v)} ( \{\rho_e\}_{e\supset v},  \{m_e\}_{e\supset v}, \{n_e\}_{e\supset v})
\ea
where we made the index structure of the intertwiners explicit.

Spin nets can be imagined as dimensional reductions to spin foams, in the sense that one considers a cut orthogonal to a spin foam edge, which now will appear as vertex. Objects which are based on faces in spin foams, are based on edges in spin nets. Spin nets were introduced in \cite{eckert} to allow numerical coarse graining investigations for 2D models, which is naturally much simpler than considering 4D models. In addition, for (standard) lattice gauge theory, a certain duality exist between 2D `Ising type' models and 4D lattice gauge theory models. In many examples statistical properties are similar between a given 4D lattice gauge model and the corresponding Ising type analogue 2D model. Spin net models have been defined based on the same principle, therefore we can hope that some of the findings on the phase structure of spin nets hold also for the 4D spin foam models.

In section \ref{interpretation} we will show that the partition function for a spin net can be understood as the partition function for a melonic spin foam with only two vertices and a large number of edges and faces (with two edges). This will allow us to interpret the renormalization flow of spin nets as a renormalization flow of spin foams.

\section{Quantum group $\text{SU}(2)_k$}
\label{quantum-group}

\subsection{Basic ingredients}
Before defining our quantum group spin net models, based on the quantum deformation of the rotation group, we need some basic tools that we summarize in this section.
We follow \cite{biedenharn} in notation and conventions, where the interested reader can find a thorough introduction to $\text{SU}(2)_k$, see also \cite{yellowbook}.

Following \cite{biedenharn} we understand the quantum group $\text{SU}(2)_k$ as the q-deformation ${\cal U}_q(su(2))$ of the universal enveloping algebra $U(su(2))$ of the Lie algebra $su(2)$. The algebra ${\cal U}_q(su(2))$ is generated by three operators $J_{\pm},J_z$ with commutation relations
\ba
[J_z , J_{\pm} ]&=& \pm J_\pm       \nn\\
~ [J_+ , J_-] &=& \frac{q^{J_z}-q^{-J_z}}{q^{1/2}-q^{-1/2}} \q .
\ea

As for the classical case, the finite dimensional representations of $\text{SU}(2)_k$ are labelled by half integers $j$ and can be defined on $(2j+1)$ dimensional representation spaces $V_j$.    The so--called quantum dimensions are given by
\ba
d_j=[2j+1]  \q ,
\ea
where 
\ba
[n]&=& \frac{q^{\frac{n}{2}}-q^{-\frac{n}{2}}  }{q^{\frac{1}{2}}-q^{-\frac{1}{2}}  }   
\ea
are the so--called quantum numbers. Here $q$ can be a root of unity or $q\in \mathbb{R}/\{0\}$.
For $q$ a root of unity, $q=\exp(\frac{2\pi}{(k+2)}i)$, the quantum numbers are periodic 
\ba
[n]&=&\frac{\sin(\frac{2\pi i n}{2k+4})}{\sin(\frac{2\pi i}{2k+4})}  \q ,
\ea
having zeros at $n=0$ and $n=k+2$. Thus $j=\frac{k}{2}$ with $d_{k/2}=1$ is the `last' representation with a strictly positive quantum dimension. Representations $j=0,\frac{1}{2},\ldots,\frac{k}{2}$ are called admissible, representations $j>\frac{k}{2}$ are of so--called quantum trace zero. The number $k\in\mathbb{N}$ is referred to as level.

Assuming two  representations $V_{j_1},V_{j_2}$ the tensor product is defined via the co--product $\Delta$. The action of the $\text{SU}(2)_k$ algebra on $V_{j_1}\otimes V_{j_2}$ is defined as
\ba
\Delta(J_\pm) &=& q^{-J_z/2} \otimes J_{\pm} \,+\, J_\pm  \otimes q^{J_z/2} \nn\\
\Delta(J_z) &=& \mathbb{I} \otimes J_z \,+\, J_z\otimes \mathbb{I} \q .
\ea 
As in the classical case we can reduce the tensor product $V_{j_1}\otimes V_{j_2}$ into a direct sum of irreducible representations and a part consisting of trace zero representations (which will be modded out). Assuming orthogonal bases $|j,m\rangle$ in the representation spaces, we can describe this by Clebsch-Gordan coefficients
\ba
|j,m\rangle &=&\sum_{m_1,m_2} {}_qC^{j_1j_2j}_{m_1m_2 m} \,|j_1m_1\rangle \otimes |j_2,m_2\rangle \q .
\ea

When coupling three admissible representations $j_I$, $j_K$ and $j_L$ in this way, the coupling coefficients are only non--vanishing if the following conditions hold:
\ba\label{couplingcond}
j_I+j_K &\geq& j_L \q\text{for all permutations} \; \{J,K,L\} \;\text{of}\; \{1,2,3\} \q ,\nn\\
j_1+j_2+j_3 &=& 0\!\mod 1 \q , \nn\\
j_1+j_2+j_3 &\leq& k \q .
\ea
Note the last condition in (\ref{couplingcond}), that is special to the quantum deformed case at root of unity. This condition signifies that  $V_{j_1}\otimes V_{j_2}$ can include trace zero parts. These trace zero parts can be modded out \cite{yellowbook}. Some equations (for instance the one defining the $[6j]$ symbol) will however hold only modulo such trace zero parts  \cite{yellowbook}. 

In particular we have the completeness relation
\ba\label{complete}
\sum_{m_3,\, j_3 \, \text{admiss.}} {}_qC^{j_1j_2j_3}_{m_1m_2m_3} \,\, {}_qC^{j_1j_2j_3}_{m'_1m'_2m_3}&=&\Pi^{j_1 j_2}_{m_1m_2\,,m'_1m'_2}
\ea
where $\Pi^{j_1 j_2}_{m_1m_2\,,m'_1m'_2}$ projects out the trace zero part in $V_{j_1}\otimes V_{j_2}$. The orthogonality relation for the Clebsch-Gordan coefficients is given as
\ba\label{ortho}
\sum_{m_1,m_2} {}_qC^{j_1j_2j}_{m_1m_2m} \,{}_qC^{j_1j_2j'}_{m_1m_2m'}\,\,=\,\,\delta_{jj'}\delta_{mm'} \theta_{j_1j_2j}
\ea
where $\theta_{j_1j_2j}=1$ if the coupling conditions (\ref{couplingcond}) are satisfied and vanishing otherwise.

When defining our spin net models we will restrict to integer $j$, that is to $\text{SO}(3)_k$ representations.

\subsection{Diagrammatic Calculus}

The quantum group introduces certain subtleties into the definition, for instance the notion of dual needs to be specified. To this end we will use diagrammatic calculus, which is just a convenient form of representing combinations of Clebsch-Gordan coefficients.  
This can indeed be seen as an embellished form of tensor network calculus. The quantum group requires to specify a special direction, which we will take as the vertical direction. All graphs can then be interpreted as representing maps from a tensor product of representation spaces of $\text{SU}(2)_k$, represented by lines coming in from below, to a tensor product of representation spaces, drawn as lines going out on top. Each of these lines carries a representation label $j$ and a magnetic index $m$. One basic example of such a map are the Clebsch-Gordan coefficients, denoted by $_q\mathcal{C}_{m_1\,m_2\,m_3}^{j_1 \, j_2 \, j_3}$\footnote{This is not the standard Clebsch-Gordan coefficient defined in \cite{biedenharn}, but it is modified by the quantum dimension: ${}_q\mathcal{C}_{m_1\,m_2\,m_3}^{j_1 \, j_2 \, j_3} = {}_q C_{m_1\,m_2\,m_3}^{j_1 \, j_2 \, j_3 } \left(\sqrt{d_{j_3}}\right)^{-1}$.}. They are interpreted as a map $V_{j_1} \otimes V_{j_2} \rightarrow V_{j_3}$, symbolizing how the spins $j_1$ and $j_2$ (with their respective magnetic indices) couple to $j_3$. We represent them graphically as follows:
\begin{equation}\label{eq:clebsch}
\begin{tikzpicture} [baseline,scale=0.75]
\draw (0,-0.75) node {$j_1$}
      (0,-0.5) -- (0.5,0) -- (1,-0.5)
      (1,-0.75) node {$j_2$}
      (0.5,0) -- (0.5,0.5) 
      (0.5,0.75) node {$j_3$};
\end{tikzpicture}
:= {}_q\mathcal{C}_{m_1\,m_2\,m_3}^{j_1 \, j_2 \, j_3} \quad .
\end{equation}
A particular version of this Clebsch-Gordan coefficient will be important later on: If we choose $j_1 = j_2 = j$ and take $j_3 = 0$, we define the `cap' as a map: $V_j\otimes V_j\rightarrow {\mathbb C}$, namely
\begin{equation} 
\begin{tikzpicture} [baseline,scale=0.75]
\draw (0,-0.75) node {$m$}
      (0,-0.5) -- (0,0) 
      (1,0) arc(-0:180:0.5)
      (0.5,0.75) node {$j$}
      (1,0) -- (1,-0.5)
      (1,-0.75) node {$m'$};
\end{tikzpicture}
:= {}_q \mathcal{C}^{j\,j\,0}_{m\,m'\,0} \sqrt{d_j} =(-1)^{j-m} q^{\frac{m}{2}} \delta_{m,-m'} \quad.
\end{equation}
From this `cap' we can similarly define a `cup' by requiring that they give the identity if we concatenate them:
\begin{equation}\label{identity}
\begin{tikzpicture}[baseline,scale=0.75]
\draw (0,-0.75) node {$m$}
      (0,-0.5) -- (0,0) 
      (1,0) arc(-0:180:0.5)
      (1,0) -- (1,-0.1)
      (1,-0.1) arc(-180:0:0.5)
      (2,-0.1) -- (2,0.6)
      (2,0.85) node {$m''$};
\end{tikzpicture}
= 
\begin{tikzpicture} [baseline,scale=0.75]
\draw (0,-0.75) node {$m$}
      (0,-0.5) -- (0,0.6)
      (0,0.85) node {$m''$};
\end{tikzpicture}
= \delta_m^{m''} \quad ,
\end{equation}
which gives:
\begin{equation}
\begin{tikzpicture}[baseline,scale=0.75]
\draw (0,0.75) node {$m$}
      (0,0.5) -- (0,0)
      (0,0) arc(-180:0:0.5)
      (1,0) -- (1,0.5)
        (0.5,-0.75) node {$j$}
      (1,0.75) node {$m'$};
\end{tikzpicture}
= (-1)^{j+m} q^{\frac{m}{2}} \delta_{m,-m'} \quad .
\end{equation}
Using these `cups' and `caps', we can construct the Clebsch-Gordan coefficients for the quantum group with inverse (here: complex conjugate) deformation parameter $\bar{q}$ by `bending up' one of the lower legs of the Clebsch-Gordan in \eqref{eq:clebsch}.
\begin{equation}
{}_{\bar{q}} \mathcal{C}_{m_1\, m_2 \, m_3}^{j_1 \, j_2 \, j_3} =
\begin{tikzpicture}[baseline,scale=0.75]
\draw (0,-0.55) -- (0,0) -- (-0.5,0.5)
      (0,0) -- (0.5,0.5)
      (0,-0.8) node {$j_3$}
      (-0.5,0.75) node {$j_2$}
      (0.5,0.75) node {$j_1$};
\end{tikzpicture}
\; = \;
\begin{tikzpicture}[baseline,scale=0.75]
\draw (-0.5,-1) -- (-0.5,-0.5) -- (0,0)
      (0,0) -- (0,0.7)
      (0,0) -- (0.5,-0.5)
      (0.5,-0.5) arc (-180:0:0.5)
      (1.5,-0.5) -- (1.5,0.7)
      (-0.5,-1.25) node {$j_3$}
      (0,0.95) node {$j_2$}
      (1.5,0.95) node {$j_1$};
\end{tikzpicture}
\; = \;
\begin{tikzpicture} [baseline,scale=0.75]
\draw (-1.5,0.7) -- (-1.5,-0.5)
      (-1.5,-0.5) arc (-180:0:0.5)
      (-0.5,-0.5) -- (0,0) -- (0,0.7)
      (0,0) -- (0.5,-0.5) -- (0.5,-1)
      (-1.5,0.95) node {$j_2$}
      (0,0.95) node {$j_1$}
      (0.5,-1.25) node {$j_3$};
\end{tikzpicture} \quad .
\end{equation}
This map can hence be interpreted as mapping $V_{j_3} \rightarrow V_{j_1} \otimes V_{j_2}$, thus it is  dual to \eqref{eq:clebsch}. With a `cap' we can `pull down' one of the legs again and arrive back at \eqref{eq:clebsch}:
\begin{equation}
\begin{tikzpicture} [baseline,scale=0.75]
\draw (-0.5,0.5) -- (0,0) -- (0,-0.7)
      (0,0) -- (0.5,0.5) 
      (1.5,0.5) arc(0:180:0.5)
      (1.5,0.5) -- (1.5,-0.7)
      (-0.5,0.75) node {$j_5$}
      (0,-0.95) node {$j_3$}
      (1.5,-0.95) node {$j_4$};
\end{tikzpicture}
\; = \;
\begin{tikzpicture} [baseline,scale=0.75]
\draw (-0.5,-0.5) -- (0,0) -- (0,0.7)
      (0,0) -- (0.5,-0.5)
      (-0.5,-0.75) node {$j_3$}
      (0.5,-0.75) node {$j_4$}
      (0,0.95) node {$j_5$};
\end{tikzpicture}
\; = \;
\begin{tikzpicture} [baseline,scale=0.75]
\draw (-1.5,-0.7) -- (-1.5,0.5) 
      (-0.5,0.5) arc(0:180:0.5)
      (-0.5,0.5) -- (0,0) -- (0,-0.7)
      (0,0) -- (0.5,0.5)
      (-1.5,-0.95) node {$j_3$}
      (0,-0.95) node {$j_4$}
      (0.5,0.75) node {$j_5$};
\end{tikzpicture} \quad .
\end{equation}
Concatenating these two maps, we obtain a map $V_{j_3} \rightarrow V_{j_3}$ proportional to the identity.
\begin{equation}
\begin{tikzpicture} [baseline,scale=0.75]
\draw (0,-1) -- (0,-0.5) -- (-0.5,0) -- (0,0.5) -- (0,1)
      (0,-0.5) -- (0.5,0) -- (0,0.5)
      (0,-1.25) node {$j_3$}
      (-0.75,0) node {$j_1$}
      (0.75,0) node {$j_2$}
      (0,1.25) node {$j_3$};
\end{tikzpicture}
\; = \;
\begin{tikzpicture}[baseline,scale=0.75]
\draw (-1,-1) -- (-0.5,-0.5) -- (-0.5,0) -- (0,0.5) -- (0,1)
      (-0.5,-0.5) -- (0,-1) arc(-180:0:0.25) -- (0.5,0) -- (0,0.5)
      (-1,-1.25) node {$j_3$}
      (-0.75,0) node {$j_1$}
      (0.75,0) node {$j_2$}
      (0,1.25) node {$j_3$};
\end{tikzpicture}
\; = \; (-1)^{j_1 + j_2 - j_3} d_{j_3}^{-1} \delta_{m_3\,m'_3} \quad.
\end{equation}

Now we have all the necessary graphical ingredients to express the $\text{SU}(2)_k$ $6j$ symbol graphically: As for $\text{SU(2)}$, the $6j$ symbol is a particular contraction of four Clebsch-Gordan coefficients, which gives a purely real number. Graphically this is expressed in a closed diagram as follows:
\begin{equation}\label{6j-def}
\begin{tikzpicture}[baseline,scale=0.6]
\draw (0,-2) -- (0,-1.5) -- (-0.5,-1) -- (-0.5,-0.5) -- (0.5,0.5) -- (0.5,1) -- (0,1.5) -- (0,2) arc(180:0:1) -- (2,-2) arc(0:-180:1)
      (0.,-1.5) -- (1,-0.5) -- (1,0) -- (0.5,0.5)
      (-0.5,-0.5) -- (-1,0) -- (-1,0.5) -- (0,1.5)
      (-0.75,-1) node {$j_1$}
      (1.3,-0.5) node {$j_2$}
      (-1.25,0) node {$j_3$}
      (0.8,1) node {$j_4$}
      (2.3,0) node{$j_5$}
      (-0.2,0.25) node {$j_6$};
\end{tikzpicture}
\; = \;
\begin{tikzpicture}[baseline,scale=0.6]
\draw (0,-2) -- (0,-1.5) -- (-1,-0.5) -- (-1,0) -- (-0.5,0.5) -- (-0.5,1) -- (0,1.5) -- (0,2) arc(180:0:1) -- (2,-2) arc(0:-180:1)
      (0.,-1.5) -- (0.5,-1) -- (0.5,-0.5) -- (1,0) -- (1,0.5) -- (0,1.5)
      (0.5,-0.5) -- (-0.5,0.5)
      (-1.25,-0.5) node {$j_1$}
      (0.8,-1) node {$j_2$}
      (1.3,0) node {$j_4$}
      (-0.75,1) node {$j_3$}
      (2.3,0) node {$j_5$}
      (0.2,0.25) node {$j_6$};
\end{tikzpicture}
\; = \; 
\left \{
\begin{matrix}
\,j_1\, & \,j_2\, & \,j_5\, \\
\,j_4\, & \,j_3\, & \,j_6\,
\end{matrix}
\right \} =\frac{(-1)^{j_1+j_2+j_3+j_4}}{\sqrt{d_{j_5}d_{j_6}}}\left[
\begin{matrix}
\, j_1 \, & \, j_2 \,  & \, j_5 \, \\
\, j_4 \, & \, j_3 \, & \, j_6 \,
\end{matrix}
\right] \quad .
\end{equation}

\section{Structure of quantum group spin net models}

\label{structurespinnets}

A spin net model can be understood as a vertex model or as a tensor network model. In both cases weights are associated to the vertices of a graph -- in the case of the tensor network model the weights, which depend on variables attached to edges, are given as tensors where the variables are now indices of these tensors. The partition function is expressed as a sum over the variables attached to the edges. For the tensor network this amounts to contracting tensors with each other according to the connectivity of the underlying graph.

We will consider a regular 4--valent graph here. As proposed in \cite{eckert,bahrdittrich}, and the results in \cite{he,bdetal13,bw1} show, coarse graining on a regular lattice might be sufficient to regain fully triangulation invariant models.\footnote{Coarse graining on a regular square lattice in a very regular way might lead to so--called CDL fixed points of the renormalization flow, that are non--isolated \cite{guwen,levinslides}. As pointed out in \cite{bw1} these do not define fully triangulation invariant models-- however each of the non--isolated families of fixed points defines one triangulation invariant model. As is noted in \cite{guwen,sherbrook}, changing the renormalization procedure, to include a so--called entanglement filtering, would reduce these families automatically to the triangulation invariant fixed point.  This could be obtained by including apart from refining and coarse graining moves also entangling moves into the renormalization algorithm \cite{refining,vidal}.}

We will introduce here  spin net models with quantum group symmetry, i.e. the vertex weights will be invariant under a certain quantum group symmetry. These are much easier to deal with in the spin representation, where the (oriented) edges carry labels $j,m,n$ which corresponds to the association of a Hilbert space
\ba\label{s1}
{\cal H}_e=\oplus_j V_j \otimes V_{j^*}
\ea
to a given edge $e$.

\begin{figure}
\begin{center}
\includegraphics[width=0.2\textwidth]{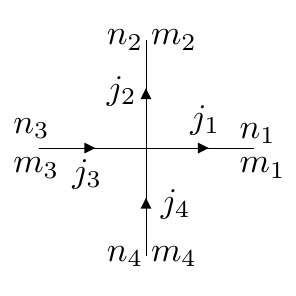}
\caption{Vertex in the square lattice. Edges are ordered anticlockwise and have fixed orientation: edges 1 and 2 are outgoing, and edges 3 and 4 are incoming.   \label{fig:vertex}}
\end{center}
\end{figure}

Here
\begin{itemize}\parskip-1mm
\item $j$ denotes an irrep of the quantum group $\text{SO}(3)_k$ and $m,n$ are labelling elements of a basis in the representation space $V_j\otimes V_{j^*}$. 
\item For a representation in $\text{SO}(3)_k$ we have  $j \in \mathbb{N}$. Additionally, we will restrict to admissible representations, namely $j \leq k/2$ for $k$ even and $j\leq (k-1)/2$ for $k$ odd.
\item $j^*$ denotes the dual representation to $j$, which for $\text{SO}(3)_k$ is unitarily equivalent to $j$. 
\item An orientation of the edges is needed as the action of the symmetry on a vertex depends on whether edges are ingoing or outgoing. We will choose an orientation and labelling of edges as in figure \ref{fig:vertex}. 
\end{itemize}

As said before, the quantum group requires to specify a special direction, which we have taken as vertical direction. For the horizontal direction we will assume an infinitely extended lattice. Alternatively one can consider periodic boundary conditions, however this requires braiding, see \cite{bw1,vidalsymm}. 
It turns out that due to the specification of the vertical direction as special, we have to amend the gluing rule for the horizontal direction. This will be explained further below in section \ref{coarse}. Modulo this subtlety the tensor network contraction can be understood as concatenation of (intertwiner) maps.

As mentioned, to implement the (quantum) group symmetry, the vertex weights have to be intertwining maps with respect to the representations involved. Such intertwining maps can be obtained by projecting an arbitrary tensor with the appropriate index structure from the `left' and the `right' with the Haar projector, which in the case of groups is given by (\ref{ghaar}). This can be expanded into Clebsch-Gordan coefficients, see \cite{bdetal13}. As then no group elements appear, this form is also the appropriate one for quantum groups. Indeed we will be defining a Haar projector with the help of diagrammatic calculus.  This will in particular help to resolve subtleties regarding the definition of the dual representation $j^*$ for the quantum group. The technical difficulties are due to the fact that the dual of a dual is actually not given by the identity, but by a phase factor. A clean definition of 'dual' is therefore crucial for a consistent definition of the models.

\subsection{Haar projector}

The four--valent Haar projector naturally splits into two three--valent Haar projectors, which are connected by a sum over a representation $j_5$. This index actually gives the intertwiner label $d$ in (\ref{shaar}), which for the four--valent projector under consideration reads
\ba\label{4Haar}
\mathcal{P}_{\left(\{m\},\{m'\}\right)}(j_1,j_2,j_3,j_4) &=& \sum_{j_5} \iota^{j_5}_{\{m\}}(j_1,j_2,j_3,j_4){\iota^*}^{j_5}_{\{m'\}}(j_1,j_2,j_3,j_4)
\quad .
\ea
The following contraction of Clebsch-Gordan coefficients, associated to the recoupling scheme $j_1\otimes j_2 \ni j_5\in j_3\otimes j_4$, gives the intertwiner  $\iota^{j_5}_{\{m\}}(j_1,j_2,j_3,j_4)$ up to normalization:
\begin{equation}
\begin{tikzpicture}[baseline,scale=0.75]
\draw (-0.5,-1) -- (0,-0.5) -- (0,0.5) -- (-0.5,1)
      (0.5,-1) -- (0,-0.5)
      (0,0.5) -- (0.5,1)
      (-0.5,-1.25) node {$j_3$}
      (0.5,-1.25) node {$j_4$}
      (0.5,1.25) node {$j_1$}
      (-0.5,1.25) node {$j_2$}
      (0.25,0) node {$j_5$};
\end{tikzpicture}
\; = \; \sum_{m_5} {}_{\bar{q}} \mathcal{C}^{j_1\,j_2\,j_5}_{m_1\,m_2\,m_5} \; {}_q \mathcal{C}^{j_3\,j_4\,j_5}_{m_3\,m_4\,m_5} \quad .
\end{equation}
Now, with the help of the previously defined `cups' and `caps', we can define its dual:
\begin{equation}\label{dualize}
\begin{tikzpicture}[baseline,scale=0.75]
\draw (-0.25,-0.75) -- (0,-0.5) -- (0,0.5) -- (-0.25,0.75) arc(0:180:0.25)
      (0.25,-0.75) -- (0,-0.5)
      (0,0.5) -- (0.25,0.75) arc(0:180:0.75)
      (-0.25,-0.75) arc(0:-180:0.25)
      (0.25,-0.75) arc(0:-180:0.75)
      (-0.75,0.5) node {$j_1$}
      (-1.25,0.5) node {$j_2$}
      (-0.75,-0.5) node {$j_4$}
      (-1.25,-0.5) node {$j_3$}
      (0.25,0) node {$j_5$};
\end{tikzpicture}
\; = \;
\begin{tikzpicture}[baseline,scale=0.75]
\draw (-0.5,0.75) -- (-0.25,1) arc(180:0:0.5) -- (0.75,-1) arc(0:-180:0.5) -- (-0.5,-0.75)
      (-0.25,1) -- (0,0.75)
      (-0.25,-1) -- (0,-0.75)
      (-0,0.5) node {$j_1$}
      (-0.5,0.5) node {$j_2$}
      (-0,-0.5) node {$j_4$}
      (-0.5,-0.5) node {$j_3$}
      (1,0) node {$j_5$};
\end{tikzpicture}
\; = \; (-1)^{2 j_5} \sum_{m_5}\; q^{m_5} \; {}_{\bar{q}} \mathcal{C}^{j_1 \, j_2 \, j_5}_{m_1 \, m_2 \, m_5} \; {}_q \mathcal{C}^{j_3 \, j_4 \, j_5}_{m_3 \, m_4 \, m_5}
\quad ,
\end{equation}
which is equal to ${\iota^*}^{j_5}_{\{m'\}}(j_1,j_2,j_3,j_4)$, again up to normalization. Substituting in Eq. \eqref{4Haar} and demanding that $\mathcal{P}$ is a projector, namely 
$\mathcal{P}^2 = \mathcal{P}$, we fix the normalization. The result is (see the proof in Appendix \ref{app-Haar})\footnote{To alleviate the notation we have omitted the magnetic indices attached to the diagrams. Here the left diagram carries $\{m\}$ indices and the right diagram carries $\{m'\}$ indices.}
\begin{equation}\label{haarig}
\mathcal{P}_{\left(\{m\},\{m'\}\right)}(j_1,j_2,j_3,j_4) = \sum_{j_5} (-1)^{j_1+j_2+j_3+j_4} d_{j_5}
\begin{tikzpicture}[baseline,scale=0.75]
\draw (-0.5,-1) -- (0,-0.5) -- (0,0.5) -- (-0.5,1)
      (0.5,-1) -- (0,-0.5)
      (0,0.5) -- (0.5,1)
      (-0.5,-1.25) node {$j_3$}
      (0.5,-1.25) node {$j_4$}
      (0.5,1.25) node {$j_1$}
      (-0.5,1.25) node {$j_2$}
      (0.25,0) node {$j_5$};
\end{tikzpicture}
\; \otimes \;
\begin{tikzpicture}[baseline,scale=0.75]
\draw (-0.5,0.75) -- (-0.25,1) arc(180:0:0.5) -- (0.75,-1) arc(0:-180:0.5) -- (-0.5,-0.75)
      (-0.25,1) -- (0,0.75)
      (-0.25,-1) -- (0,-0.75)
      (-0,0.5) node {$j_1$}
      (-0.5,0.5) node {$j_2$}
      (-0,-0.5) node {$j_4$}
      (-0.5,-0.5) node {$j_3$}
      (1,0) node {$j_5$};
\end{tikzpicture}\q .
\end{equation}
Note that for the Haar projector (\ref{haarig}) we have the same intertwiner label $j_5$ appearing for the left and right (dualized) copy. This will actually change for the spin net models introduced in (\ref{split1}). There basis elements $j_5\neq j'_5$ appear in the recoupling scheme.

The definition of this Haar projector corresponds to a particular recoupling scheme: $j_1\otimes j_2 \ni j_5\in j_3\otimes j_4$, associated to a vertical splitting of the 4--valent vertex in two three--valent ones.
Similarly we can define the Haar projector for the recoupling scheme: $j_2\otimes j_3 \ni j_6\in j_4\otimes j_1$, which leads to a horizontal splitting of the 4--valent vertex. Instead of repeating the same procedure, we can derive it from the previous Haar projector. First we define:
\begin{equation}\label{eq:other-splitting}
\begin{tikzpicture}[baseline,scale=0.75]
\draw (-1,-0.75) -- (-1,-0.25) -- (-0.5,0.25) -- (-0.5,0.75)
      (-1,-0.25) -- (-1.5,0.25) -- (-1.5,0.75)
      (-0.5,0.25) -- (0,-0.25) -- (0,-0.75)
      (-1,-1) node {$j_3$}
      (-1.5,1) node {$j_2$}
      (0,-1) node {$j_4$}
      (-0.5,1) node {$j_1$}
      (-0.8,0.25) node {$j_6$};
\end{tikzpicture}
\; = \;
\begin{tikzpicture}[baseline,scale=0.75]
\draw (1,-0.75) -- (1,-0.25) -- (0.5,0.25) -- (0.5,0.75)
      (1,-0.25) -- (1.5,0.25) -- (1.5,0.75)
      (0.5,0.25) -- (0,-0.25) -- (0,-0.75)
      (0,-1) node {$j_3$}
      (1,-1) node {$j_4$}
      (0.5,1) node {$j_2$}
      (1.5,1) node {$j_1$}
      (0.82,0.2) node {$j_6$};
\end{tikzpicture}
\; =: \;
\begin{tikzpicture}[baseline,scale=0.75]
\draw (-1,0.5) -- (-0.5,0) -- (0,0) -- (0.5,0.5)
      (-0.5,0) -- (-1,-0.5)
      (0,0) -- (0.5,-0.5)
      (-1,-0.75) node {$j_3$}
      (-1,0.75) node {$j_2$}
      (0.5,0.75) node {$j_1$}
      (0.5,-0.75) node {$j_4$}
      (-0.25,0.25) node {$j_6$};
\end{tikzpicture}
\quad .
\end{equation}
These two different splittings (or recoupling schemes) are not independent. It is straightforward to realize that  (see proof in Appendix \ref{app-recouplings})
\begin{align} \label{eq:rel-splitting}
\begin{tikzpicture}[baseline,scale=0.75]
\draw (-1,0.5) -- (-0.5,0) -- (0,0) -- (0.5,0.5)
      (-0.5,0) -- (-1,-0.5)
      (0,0) -- (0.5,-0.5)
      (-1,-0.75) node {$j_3$}
      (-1,0.75) node {$j_2$}
      (0.5,0.75) node {$j_1$}
      (0.5,-0.75) node {$j_4$}
      (-0.25,0.25) node {$j_6$};
\end{tikzpicture}
\; & = \; \sum_{j_5}  \sqrt{\frac{d_{j_5}}{d_{j_6}}} 
\left[
\begin{matrix}
\, j_1 \, & \, j_2 \, & \, j_5 \, \\
\, j_3 \, & \, j_4 \, & \, j_6 \,
\end{matrix}
\right] 
\begin{tikzpicture}[baseline,scale=0.75]
\draw (-0.5,-0.75) -- (0,-0.25) -- (0,0.25) -- (-0.5,0.75)
      (0.5,-0.75) -- (0,-0.25)
      (0,0.25) -- (0.5,0.75)
      (-0.75,-0.75) node {$j_3$}
      (0.75,-0.75) node {$j_4$}
      (0.75,0.75) node {$j_1$}
      (-0.75,0.75) node {$j_2$}
      (0.25,0) node {$j_5$};
\end{tikzpicture}
\quad .
\end{align}
Similarly one can derive an analogue relation to arrive back at the previous splitting.
\begin{equation}
\begin{tikzpicture}[baseline,scale=0.75]
\draw (-0.5,-0.75) -- (0,-0.25) -- (0,0.25) -- (-0.5,0.75)
      (0.5,-0.75) -- (0,-0.25)
      (0,0.25) -- (0.5,0.75)
      (-0.75,-0.75) node {$j_3$}
      (0.75,-0.75) node {$j_4$}
      (0.75,0.75) node {$j_1$}
      (-0.75,0.75) node {$j_2$}
      (0.25,0) node {$j_5$};
\end{tikzpicture}
\; = \;
\sum_{j_6} \sqrt{\frac{d_{j_6}}{d_{j_5}}}
\left[
\begin{matrix}
\, j_1 \, & \, j_2 \, & \, j_5 \, \\
\, j_3 \, & \, j_4 \, & \, j_6 \,
\end{matrix}
\right]
\begin{tikzpicture}[baseline,scale=0.75]
\draw (-1,0.5) -- (-0.5,0) -- (0,0) -- (0.5,0.5)
      (-0.5,0) -- (-1,-0.5)
      (0,0) -- (0.5,-0.5)
      (-1,-0.75) node {$j_3$}
      (-1,0.75) node {$j_2$}
      (0.5,0.75) node {$j_1$}
      (0.5,-0.75) node {$j_4$}
      (-0.25,0.25) node {$j_6$};
\end{tikzpicture}
\quad ,
\end{equation}
because of the orthogonality relation of the $[6j]$ symbol:
\begin{equation}
\sum_{j_5} 
\left[
\begin{matrix}
\, j_1 \, & \, j_2 \, & \, j_5 \, \\
\, j_3 \, & \, j_4 \, & \, j_6 \,
\end{matrix}
\right]
\left[
\begin{matrix}
\, j_1 \, & \, j_2 \, & \, j_5 \, \\
\, j_3 \, & \, j_4 \, & \, j'_6 \,
\end{matrix}
\right]
= \delta_{j_6,j'_6} \cdot (\text{admissibility cond.})
\quad .
\end{equation}
Now by dualizing \eqref{eq:other-splitting}, one can analogously define a Haar projector for the horizontal splitting:
\begin{equation}
\mathcal{P}_{\left(\{m\},\{m'\}\right)}(j_1,j_2,j_3,j_4)= \sum_{j_6} (-1)^{j_1+j_2+j_3+j_4} d_{j_6}
\begin{tikzpicture}[baseline,scale=0.75]
\draw (-1,0.5) -- (-0.5,0) -- (0,0) -- (0.5,0.5)
      (-0.5,0) -- (-1,-0.5)
      (0,0) -- (0.5,-0.5)
      (-1,-0.75) node {$j_3$}
      (-1,0.75) node {$j_2$}
      (0.5,0.75) node {$j_1$}
      (0.5,-0.75) node {$j_4$}
      (-0.25,0.25) node {$j_6$};
\end{tikzpicture}
\; \otimes \;
\begin{tikzpicture}[baseline,scale=0.75]
\draw (-1,0.5) -- (-0.5,0) -- (0,0) -- (0.5,0.5)
      (-0.5,0) -- (-1,-0.5)
      (0,0) -- (0.5,-0.5)
        (0.5,0.5) arc(0:180:1.25)
      (-1,0.5) arc(0:180:0.25)
      (-1,-0.5) arc(0:-180:0.25)
      (0.5,-0.5) arc(0:-180:1.25)
      (-1.5,-0.25) node {$j_4$}
      (-1.5,0.25) node {$j_1$}
      (-2,0.25) node {$j_2$}
      (-2,-0.25) node {$j_3$}
      (-0.25,0.25) node {$j_6$};
\end{tikzpicture}
\quad .
\end{equation}

\subsection{Vertex weights}

Attached to each vertex, we consider weights $t(\{j\},\{m\},\{n\})$. The partition function for the model is then (basically) defined by summing over all indices attached to the edges. As mentioned previously, there is a subtlety of how to contract the horizontal edges -- here one has to include a certain phase. This will be explained in detail in section \ref{coarse}.

With the Haar projector ${\mathcal P}$ at hand it is straightforward to characterize the vertex weights, by the condition
\ba\label{s2}
t(\{j\},\{m\},\{n\})\,=\sum_{\{m'\},\{n'\}} \mathcal{P}_{\left(\{m\},\{m'\}\right)}(\{j\}) \,\,t(\{j\},\{m'\},\{n'\}) \,\, \mathcal{P}_{\left(\{n'\},\{n\}\right)}(\{j\}) \q ,
\ea
namely $t(\{j\},\{m\},\{n\})$ is invariant under the action of $\mathcal{P}$ from the left and from the right. This gauge invariance is the quantum group symmetry we referred to at the beginning of this section. It implies that if $j_1\otimes j_2\otimes j_3\otimes j_4$ does not couple to the identity, then the tensor $t(\{j\},\{m\},\{n\})$ vanishes. To exploit this quantum group symmetry and avoid redundancies in the representation (\ref{s2}), we will reexpress the vertex weights in the intertwiner basis. As we will see this is advantageous for the coarse graining algorithm.

Indeed, by using the intertwiner basis, we can write \footnote{Recall that, for the sake of clarity in the notation, we have omitted the magnetic indices in the diagrams. In Eqs. \eqref{split1}-\eqref{split2}, the diagrams on the left carry $\{m\}$ indices, while the diagrams on the right carry $\{n\}$ indices.}
\begin{align}
t(\{j\},\{m\},\{n\})& = \sum_{j_5,j'_5} \hat{t}^{(j_5,j'_5)}_1(j_1,j_2;j_3,j_4) \; d_{j_5} d_{j'_5}
\begin{tikzpicture}[baseline,scale=0.75]
\draw (-0.5,-1) -- (0,-0.5) -- (0,0.5) -- (-0.5,1)
      (0.5,-1) -- (0,-0.5)
      (0,0.5) -- (0.5,1)
      (-0.5,-1.25) node {$j_3$}
      (0.5,-1.25) node {$j_4$}
      (0.5,1.25) node {$j_1$}
      (-0.5,1.25) node {$j_2$}
      (0.25,0) node {$j_5$};
\end{tikzpicture}
\; \otimes \;
\begin{tikzpicture}[baseline,scale=0.75]
\draw (-0.5,0.75) -- (-0.25,1) arc(180:0:0.5) -- (0.75,-1) arc(0:-180:0.5) -- (-0.5,-0.75)
      (-0.25,1) -- (0,0.75)
      (-0.25,-1) -- (0,-0.75)
      (-0,0.5) node {$j_1$}
      (-0.5,0.5) node {$j_2$}
      (-0,-0.5) node {$j_4$}
      (-0.5,-0.5) node {$j_3$}
      (1,0) node {$j'_5$};
\end{tikzpicture}\label{split1}\\
&= \sum_{j_6,j'_6} \hat{t}^{(j_6,j'_6)}_2(j_1,j_4;j_2,j_3) \; d_{j_6} d_{j'_6}
\begin{tikzpicture}[baseline,scale=0.75]
\draw (-1,0.5) -- (-0.5,0) -- (0,0) -- (0.5,0.5)
      (-0.5,0) -- (-1,-0.5)
      (0,0) -- (0.5,-0.5)
      (-1,-0.75) node {$j_3$}
      (-1,0.75) node {$j_2$}
      (0.5,0.75) node {$j_1$}
      (0.5,-0.75) node {$j_4$}
      (-0.25,0.25) node {$j_6$};
\end{tikzpicture}
\; \otimes \;
\begin{tikzpicture}[baseline,scale=0.75]
\draw (-1,0.5) -- (-0.5,0) -- (0,0) -- (0.5,0.5)
      (-0.5,0) -- (-1,-0.5)
      (0,0) -- (0.5,-0.5)
        (0.5,0.5) arc(0:180:1.25)
      (-1,0.5) arc(0:180:0.25)
      (-1,-0.5) arc(0:-180:0.25)
      (0.5,-0.5) arc(0:-180:1.25)
      (-1.5,-0.25) node {$j_4$}
      (-1.5,0.25) node {$j_1$}
      (-2,0.25) node {$j_2$}
      (-2,-0.25) node {$j_3$}
      (-0.25,0.25) node {$j'_6$};
\end{tikzpicture}\label{split2}
\quad .
\end{align}

These equations are actually two different changes of basis according with the two recoupling schemes. Let us look e.g. at the first equation: On the right hand side the only relevant information is the tensor $\hat{t}^{(j_5,j'_5)}_1(j_1,j_2;j_3,j_4)$, which describes how the representations $j_1$ and $j_2$ couple to (the pair) ($j_5,j'_5$) and how $j_3$ and $j_4$ also couple to that pair \footnote{Recall that every edge carries two representations, $j$ and $j^*$. Here we explicitly allow them to couple to different representations $j'_5 \neq j^*_5$.}. Note that $\hat{t}_1$ does not depend on the magnetic indices, but only on the representation labels and their recoupling into the representations $(j_5,j'_5)$; hence $\hat{t}_1$ provides the vertex weight expressed in the intertwiner basis. In the same way, $\hat{t}_2$ is the vertex weight expressed in the intertwiner basis associated with the other recoupling scheme, in which the pairs $j_1$ and $j_4$, and $j_2$ and $j_3$, both couple to the pair $(j_6,j'_6)$ .

 Using the relations \eqref{eq:rel-splitting} it is straightforward to get the relation between the tensors in the two recoupling schemes:
\begin{equation}\label{splittingrelation}
\hat{t}^{(j_6,j'_6)}_2(j_1,j_4;j_2,j_3) = \sum_{j_5,j'_5} \hat{t}^{(j_5,j'_5)}_1(j_1,j_2;j_3,j_4) \sqrt{\frac{[2 j_5 + 1][2 j'_5 + 1]}{[2 j_6 + 1][2 j'_6 + 1]}}
\left[
\begin{matrix}
\, j_1 \, & \, j_2 \, & \, j_5 \, \\
\, j_3 \, & \, j_4 \, & \, j_6 \,
\end{matrix}
\right]
\left[
\begin{matrix}
\, j_1 \, & \, j_2 \, & \, j'_5 \, \\
\, j_3 \, & \, j_4 \, & \, j'_6 \,
\end{matrix}
\right] \quad .
\end{equation}

Via the inverse transformation any choice of tensor $\hat t$ leads to a vertex weight $t$ satisfying (\ref{s2}).  Additionally one might want to demand certain symmetries, for instance under a reflection of the underlying four--valent vertex, etc.. This translates into certain conditions on the form of $\hat t_1$. In section \ref{results} we will explicitly define the models that we will analyze, and we will discus the extra symmetries that they enjoy.

\section{The coarse graining algorithm}
\label{coarse}

As we have motivated in the previous section, the previously introduced quantum group spin net models can be cast into a tensor network form, to which one can apply tensor network renormalization \cite{levin,guwen} as a coarse graining procedure. In the following we will describe this algorithm.

We consider our quantum group spin net models defined on the square lattice. To each four-valent vertex of the lattice we attach the tensors
\begin{equation}\label{initialt}
t(\{j\},\{m\},\{n\})=
\begin{tikzpicture}[baseline]
\draw (-0.6,0.6) -- (0,0) -- (0.6,0.6)
     (-0.6,-0.6) -- (0,0) -- (0.6,-0.6)
       (0.5,0.2) node {$j_1$}
       (0.35,0.7) node {$m_1$}
       (0.85,0.7) node {$n_1$}
      (-0.5,0.2) node {$j_2$}
       (-0.9,0.7) node {$m_2$}
       (-0.4,0.7) node {$n_2$}
        (-0.5,-0.2) node {$j_3$}
       (-0.9,-0.7) node {$m_3$}
       (-0.4,-0.7) node {$n_3$}
 (0.5,-0.2) node {$j_4$}
       (0.35,-0.7) node {$m_4$}
       (0.85,-0.7) node {$n_4$};
\end{tikzpicture}
\quad .
\end{equation}
To complete the definition of our models, we require an additional ingredient in the horizontal edges: to every horizontal edge, with labels $\{j_a,m_a,n_a\}$, we attach the `swirl'
\begin{equation}\label{phase}
(-1)^{2j_a} q^{-n_a} \delta_{n_a,n'_a}=
\begin{tikzpicture}[baseline,scale=0.75]
 \draw (0,0.6) arc(0:180:0.5)
 (0,0.6)--(0,-0.35)
 (0,-0.35) arc(0:-180:0.5)
  (-0.65,0.45) node {$n_a$}
       (-0.65,-0.2) node {$n'_a$}
       (0.3,0.25) node {$j_a$};
\end{tikzpicture}
\quad .
\end{equation}
This extra factor will be justified when explaining the contraction scheme. We will see that, thanks to it, under coarse graining the contributions coming from the $\{n\}$ indices (which undergo a dualization) will behave in the same way as those coming from the indices $\{m\}$. As a result, the contraction carried out to coarse grain will lead to a flow equation with a very nice compact form, which will be given in terms of the recoupling symbols of the group.

The partition function for the model then reads
\begin{equation}\label{Z}
Z = \sum_{a,\dots} \dots \; t(\{j\},\{m\},\{n\})_{abcd}\; (-1)^{2j_a} q^{-n_a} \; t(\{j\},\{m\},\{n\})_{b'c'ad'} \dots \quad ,
\end{equation}
namely it is given by the contraction, according with the connectivity of the lattice, of all the tensors and the extra factors in the horizontal edges.

\begin{figure}
\begin{center}
\includegraphics[width=0.5\textwidth]{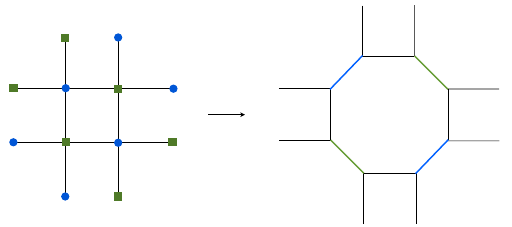}

\caption{Rewriting of the four--valent tensor network as a three--valent tensor network.  \label{fig:split-lattice}}
\end{center}
\end{figure}

\begin{figure}
\begin{center}
\includegraphics[width=0.5\textwidth]{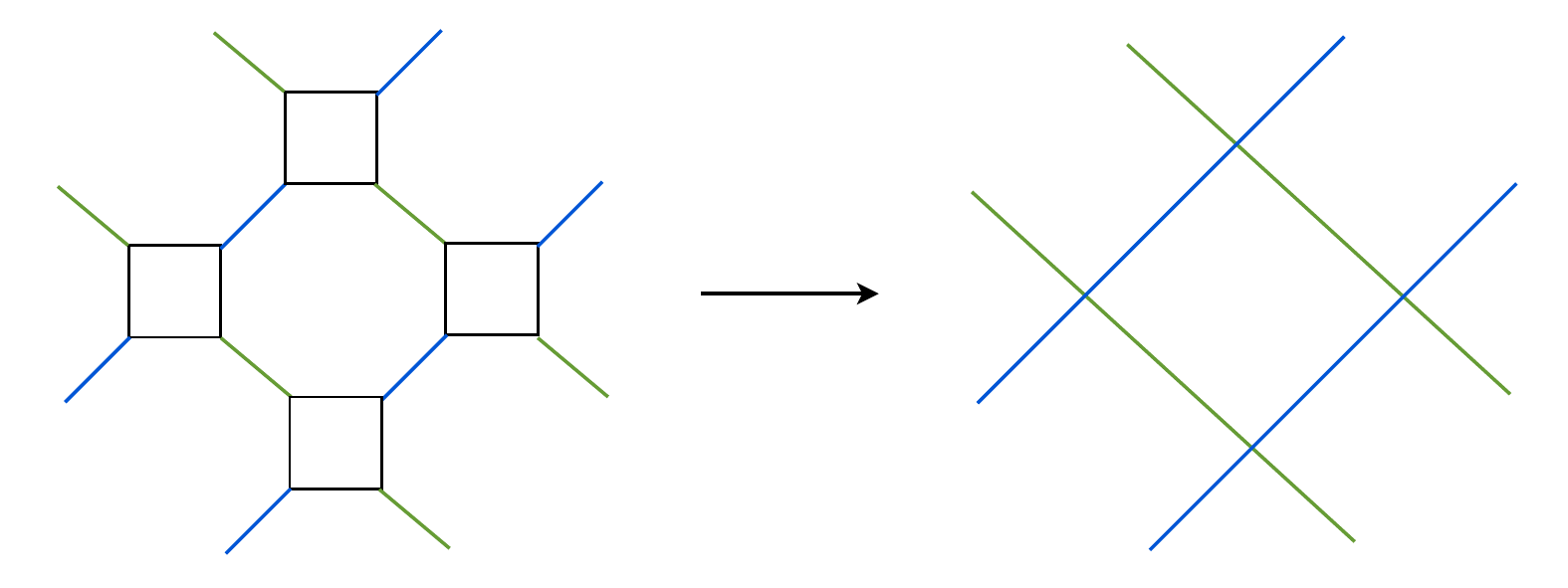}
\caption{The contraction of the three--valent tensor network yields a coarser (rotated) square lattice.  \label{fig:effective-lattice}}
\end{center}
\end{figure}

In order to coarse grain this tensor network we consider the algorithm employed in  \cite{guwen}, which is the adaptation to the square lattice of the original algorithm introduced in  \cite{levin} for a hexagonal lattice. The idea is simple: we first rewrite the four--valent tensor network into a three--valent tensor network as depicted in Figure  \ref{fig:split-lattice}. In this way we can contract the three-valent tensors into new effective four--valent tensors forming a coarser rotated lattice, as in Figure \ref{fig:effective-lattice}. 

The splitting carried out in Figure \ref{fig:split-lattice} is achieved  by regarding the four--valent tensors as matrices and employing a singular value decomposition. We adapt this singular value decomposition to the recoupling schemes that we discussed in previous section, obtaining an algorithm that preserves the group symmetry properties of our tensors. This symmetry preserving algorithm was already introduced in  \cite{bdetal13}, where we discussed its advantages. Let us describe how it works for the current quantum group spin net models.

\subsection{Symmetry preserving algorithm}

As explained in the previous section (see Eqs \eqref{split1}-\eqref{split2}), given the tensor $t(\{j\},\{m\},\{n\})$ we can perform a change of basis to the recoupling basis, defining either $ \hat{t}^{(j_5,j'_5)}_1(j_1,j_2;j_3,j_4)$, associated to the recoupling scheme $j_1\otimes j_2\ni \{j_5,j'_5\}\in j_3\otimes j_4$, or $ \hat{t}^{(j_6,j'_6)}_2(j_1,j_2;j_3,j_4)$, associated to the recoupling scheme $j_2\otimes j_3\ni \{j_6,j'_6\}\in j_4\otimes j_1$. In the recoupling basis the tensor is then block-diagonal, where blocks are labelled by pairs of representations $(j,j')$ associated to the recoupling scheme. We can regard these blocks as matrices and perform a singular value decomposition, that we truncate to certain cut-off $\mu$:
\begin{align}
\hat{t}^{(j_5,j'_5)}_1(j_1,j_2;j_3,j_4)&\equiv (M_1)^{(j_5,j'_5)}_{A=\{j_1,j_2\},B=\{j_3,j_4\}}\simeq\sum_{i=1}^{\mu(j_5,j'_5)} (U_1)^{(j_5,j'_5)}_{A=\{j_1,j_2\},i}(\lambda_1)^{(j_5,j'_5)}_{ii}(V_1)^{(j_5,j'_5)}_{i,B=\{j_3,j_4\}}\quad ,
\end{align}
\begin{align}
\hat{t}^{(j_6,j'_6)}_2(j_1,j_2;j_3,j_4)&\equiv (M_2)^{(j_6,j'_6)}_{A=\{j_2,j_3\},B=\{j_4,j_1\}}\simeq\sum_{i=1}^{\mu(j_6,j'_6)} (U_2)^{(j_6,j'_6)}_{A=\{j_2,j_3\},i}(\lambda_2)^{(j_6,j'_6)}_{ii}(V_2)^{(j_6,j'_6)}_{i,B=\{j_4,j_1\}}\quad .
\end{align}
Without the cut--off the dimension of the effective index $i$, the so--called \emph{bond} dimension, would grow exponentially when iterating the coarse graining procedure. Thanks to this cut--off we render the coarse graining procedure a feasible task. Note that we have control on the error made by imposing this truncation: The larger the cut--off the better the approximation.

\begin{figure}
\begin{center}
\includegraphics[width=0.75\textwidth]{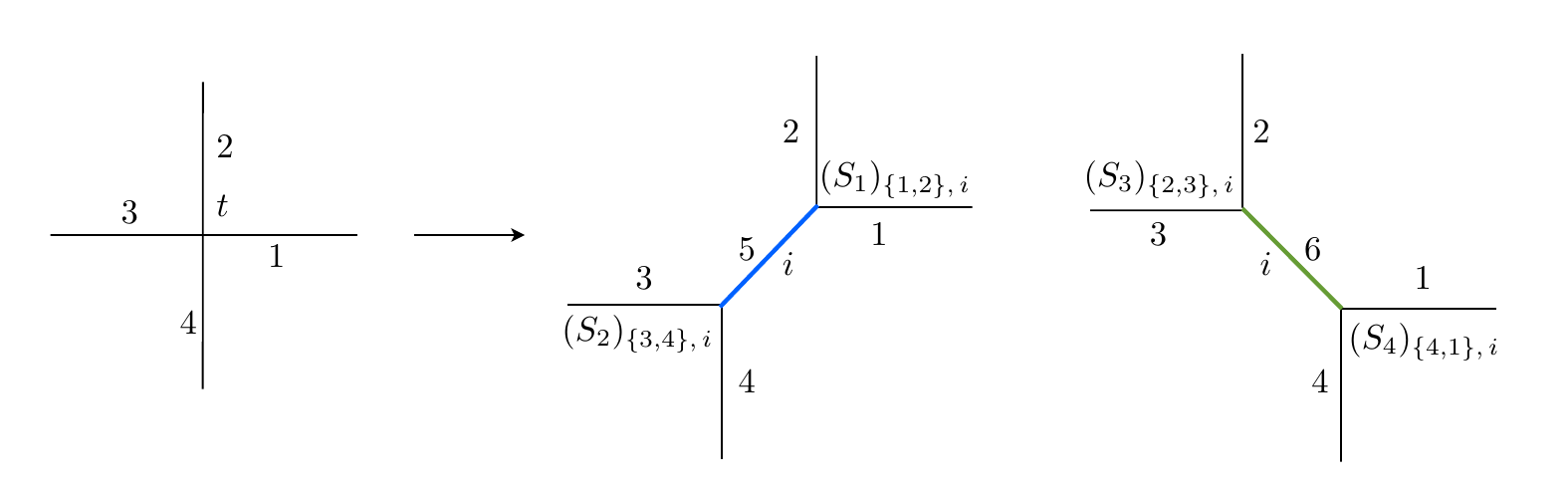}
\caption{Splitting of the tensor $t(\{j\},\{m\},\{n\})$ according with the two different recoupling schemes, as expressed in equations \eqref{splitting1}-\eqref{splitting2}.\label{fig:split-vertices}}
\end{center}
\end{figure}

The above singular value decompositions allow us to split the four--valent tensors $t(\{j\},\{m\},\{n\})$  in two three--valent tensors. We can carry out two different splittings, according with the two different recoupling schemes. For vertices located in odd positions (blue circles in Fig. \ref{fig:split-lattice}) we write
\begin{equation}\label{splitting1}
t(\{j\},\{m\},\{n\})=\sum_{j_5,j'_5}\sum_{m_5,n_5}\sum_{i=1}^{\mu(j_5,j'_5)} (S_1)^{j_5,j'_5}_{m_5,n_5}(\{I\}_{\{1,2\}},i)  (S_2)^{j_5,j'_5}_{m_5,n_5}(\{I\}_{\{3,4\}},i)\q,
\end{equation}
while for vertices located in even positions (green squares in Fig. \ref{fig:split-lattice}) we have
\begin{equation}\label{splitting2}
t(\{j\},\{m\},\{n\})=\sum_{j_6,j'_6}\sum_{m_6,n_6} \sum_{i=1}^{\mu(j_6,j'_6)}(S_3)^{j_6,j'_6}_{m_6,n_6}(\{I\}_{\{2,3\}},i)  (S_4)^{j_6,j'_6}_{m_6,n_6}(\{I\}_{\{4,1\}},i)\q .
\end{equation}
Here we have have simplified the notation by introducing $\{I\}_{\{a,b\}}=\{j_a,m_a,n_a,j_b,m_b,n_b\}$. 
In the above splittings, which are sketched in Figure \ref{fig:split-vertices}, we have defined the following objects
\begin{align} \label{S1}
(S_1)^{j_5,j'_5}_{m_5,n_5}(\{I\}_{\{1,2\}},i)= &\sqrt{ d_{j_5}d_{j'_5}(\lambda_1)^{(j_5,j'_5)}_{ii}}
(U_1)^{(j_5,j'_5)}_{\{j_1,j_2\},i}\begin{tikzpicture}[baseline,scale=0.75]
\draw  (0,-0.75) -- (0,0) -- (-0.5,0.5)
      (0,0) -- (0.5,0.5)
         (0.5,0.75) node {$j_1$}
      (-0.5,0.75) node {$j_2$}
      (0.25,-0.5) node {$j_5$};
\end{tikzpicture}
\; \otimes \;
\begin{tikzpicture}[baseline,scale=0.75]
\draw  (0,-0.5) -- (0,0) -- (-0.25,0.25) arc(0:180:0.25)
      (0,0) -- (0.25,0.25) arc(0:180:0.75)
      (0,-0.5) arc(0:-180:0.25)
      (-0.75,0) node {$j_1$}
      (-1.25,0) node {$j_2$}
      (-0.75,-0.5) node {$j'_5$};
\end{tikzpicture}
\q ,\\ \label{S2}
(S_2)^{j_5,j'_5}_{m_5,n_5}(\{I\}_{\{3,4\}},i)= &\sqrt{ d_{j_5}d_{j'_5}(\lambda_1)^{(j_5,j'_5)}_{ii}}
(V_1)^{(j_5,j'_5)}_{i,\{j_3,j_4\}}
\begin{tikzpicture}[baseline,scale=0.75]
\draw (-0.5,-0.5) -- (0,0) -- (0,0.75)
      (0.5,-0.5) -- (0,0)
      (-0.5,-0.75) node {$j_3$}
      (0.5,-0.75) node {$j_4$}
      (0.25,0.5) node {$j_5$};
\end{tikzpicture}
\; \otimes \;
\begin{tikzpicture}[baseline,scale=0.75]
\draw (-0.25,-0.25) -- (0,0) -- (0,0.5)
      (0.25,-0.25) -- (0,0)
      (-0.25,-0.25) arc(0:-180:0.25)
      (0.25,-0.25) arc(0:-180:0.75)
            (0.5,0.5) arc(0:180:0.25)
      (-0.75,0) node {$j_4$}
      (-1.25,0) node {$j_3$}
      (0.5,0.2) node {$j'_5$};
\end{tikzpicture}
\quad ,\\
\label{S3}
(S_3)^{j_6,j'_6}_{m_6,n_6}(\{I\}_{\{2,3\}},i)= &\sqrt{ d_{j_6}d_{j'_6}(\lambda_2)^{(j_6,j'_6)}_{ii}}
(U_2)^{(j_6,j'_6)}_{\{j_2,j_3\},i}
\begin{tikzpicture}[baseline,scale=0.75]
\draw (-1.25,0.5) -- (-0.75,0) --(0,-0.3)
      (-0.75,0) -- (-1.25,-0.5)
      (-1,-0.75) node {$j_3$}
      (-1,0.75) node {$j_2$}
      (-0.25,0.25) node {$j_6$};
\end{tikzpicture} 
\; \otimes \;
\begin{tikzpicture}[baseline,scale=0.75]
\draw (-0.75,0.3) -- (0,0) -- (0.5,0.5)
      (0,0) -- (0.5,-0.5)
        (0.5,0.5) arc(0:180:1)
      (0.5,-0.5) arc(0:-180:1)
      (-1.5,0.25) node {$j_2$}
      (-1.5,-0.25) node {$j_3$}
      (-0.5,-0.2) node {$j'_6$};
\end{tikzpicture}
\q ,\\ \label{S4}
(S_4)^{j_6,j'_6}_{m_6,n_6}(\{I\}_{\{4,1\}},i)= &\sqrt{ d_{j_6}d_{j'_6}(\lambda_2)^{(j_6,j'_6)}_{ii}}
(V_2)^{(j_6,j'_6)}_{i,\{j_4,j_1\}}\;
\begin{tikzpicture}[baseline,scale=0.75]
\draw (-0.75,0.25) -- (0,0) -- (0.5,0.5)
      (0,0) -- (0.5,-0.5)
      (0.5,0.75) node {$j_1$}
      (0.5,-0.75) node {$j_4$}
      (-0.5,-0.1) node {$j_6$};
\end{tikzpicture}
\; \otimes \;
\begin{tikzpicture}[baseline,scale=0.75]
\draw (-1.25,0.5) -- (-0.75,0) -- (0,-0.3) 
      (-0.75,0) -- (-1.25,-0.5)
      (-1.25,0.5) arc(0:180:0.25)
      (-1.25,-0.5) arc(0:-180:0.25)
      (-1.5,-0.25) node {$j_4$}
      (-1.5,0.25) node {$j_1$}
      (-0.25,0.25) node {$j'_6$};
\end{tikzpicture}
\quad .
\end{align}
In the next section we will use these expressions when computing the contraction that yields the new effective tensor. For that purpose it is convenient to use the following identities, which can be easily proven with the help of the identity ${}_q\mathcal{C}_{m_1\,m_2\,m_3}^{j_1 \, j_2 \, j_3} =(-1)^{j_1+j_2-j_3}{}_{\bar{q}}\mathcal{C}_{-m_1\,-m_2\,-m_3}^{\;\; \; j_1 \q\; j_2 \q\, j_3}$:
\begin{align}
\begin{tikzpicture}[baseline,scale=0.75]
\draw  (0,-0.5) -- (0,0) -- (-0.25,0.25) arc(0:180:0.25)
      (0,0) -- (0.25,0.25) arc(0:180:0.75)
      (0,-0.5) arc(0:-180:0.25)
      (-0.75,0) node {$j_1$}
      (-1.25,0) node {$j_2$}
      (-0.75,-0.5) node {$j'_5$};
\end{tikzpicture}
=
\begin{tikzpicture}[baseline,scale=0.6]
\draw (-0.5,-0.25) -- (0,0.25) -- (0,0.75)
      (0.5,-0.25) -- (0,0.25)
       (1,0.75) arc(0:180:0.5)
       (1,0.75)--(1,-0.75)
       (1,-0.75) arc(0:-180:0.5)
      (-0.5,-0.6) node {$j_2$}
      (0.5,-0.6) node {$j_1$}
      (1.3,0) node {$j'_5$};
\end{tikzpicture}
\qquad&;\qquad
\begin{tikzpicture}[baseline,scale=0.75]
\draw (-0.25,-0.25) -- (0,0) -- (0,0.5)
      (0.25,-0.25) -- (0,0)
      (-0.25,-0.25) arc(0:-180:0.25)
      (0.25,-0.25) arc(0:-180:0.75)
            (0.5,0.5) arc(0:180:0.25)
      (-0.75,0) node {$j_4$}
      (-1.25,0) node {$j_3$}
      (0.5,0.2) node {$j'_5$};
\end{tikzpicture}
=
\begin{tikzpicture}[baseline,scale=0.75]
\draw  (0,-0.75) -- (0,0) -- (-0.5,0.5)
      (0,0) -- (0.5,0.5)
         (0.5,0.75) node {$j_4$}
      (-0.5,0.75) node {$j_3$}
      (0.25,-0.5) node {$j'_5$};
\end{tikzpicture}
\q \label{right1},\\\label{right2}
\begin{tikzpicture}[baseline,scale=0.6]
\draw (-0.75,0.3) -- (0,0) -- (0.5,0.5)
      (0,0) -- (0.5,-0.5)
        (0.5,0.5) arc(0:180:1)
      (0.5,-0.5) arc(0:-180:1)
      (-1.5,0.25) node {$j_2$}
      (-1.5,-0.25) node {$j_3$}
      (-0.5,-0.2) node {$j'_6$};
\end{tikzpicture}
=
\begin{tikzpicture}[baseline,scale=0.6]
\draw (-0.5,-0.5) -- (0,0.25) -- (0,1)
      (0.5,-0.5) -- (0,0.25)
       (1,1) arc(0:180:0.5)
       (1,1)--(1,0.75)
        (1,0.75) arc(0:-180:0.25)
       (0.5,-0.5) arc(0:-180:1)
      (-0.5,-0.7) node {$j_2$}
      (-1.75,-0.7) node {$j'_6$}
      (1,0.2) node {$j_3$};
\end{tikzpicture}
\qquad&;\qquad
\begin{tikzpicture}[baseline,scale=0.75]
\draw (-1.25,0.5) -- (-0.75,0) -- (0,-0.3) 
      (-0.75,0) -- (-1.25,-0.5)
      (-1.25,0.5) arc(0:180:0.25)
      (-1.25,-0.5) arc(0:-180:0.25)
      (-1.5,-0.25) node {$j_4$}
      (-1.5,0.25) node {$j_1$}
      (-0.25,0.25) node {$j'_6$};
\end{tikzpicture}
=
\begin{tikzpicture}[baseline,scale=0.6]
\draw  (0,-0.75) -- (0,0) -- (-0.5,0.5)
      (0,0) -- (0.5,0.5)
       (-0.5,0.5) arc(0:180:0.5)
        (1,-0.5) arc(0:180:0.25)
      (1,-0.75) arc(0:-180:0.5)
      (1,-0.75)--(1,-0.5)
         (0.5,0.75) node {$j_4$}
      (-1,0.5) node {$j'_6$}
      (0.5,-0.8) node {$j_1$};
\end{tikzpicture}
\q .
\end{align}
In view of expression \eqref{right2}, we can already clarify why we need the extra factor \eqref{phase} associated with the horizontal edges, here labeled by 1 and 3, namely the factors $(-1)^{2j_1}q^{-n_1} \; (-1)^{2j_3}q^{-n_3}$: Their role is to remove the swirl (double arc) that is attached in the equalties \eqref{right2} to the edges with labels $j_1$ and $j_3$.

\subsection{Derivation of the flow equation}

\begin{figure}
\begin{center}
\includegraphics[width=0.45\textwidth]{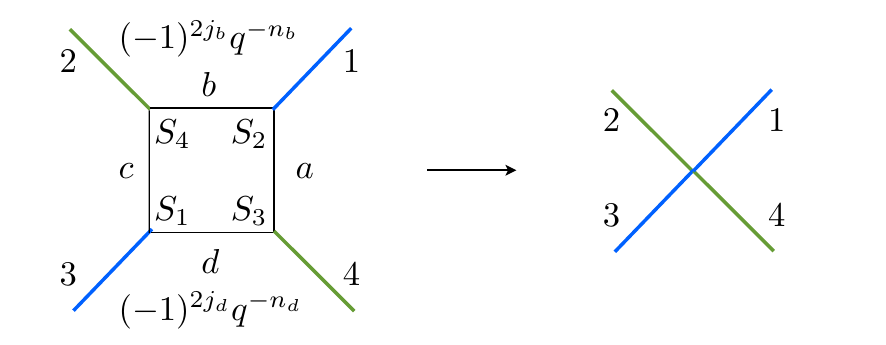}
\caption{The new tensor $t(\{j\},\{m\},\{j'\},\{n\}, \{i\})_{1,2,3,4}$ is the result of contracting $(S_1)^{j_3,j'_3}_{m_3,n_3}(\{I\}_{\{d,c\}},i_3)$, 
$(S_2)^{j_1,j'_1}_{m_1,n_1}(\{I\}_{\{b,a\}},i_1)$,
$(S_3)^{j_4,j'_4}_{m_4,n_4}(\{I\}_{\{a,d\}},i_4)$, and 
$(S_4)^{j_2,j'_2}_{m_2,n_2}(\{I\}_{\{b,c\}},i_2)$, together with the factors $(-1)^{2j_b}q^{-n_b} \; (-1)^{2j_d}q^{-n_d}$ associated to horizontal edges. \label{fig:eff-vertex}}
\end{center}
\end{figure}

With the splitting of the four--valent tensors into three-valent ones, it is now straightforward to get the new effective four--valent tensor by contraction, as depicted in Figure  \ref{fig:eff-vertex}. We note that the effective edges now carry labels $I'=\{j,m,j',n,i\}$, with $i$ running from 1 to $\mu(j,j')$, so that the Hilbert space associated to every edge gets a more general structure than that of the original Hilbert space \eqref{s1}, namely
\be
\mathcal H_e= \oplus_{j,j'} \mu(j,j') V_j\otimes V_{j'}.
\ee
In consequence, the effective tensor also gets more general than \eqref{initialt}, as its structure is of the form
\be
t(\{j\},\{m\},\{j'\},\{n\}, \{i\})=
\begin{tikzpicture}[baseline]
\draw (-0.5,0.5) -- (0,0) -- (0.5,0.5)
     (-0.5,-0.5) -- (0,0) -- (0.5,-0.5)
       (0.7,0.5) node {$I'_1$}
      (-0.7,0.5) node {$I'_2$}
        (-0.7,-0.5) node {$I'_3$}
 (0.7,-0.5) node {$I'_4$};
\end{tikzpicture}
\quad ;\qquad I'_a=\{j_a,m_a,j'_a,n_a,i_a\} \quad .
\ee
Thus we have in particular now the possibility $j\neq j'$ also for the `outer' edges. In addition multiplicity labels $i$ (whose range can depend on the representation labels $(j,j')$) can appear.

This new tensor possesses the quantum group symmetry of the original vertex weight, namely it is invariant under the left action of $\mathcal{P}_{\left(\{m\},\{m'\}\right)}(\{j\})$ and under the right action of $\mathcal{P}_{\left(\{n'\},\{n\}\right)}(\{j'\})$.
Our symmetry preserving algorithm gives us directly the tensor in the recoupling basis. Let us compute the new block $\hat{t}_1^{(j_5,j'_5)}(J_1,J_2;J_3,J_4)$, with $J_a=\{j_a,j'_a,i_a\}$, :
\begin{align}
\hat{t}_1^{(j_5,j'_5)}(J_1,J_2;J_3,J_4)&=\sum_{a,b,c,d}\sum_{\{m\}}\sum_{\{n\}}(-1)^{j_1+j_2,+j_3,+j_4}(-1)^{j'_1+j'_2,+j'_3,+j'_4}
\begin{tikzpicture}[baseline,scale=0.75]
\draw (-0.5,0.75) -- (-0.25,1) arc(180:0:0.5) -- (0.75,-1) arc(0:-180:0.5) -- (-0.5,-0.75)
      (-0.25,1) -- (0,0.75)
      (-0.25,-1) -- (0,-0.75)
      (-0,0.5) node {$j_1$}
      (-0.5,0.5) node {$j_2$}
      (-0,-0.5) node {$j_4$}
      (-0.5,-0.5) node {$j_3$}
      (1,0) node {$j_5$};
\end{tikzpicture}
\; \otimes \;
\begin{tikzpicture}[baseline,scale=0.75]
\draw (-0.5,-1) -- (0,-0.5) -- (0,0.5) -- (-0.5,1)
      (0.5,-1) -- (0,-0.5)
      (0,0.5) -- (0.5,1)
      (-0.5,-1.25) node {$j'_3$}
      (0.5,-1.25) node {$j'_4$}
      (0.5,1.25) node {$j'_1$}
      (-0.5,1.25) node {$j'_2$}
      (0.25,0) node {$j'_5$};
\end{tikzpicture}
\nonumber\\
&\times
(-1)^{2j_b}q^{-n_b}
(S_2)^{j_1,j'_1}_{m_1,n_1}(\{I\}_{\{b,a\}},i_1)
(S_4)^{j_2,j'_2}_{m_2,n_2}(\{I\}_{\{b,c\}},i_2)
\nonumber\\
&\times
(-1)^{2j_d}q^{-n_d}
(S_1)^{j_3,j'_3}_{m_3,n_3}(\{I\}_{\{d,c\}},i_3)
(S_3)^{j_4,j'_4}_{m_4,n_4}(\{I\}_{\{a,d\}},i_4)\quad .
\end{align}

Substituting in this equation the expressions \eqref{S1}-\eqref{S4}, we can contract the involved Clebsch-Gordan coefficients. It is straightforward to check that the contraction of those coefficients carrying indices $\{j,m\}$ (left part of the diagrams) gives the following $9j$-symbol:
\begin{align}\label{9j}
\begin{tikzpicture}[baseline,scale=0.75]
\draw (-1,-1) -- (-0.5,-0.5) -- (-0.5,0.5) -- (-1,1) -- (0,1.5) arc(180:0:0.75) -- (1.5,-1.5) arc(0:-180:0.75) -- (-1,-1)
      (0,-1.5) -- (1,-1) -- (0.5,-0.5) -- (0.5,0.5) -- (1,1) -- (0,1.5)
      (-0.5,-0.5) -- (0.5,-0.5)
      (-0.5,0.5) -- (0.5,0.5)
      (0,-0.75) node {$j_d$}
      (-0.75,0) node {$j_c$}
      (0.75,0) node {$j_a$}
      (0,0.75) node {$j_b$}
      (-1.25,-1.) node {$j_3$}
      (1.25,-1.) node {$j_4$}
      (1.25,1.) node {$j_1$}
      (-1.25,1.) node {$j_2$}
      (1.75,0) node {$j_5$};
\end{tikzpicture}& 
=\frac{(-1)^{j_c + j_a + j_5} (-1)^{j_1+j_2+j_3+j_4} }{d_{j_5}\sqrt{d_{j_b} d_{j_d}}}
\left[
\begin{matrix}
\, j_c \, & \, j_a \,  & \, j_5 \, \\
\, j_1 \, & \, j_2 \, & \, j_b \,
\end{matrix}
\right] 
\left[
\begin{matrix}
\, j_c \, & \, j_a \,  & \, j_5 \, \\
\, j_4 \, & \, j_3 \, & \, j_d \,
\end{matrix}
\right] \quad .
\end{align}
This equality is shown in Appendix \ref{app9j}.

As for the right side of the diagrams, which come from a dualization and that are formed by Clebsch-Gordan coefficients that carry indices $\{j', n\}$, they also contract to the $9j$-symbol, which is obvious once we consider the identities \eqref{right1}-\eqref{right2}, and we remove the swirls with the extra factors  $(-1)^{2j_b}q^{-n_b} \; (-1)^{2j_d}q^{-n_d}$. As we anticipated before, without the introduction of these extra factors for the horizontal edges, we would not get the $9j$-symbol when contracting the Clebsch-Gordan coefficients that come from the dual contributions.

In conclusion, the new effective tensor in the recoupling scheme 1 reads
\begin{align}\label{flow}
\hat{t}_1^{(j_5,j'_5)}(J_1,J_2;J_3,J_4)&=\sum_{a,b,c,d}\sum_{\{m\}}\sum_{\{n\}}
\frac{(-1)^{j_c + j_a + j_5} (-1)^{j'_c + j'_a + j'_5} }{d_{j_5}d_{j'_5}\sqrt{d_{j_b}d_{j'_b} d_{j_d}d_{j'_d}}}
\sqrt{ d_{j_1}d_{j'_1}d_{j_2}d_{j'_2}d_{j_3}d_{j'_3}d_{j_4}d_{j'_4}}\nonumber\\
&\times
\sqrt{(\lambda_1)^{(j_1,j'_1)}_{i_1i_1}\,
(\lambda_2)^{(j_2,j'_2)}_{i_2i_2}\, (\lambda_1)^{(j_3,j'_3)}_{i_3i_3}\,(\lambda_2)^{(j_4,j'_4)}_{i_4i_4}}
\nonumber\\
&\times
(V_1)^{(j_1,j'_1)}_{i_1,\{j_b,j'_b,j_a,j'_a\}}
(V_2)^{(j_2,j'_2)}_{i_2,\{j_b,j'_b,j_c,j'_c\}}
(U_1)^{(j_3,j'_3)}_{\{j_d,j'_d,j_c,j_c'\},i_3}
(U_2)^{(j_4,j'_4)}_{\{j_a,j'_a,j_d,j'_d\},i_4}\nonumber\\
&\times
\left[
\begin{matrix}
\, j_c \, & \, j_a \,  & \, j_5 \, \\
\, j_1 \, & \, j_2 \, & \, j_b \,
\end{matrix}
\right] 
\left[
\begin{matrix}
\, j_c \, & \, j_a \,  & \, j_5 \, \\
\, j_4 \, & \, j_3 \, & \, j_d \,
\end{matrix}
\right] \left[
\begin{matrix}
\, j'_c \, & \, j'_a \,  & \, j'_5 \, \\
\, j'_1 \, & \, j'_2 \, & \, j'_b \,
\end{matrix}
\right] 
\left[
\begin{matrix}
\, j'_c \, & \, j'_a \,  & \, j'_5 \, \\
\, j'_4 \, & \, j'_3 \, & \, j'_d \,
\end{matrix}
\right] 
\quad ,
\end{align}
where $j_i=j'_i$  with  $i\in\{a,b,c,d\}$ for the first iteration, while they are a priori independent for subsequent iterations.
The blocks $\hat{t}_2^{(j_6,j'_6)}(J_1,J_4;J_2,J_3)$ for the second splitting can in general be obtained from $\hat{t}_1^{(j_5,j'_5)}(J_1,J_2;J_3,J_4)$ by convoluting it with $6j$-symbols, as in equation  \eqref{splittingrelation}. Nevertheless, as we will further comment in section \ref{results}, the models that we will consider have the symmetry $\hat{t}_2^{(j_6,j'_6)}(J_1,J_4;J_2,J_3)=\hat{t}_1^{(j_6,j'_6)}(J_1,J_4;J_2,J_3)$, which is preserved under coarse graining, hence this step of computing $\hat t_2$ is actually not necessary.

The above equation \eqref{flow} can be regarded as the {\it coarse graining flow equation} for $\hat t_1$, in the sense that the factors $\{\lambda, U, V\}$ are determined by the tensor in the previous step of the iteration procedure. By choosing initial data, we can analyze the behaviour of the vertex weights under this coarse-graining flow. In the next section we carry out this analysis for a particular set of models, determining the fixed points and phases of this flow, as well as phase transitions.

Note that this flow equation is still quite challenging numerically. The coarse grained tensor $\hat t$ will have in general $\chi^5$ components where even in the simplest (useful) approximation $\chi=(j_{max}+1)^2$, with $j_{max}=k/2$ for $k$ even and $j_{max}=(k-1)/2$ for $k$ odd. To save and compute more efficiently with these tensors we developed the method of super indices. These super indices summarize combinations of indices, and keep only those that actually lead to non--vanishing entries due to the coupling rule. This leads already to a huge saving for the $[6j]$ symbol -- instead of $(j_{max}+1)^6$ entries one only saves the components non--vanishing due to coupling rules. 

\subsection{Geometric interpretation of the flow equation}

Here we wish to point out that the flow equation (\ref{flow}) for the 2D spin nets has an interesting 3D geometric interpretation. A similar 3D interpretation of the 2D models has also been noticed in \cite{bw1}.
The 3D geometrical interpretation is due to the $[6j]$ symbols appearing in (\ref{flow}). The 3D spin foam models without and with (positive) cosmological constant are given by the Ponzano--Regge \cite{ponzanoregge} and the Turaev--Viro model \cite{turaev} respectively. These models are topological, that is triangulation invariant, and built by associating $[6j]$ symbols to the tetrahedra. Similarly we can associate tetrahedra to the $[6j]$ symbols appearing in (\ref{flow}). The edges of the tetrahedra are labelled by the spins appearing in the $[6j]$ symbol, so that the coupling conditions of the $[6j]$ symbol reflect the triangle inequalities for the tetrahedron. The summation over common $j$  and $j'$ is then interpreted as gluing of edges. 

\begin{figure}[h!]
\begin{center}
\includegraphics[scale=1]{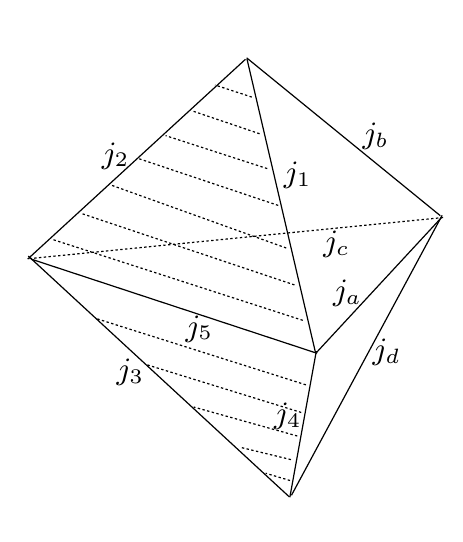}
\caption{Double pyramid with triangles  $T_1=\{(j_1,j_2,j_5),(j_3,j_4,j_5)\}$, those filled with doted lines, and triangles  $T_2=\{(j_1,j_a,j_b),(j_2,j_b,j_c),(j_3,j_c,j_d),(j_4,j_a,j_d)\}$.\label{pyramid}}
\end{center}
\end{figure}

This gluing of tetrahedra appears in stages:
\begin{itemize}
\item For the left and right copy separately (i.e. for unprimed and primed spins) we have the gluing of two tetrahedra to a double pyramid. Here the gluing is along the triangle with edges labelled by $j_a,j_c$ and $j_5$ (and similarly for the primed indices).
\item The triangles of the double pyramid, i.e. for the left copy, are given by $T_1=\{(j_1,j_2,j_5),(j_3,j_4,j_5)\}$ as well as $T_2=\{(j_1,j_a,j_b),(j_2,j_b,j_c),(j_3,j_c,j_d),(j_4,j_a,j_d)\}$, as shown in Figure \ref{pyramid}. One will notice that $T_1$ corresponds to the indices of the new effective tensor, whereas the triangles in $T_2$ are glued with the appropriate $U$ and $V$'s from the singular value decomposition. This singular value decomposition results in three--valent tensors, dual to triangles, and we can indeed interpret this as gluing triangle amplitudes onto the pyramides. Note, that for some families of intertwiner fixed point models described in the next section \ref{results} these amplitudes themselves are given by $[6j]$ symbols. Hence we can interpret this case as gluing further tetrahedra onto the pyramids.

The gluing of the triangle amplitudes is also the place where a coupling between the left and right copy, i.e. primed and unprimed spins occur. We can interpret this as a gluing of the two double pyramids via the triangle amplitudes, that depend on both primed and unprimed indices. However, in the case of factorizing models, described in section \ref{results}, all quantities factorize with respect to primed and unprimed spins, and so does the coarse graining flow.

\end{itemize}

This geometric interpretation, in particular in the case that the `triangle amplitudes' themselves correspond to geometric objects, might help to obtain an intuition about the coarse graining flow. For instance, if only $[6j]$ symbols (with appropriate face and edge factors and signs) are involved, the complex will describe a Turaev--Viro model. This is invariant under changes of the triangulation, which will explain some of the occurring fixed points.

This geometric interpretation might also help to approximate the coarse graining flow equation by a simpler equation. For instance it seems possible to consider instead of the flow equation related to the double pyramid a flow equation related to just one tetrahedron. This applies if $\hat{t}_1=\hat{t}_2$ in the sense explained below equation (\ref{initt}), which is preserved under coarse graining. Preliminary numerical investigations have shown that this leads indeed to the same phase diagram as the more complicated flow equation. We will investigate this possibility further as it will allow us to consider more easily  the replacement of the structure group $SU(2)_k$ with i.e.\ $SU(2)_k \times SU(2)_k$, as appear for the 4D gravitational models.

\section{Fixed point intertwiners as initial data and coarse graining results}\label{results}

The space of possible tensors $\hat{t}_1$, and therefore spin net models is huge. One therefore needs to add further selection principles to arrive at a suitable phase space. For instance \cite{bahretal12,bdetal13} introduces and uses the so--called $E$--function parametrization, which for the spin foams encompasses all current models. It also allows an interpretation as imposing simplicity constraints, which are encoded in the choice of the $E$--function. However there are several problems with this parametrization. First of all, it turns out that for a quantum group, models based on $E$--functions have not the expected\footnote{Such symmetries are realized in the $S_3$ group case \cite{bdetal13}.} symmetries, for instance under changing the recoupling scheme, via a 2--2 move. Furthermore, the models mostly flow to trivial phases and a fine tuning is necessary to arrive at a non--trivial fixed point, i.e. for which simplicity constraints are realized.

Here we will introduce another parametrization, which can be understood as an application of Reisenberger's construction outlined in section \ref{sec:reisenberger}. This parametrization is based on fixed points (i.e. triangulation invariant instances)  of so--called intertwiner models, that have been introduced and constructed in \cite{bw1,bw2}.

We will use  specific families of these fixed points for intertwiner models here, for a specification of all possible fixed points, see \cite{bw1,bw2}. In this section we will describe them briefly and explain that they do not define fixed points of the here considered spin net models, but, moreover, flow to non trivial fixed points, i.e. fixed points beyond the degenerate or analogue $BF$ phase. Hence they define new phases and by linear combination we can additionally study phase diagrams.

\subsection{Description of the  fixed point intertwiners}

The {\it fixed point intertwiners} can be written as a modification of the Clebsch-Gordan coefficients\footnote{This modification only depends on the triple of representation labels, not on the magnetic indices $m$. The vertices define therefore still $SU(2)_k$ intertwining maps.}, which we diagrammatically denote as `fat' vertices introduced in \cite{bw1}:
\begin{equation} \label{eq:fat-clebsch}
\begin{tikzpicture} [baseline,scale=0.75]
\draw (0,-0.75) node {$j_1$}
      (0,-0.5) -- (0.5,0) -- (1,-0.5)
      (1,-0.75) node {$j_2$}
      (0.5,0) -- (0.5,0.5) 
      (0.5,0.75) node {$j_3$};
\draw[fill] (0.5,0) circle [radius=0.1];
\end{tikzpicture}
:= a(j_1,j_2,j_3) \; {}_q\mathcal{C}_{m_1\,m_2\,m_3}^{j_1 \, j_2 \, j_3} \quad , \quad 
\begin{tikzpicture}[baseline,scale=0.75]
\draw (0,-0.55) -- (0,0) -- (-0.5,0.5)
      (0,0) -- (0.5,0.5)
      (0,-0.8) node {$j_3$}
      (-0.5,0.75) node {$j_2$}
      (0.5,0.75) node {$j_1$};
\draw[fill] (0,0) circle [radius=0.1];
\end{tikzpicture}
\; = \; a'(j_1,j_2,j_3) \;{}_{\bar{q}} \mathcal{C}_{m_1\, m_2 \, m_3}^{j_1 \, j_2 \, j_3}  \quad .
\end{equation}
At first sight such a modification appears to be an arbitrary choice and indeed without further requirements it allows for a parametrization with a plethora of parameters. Nevertheless, as it has been shown in \cite{bw1}, one can uniquely fix these factors by (a) requiring invariance under planar (i.e. $2-2$, $3-1$ and $1-3$) Pachner moves, i.e. triangulation independence with respect to the dual triangulation, and (b) specifying the allowed representations, i.e. the set of labels $j$ which do not lead to vanishing weights $a(j,\cdot,\cdot)$ and $a'(j,\cdot,\cdot)$. These special factors define the before mentioned {\it fixed point intertwiners}, of which we will present a few in the following subsections.

Note that this class of tensors leads in general to complex weights. Fortunately this is not a problem at all for the tensor network algorithm that we will be using. To our knowledge it gives a first example where the tensor network algorithm is applied to a statistical system with complex weights. 

Thus the models that we will analyze have initial tensors of the form
\begin{align}\label{idata}
{t}(\{j\},\{m\},\{n\})& = \sum_{j_5,j'_5}\frac1{d_{j_5}d_{j'_5}}
\begin{tikzpicture}[baseline,scale=0.75]
\draw (-0.5,-1) -- (0,-0.5) -- (0,0.5) -- (-0.5,1)
      (0.5,-1) -- (0,-0.5)
      (0,0.5) -- (0.5,1)
      (-0.5,-1.25) node {$j_3$}
      (0.5,-1.25) node {$j_4$}
      (0.5,1.25) node {$j_1$}
      (-0.5,1.25) node {$j_2$}
      (0.25,0) node {$j_5$};
\draw[fill] (0,-0.5) circle [radius=0.1]
            (0,0.5) circle [radius=0.1];
\end{tikzpicture}
\; \otimes \;
\begin{tikzpicture}[baseline,scale=0.75]
\draw (-0.5,0.75) -- (-0.25,1) arc(180:0:0.5) -- (0.75,-1) arc(0:-180:0.5) -- (-0.5,-0.75)
      (-0.25,1) -- (0,0.75)
      (-0.25,-1) -- (0,-0.75)
      (-0,0.5) node {$j_1$}
      (-0.5,0.5) node {$j_2$}
      (-0,-0.5) node {$j_4$}
      (-0.5,-0.5) node {$j_3$}
      (1,0) node {$j'_5$};
\draw[fill] (-0.25,1.) circle [radius=0.1]
            (-0.25,-1.) circle [radius=0.1];
            \end{tikzpicture}
            \nonumber\\
           & = \sum_{j_6,j'_6}\frac1{d_{j_6}d_{j'_6}}
\begin{tikzpicture}[baseline,scale=0.75]
\draw (-1,0.5) -- (-0.5,0) -- (0,0) -- (0.5,0.5)
      (-0.5,0) -- (-1,-0.5)
      (0,0) -- (0.5,-0.5)
      (-1,-0.75) node {$j_3$}
      (-1,0.75) node {$j_2$}
      (0.5,0.75) node {$j_1$}
      (0.5,-0.75) node {$j_4$}
      (-0.25,-0.3) node {$j_6$};
      \draw[fill] (-0.5,0) circle [radius=0.1]
            (0,0) circle [radius=0.1];
\end{tikzpicture}
\; \otimes \;
\begin{tikzpicture}[baseline,scale=0.75]
\draw (-1,0.5) -- (-0.5,0) -- (0,0) -- (0.5,0.5)
      (-0.5,0) -- (-1,-0.5)
      (0,0) -- (0.5,-0.5)
        (0.5,0.5) arc(0:180:1.25)
      (-1,0.5) arc(0:180:0.25)
      (-1,-0.5) arc(0:-180:0.25)
      (0.5,-0.5) arc(0:-180:1.25)
      (-1.5,-0.25) node {$j_4$}
      (-1.5,0.25) node {$j_1$}
      (-2,0.25) node {$j_2$}
      (-2,-0.25) node {$j_3$}
      (-0.25,-0.3) node {$j'_6$};
        \draw[fill] (-0.5,0) circle [radius=0.1]
            (0,0) circle [radius=0.1];
\end{tikzpicture}
\quad .
\end{align}

We note that
\ba\label{initt}
\hat{t}_1^{(j_5,j'_5)}(j_1,j_2;j_3,j_4)&=\frac{a'(j_1,j_2,j_5)a(j_3,j_4,j_5)a(j_2,j_1,j'_5)a'(j_4,j_3,j'_5)}{(d_{j_5}d_{j'_5})^2}\q ,\\
\hat{t}_2^{(j_6,j'_6)}(j_2,j_3;j_1,j_4)&=\frac{a'(j_1,j_6,j_4)a(j_3,j_6,j_2)a(j_4,j'_6,j_1)a'(j_2,j'_6,j_3)}{(d_{j_6}d_{j'_6})^2}\q .
\ea

As indicated already, due to the fixed point property of the intertwiners, the amplitudes for initial tensor (and as it happens also for the coarse grained tensors) are the same in the two recoupling bases, namely the tensors verify the symmetry property $\hat{t}_1^{(j_5,j'_5)}(j_1,j_2;j_3,j_4)=\hat{t}_2^{(j_5,j'_5)}(j_1,j_2;j_3,j_4)$. This is indeed a very physical requirement, imposing that the weights are the same if expanded in the two possible (planar) recoupling schemes. A asymmetry would specify a special direction, which we avoid by choosing the fixed point intertwiners as initial data. This equality of the tensor components in the two recoupling schemes will be preserved under the coarse graining flow.

Further symmetries that our tensors verify are 
\ba
\hat{t}_1^{(j_5,j'_5)}(j_1,j_2;j_3,j_4)&=&\hat{t}_1^{(j_5,j'_5)}(j_3,j_4;j_1,j_2)=\hat{t}_1^{(j_5,j'_5)}(j_2,j_1;j_4,j_3)  \q,\nn\\
 \hat{t}_1^{(j_5,j'_5)}(j_1,j_2;j_3,j_4) &=& \hat{t}_1^{(j'_5,j_5)}(j_1,j_2;j_3,j_4) \q ,
\ea
which represent discrete rotation and reflection symmetries.

Note that in (\ref{idata}) the spin labels at the outer edges are the same in the left and right copy, whereas the intertwiner labels $j_5,j'_5$ or $j_6,j'_6$ can a priori differ. (They do not for models analogous to standard gauge theories, which are based on the Haar intertwiner.) This might actually change under the renormalization flow: we allow generalized tensors carrying two spin labels $j,j'$ per edge. As we will see, in the extreme case we will obtain factorizing models as fixed points of the renormalization flow. In this case the sum over $j$ and $j'$ can be performed independently and the partition function is equal to a square of an intertwiner model partition function. In cases in which that happens, we do regain the square of the model associated to the intertwiner we started with. 

To characterize the fixed points we will give the non--vanishing intertwiner channels. I.e. an intertwiner channel $(j,j')$ means that there are non--vanishing tensor components with $j_5=j$ and $j_5'=j'$. Due to the symmetry under changing the recoupling scheme this holds also for $j_6=j$ and $j'_6=j'$.  We will speak of an `excited' representation, if this representation is allowed by the simplicity constraints encoded in the model, that is if $j\in {\cal S}_1$ in the sense of section \ref{sec:reisenberger}. I.e.\  the representation does not lead to a vanishing weight for the initial model.

\subsubsection{Only $j=0$ and $j_{max}$ excited}

A straightforward but exceptional example, which can be defined for even $k$, is to require that only representations $j=0$ and $j_{max} = \frac{k}{2}$ are excited. There are only two non--vanishing $a$-factors (and permutations) appearing, namely:
\begin{equation}
a(0,0,0) = a'(0,0,0) = 1 \quad , \quad a(j_{max},j_{max},0) = a'(j_{max},j_{max},0) = 1 \quad .
\end{equation}
This is because $j_{max}$ is of quantum dimension one and additionally has to satisfy the coupling rule $j_{max} \otimes j_{max} \equiv 0$. Also note that the factors $a$ are invariant under permutations of their arguments.

If we plug in these  fixed point intertwiners as initial data of our spin net models, we do not observe a flow; these initial data already define a fixed point. It only has two excited intertwiner channels, namely $(0,0)$ and $(j_{max},j_{max})$. In fact one can easily see that the model is equivalent to the Ising model at zero temperature, but in a `spin representation', equivalent to a (analogue) $BF$ spin net model on $\mathbb{Z}_2$ \cite{sffinite,eckert}. That is $j_{max}$ corresponds to the non--trivial representation of ${\mathbb Z}_2$ and there has always to meet an even number of such non--trivial representations at every vertex.

\subsubsection{Only $j=0$ and $j=j_1<j_{max}$ excited}

We study here two examples of models based on fixed point intertwiners that have only two representations excited, the trivial representation and  a $j=j_1$ different from $j_{max}$.
The two examples are the following:
\begin{itemize}
\item $k=6, j_1=2$

The fixed point intertwiner is defined by the amplitudes
\begin{equation}
a(0,0,0) = a'(0,0,0) = 1 \, , \q a(2,2,2) = a'(2,2,2) = \sqrt{d_2(d_2-1)} \, , \q a(2,2,0) = a'(2,2,0) =  \sqrt{d_2}   ,
\end{equation}
and permutations of them, with $a$ being invariant under permutations of its arguments.
\item $k=10, j_1=3$

The fixed point intertwiner is defined by the amplitudes
\begin{equation}
a(0,0,0) = a'(0,0,0) = 1 \, , \q a(3,3,3) = a'(3,3,3) = i \sqrt{d_3(d_3-1)} \, , \q a(3,3,0) = a'(3,3,0) =  \sqrt{d_3}   ,
\end{equation}
and permutations of them, with $a$ being invariant under permutations of its arguments.
\end{itemize}

Taking these amplitudes as initial data for our spin net models, the resulting model flows to a fixed point with only contributing intertwiner channels $(0,0)$, $(0,j_1)$, $(j_1,0)$ and ($j_1,j_1)$. This fixed point is factorizing:  the partition function is the product of two partition functions, where each factor corresponds to the partition function of the intertwiner model we started with. 



\subsubsection{Only even representations excited}
\label{even}

For a quantum group with $k$ multiple of 4, one can construct   fixed point intertwiners by only allowing representation with even $j$ excited \cite{bw2}. This fixed point intertwiners are characterized by the following amplitudes:
\begin{align}
a(j_1,j_2,j_3) = a'(j_1,j_2,j_3) = & \sqrt{(-)^{J-j_1}} \sqrt{(-)^{J-j_2}} \sqrt{(-)^{J-j_3}} (-)^{2J - j_1 - j_2} \times \nonumber \\
& \times \sqrt{\frac{[2j_1 + 1] [2j_2 + 1]}{[J+1]}}
\left[
\begin{matrix}
\, j_1 \, & \, j_2 \, & \, j_3 \, \\
\, \frac{k}{4} \, & \, \frac{k}{4} \, & \, \frac{k}{4} \, 
\end{matrix}
\right] \quad,
\end{align}
if $j_1$, $j_2$ and $j_3$ are all even, otherwise the amplitude vanishes.

Starting with these intertwiners, we analyzed cases $k=8$ and $k=12$. 

For $k=8$ we obtain the following fixed point under coarse graining: only intertwiner channels $(0,0)$, $(0,4)$, $(4,0)$, $(4,4)$, and $(2,2)$ are excited. This fixed point is not factorizing, as the channels $(0,0)$ and $(2,2)$ appear, but not for instance the channel $(0,2)$.  This constitutes an example of a so-called `mixed' fixed point. I.e. non--diagonal entries appear but the fixed point is not of factorizing type. These type of fixed points will play a crucial role in the interpretation of the fixed points as phases for spin foams.

In contrast, for $k=12$ we do find the expected factorizing fixed point with all the intertwiner channels labeled by two even representations excited. 

\subsubsection{Maximal spin $J$}
\label{maximalJ}

At last we introduce a whole family of  fixed point intertwiners for each level $k$. A member of this family is labelled by an integer spin $J \leq k/2$. For a given $J$ all representations $j\leq J$ are excited, where $J$ is independent  (apart from the bound) of the maximal spin $\frac{k}{2}$ of the quantum group. In fact these fixed points generalize to the classical group $SU(2)$ and provide a cut--off in spins there. (The coarse graining flow could still lead to arbitrary large spins.)  Here the simplicity constraints can be interpreted as forbidding spins larger than $J$.

The $a$ factors are given by:
\begin{align}
a^J_{CDL}(j_1,j_2,j_3) = a'^J_{CDL}(j_1,j_2,j_3) = & \sqrt{(-)^{J-j_1}} \sqrt{(-)^{J-j_2}} \sqrt{(-)^{J-j_3}} (-)^{2J - j_1 - j_2} \times \nonumber \\
& \times \sqrt{\frac{[2j_1 + 1] [2j_2 + 1]}{[J+1]}}
\left[
\begin{matrix}
\, j_1 \, & \, j_2 \, & \, j_3 \, \\
\, \frac{J}{2} \, & \, \frac{J}{2} \, & \, \frac{J}{2} \, 
\end{matrix}
\right] \quad ,
\end{align}
Note that when $J=k/2$, and $j_1$, $j_2$ and $j_3$ are all even, the amplitudes coincide with those of case \ref{even}, but now also odd representations are allowed.

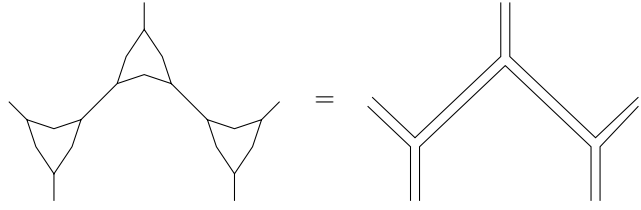
\begin{figure}
\begin{center}
\begin{tikzpicture}[baseline,scale=1.2]
\draw (0,-0.5) -- (0.2,-0.3) -- (0.3,0.) -- (0.5,0.3) -- (0.7,0.) -- (0.8,-0.3) -- (0.5,-0.2) -- (0.2,-0.3)
      (0.5,0.6) -- (0.5,0.3) 
      (1,-0.5) -- (0.8,-0.3)
      (1,-0.5) -- (1.2,-0.7)
      (0,-0.5) -- (-0.2,-0.7) -- (-0.5,-0.8) -- (-0.8,-0.7) -- (-0.7,-1.0) -- (-0.5,-1.3) -- (-0.3,-1.0) -- (-0.2,-0.7)
      (-0.8,-0.7) -- (-1,-0.5)
      (-0.5,-1.3) -- (-0.5,-1.6)
      (2,-0.5) -- (1.8,-0.7) -- (1.5,-0.8) -- (1.2,-0.7) -- (1.3,-1.0) -- (1.5,-1.3) -- (1.7,-1.0) -- (1.8,-0.7)
      (1.5,-1.3) -- (1.5,-1.6)
      (4.45,0.6) -- (4.45,0) -- (3.5,-0.9) -- (3.025,-0.45)
      (4.55,0.6) -- (4.55,0) -- (5.5,-0.9) -- (5.975,-0.45)
      (4.5,-0.1) -- (3.55,-1) -- (3.55,-1.6)
      (3.45,-1.6) -- (3.45,-1) -- (2.975,-0.55)
      (4.5,-0.1) -- (5.45,-1) -- (5.45,-1.6)
      (5.55,-1.6) -- (5.55,-1) -- (5.975,-0.55)
      (2.5,-0.5) node {$=$};
\end{tikzpicture}
\caption{Diagrammatic explanation of the `CDL' structure. With identity \eqref{eq:identity1} one can replace the intermediate edges by double lines (we ignore normalization factors associated to the edges), which turn the tensor network into a ribbon graph.  This ribbon graph structure gives a fixed point under renormalization flow, as the strands going around inner faces form closed loops, which just give constant factors and hence drop out. The remaining graph gives a rescaled version of the original graph. \label{fig:CDL}}
\end{center}
\end{figure}

To understand why the $a^J_{CDL}$ define fixed points intertwiner models it is instructive to interpret them diagrammatically \cite{bw1}. Ignoring normalization (edge) factors, one three--valent intertiner is given as
\begin{equation}
\begin{tikzpicture} [baseline]
\draw (0,-0.75) node {$j_1$}
      (0,-0.5) -- (0.2,-0.3) -- (0.3,0.) -- (0.5,0.3) -- (0.7,0.) -- (0.8,-0.3) -- (0.5,-0.2) -- (0.2,-0.3)
      (0.5,0.6) -- (0.5,0.3) 
      (1,-0.5) -- (0.8,-0.3)
      (1,-0.75) node {$j_2$}
      (0.5,0.75) node {$j_3$}
      (0,0) node {$\frac{J}{2}$}
      (1,0) node {$\frac{J}{2}$}
      (0.5,-0.55) node {$\frac{J}{2}$};
\end{tikzpicture} \quad .
\end{equation}
If one glues several of these vertices together, one can sum over the intermediate spins and replace the single line by a double line (see identity \eqref{eq:identity1}). This explains our notation $a^J_{CDL}$ denoting these tensors as `corner double line', see also figure \ref{fig:CDL}; for such tensor networks it is straightforward to see that they are fixed points of the coarse graining algorithm for intertwiner models \cite{bw1}.

The initial tensor for the spin net model can then be represented as (again modulo edge normalization factors)
\ba
t(\{j\},\{m\},\{n\})\sim \sum_J
\begin{tikzpicture}[baseline,scale=1.2]
\draw (-0.5,0.5)--(-0.25,0.25)--(0.25,0.25)--(0.5,0.5)
(-0.25,0.25)--(-0.25,-0.25)
(0.25,0.25)--(0.25,-0.25)
(-0.5,-0.5)--(-0.25,-0.25)--(0.25,-0.25)--(0.5,-0.5)
(0,0.5) node  {$\frac{J}{2}$}
(0,-0.5) node  {$\frac{J}{2}$}
(-0.4,0) node  {$\frac{J}{2}$}
(0.4,0) node  {$\frac{J}{2}$}
(0.5,0.7) node {$j_1$}
(-0.5,0.7) node {$j_2$}
(-0.5,-0.7) node {$j_3$}
(0.5,-0.7) node {$j_4$};
\end{tikzpicture}
\; \otimes \;
\begin{tikzpicture}[baseline,scale=1.2]
\draw (-0.5,0.5)--(-0.25,0.25)--(0.25,0.25)--(0.5,0.5)
(-0.25,0.25)--(-0.25,-0.25)
(0.25,0.25)--(0.25,-0.25)
(-0.5,-0.5)--(-0.25,-0.25)--(0.25,-0.25)--(0.5,-0.5)
(0,0.5) node  {$\frac{J}{2}$}
(0,-0.5) node  {$\frac{J}{2}$}
(-0.4,0) node  {$\frac{J}{2}$}
(0.4,0) node  {$\frac{J}{2}$}
  (-0.5,0.5) arc(0:180:0.2)
   (-0.5,-0.5) arc(0:-180:0.2)
   (0.5,0.5) arc(0:180:0.9)
   (0.5,-0.5) arc(0:-180:0.9)
(-1.25,0.35) node {$j_1$}
(-0.9,0.35) node {$j_2$}
(-0.9,-0.35) node {$j_3$}
(-1.25,-0.35) node {$j_4$};
\end{tikzpicture}
\q .
\ea
The symmetry of the graphical representation under rotations of $\frac{\pi}{2}$  explains the invariance of the tensor under change of recoupling basis from $(j_5,j_5')$ to $(j_6,j_6')$.

At this point the thoughtful reader may wonder, why we choose `fixed points' as the initial data of a statistical system, which we intend to coarse grain. In order to guard against confusion we strongly stress that the previously defined $a^J_{CDL}$ do {\it not} define fixed points of the here considered spin net models. To explain why, consider the definition of the initial tensor 
given in Eq. \eqref{idata}.
Even though it is written in this product form, the intertwiners for one copy and its dual do not decouple: During the contraction of the outer edges in the coarse graining algorithm \eqref{flow}, the summations over the primed and unprimed representation labels cannot be taken individually.  Thus already the first coarse graining step will result in a different model from the initial one and  the system  `flows' away from the initial condition. 

There are however fixed points of the full spin net models, for which both copies indeed decouple and are thus denoted as `factorizing'. As we will explain below, most of the initial $a^J_{CDL}$ models actually flow to their `corresponding' factorizing fixed point, meaning that the fixed point shares the same maximal spin $J$ with the initial models. We have seen such a behaviour already for the $k=12$ example in section \ref{even}.

After the introduction and justification of the initial data let us focus on the results: Remarkably every single initial fixed point intertwiner  model flows to a different fixed point and furthermore the flow in different quantum groups follows a clear pattern, which we summarize in table \ref{tab:fixedpoints}\footnote{The pattern is confirmed for quantum groups SU$(2)_k$ with $k=\{4,5,\dots,10\}$. For $k=4$, the initial model for $J=1$ flows to the factorizing fixed point, not the analogue $BF$ one.}. Let us briefly describe the different types of fixed points:
\begin{table}
\begin{center}
\begin{tabular}{|c || c | c |}
\hline
$J$ & $k$ even & $k$ odd \\
\hline
$J=0$ & degenerate & degenerate\\
\hline
$J=1$ & factorizing $J=1$ & factorizing $J=1$ \\
\hline
$J=2$ & factorizing $J=2$ & factorizing $J=2$ \\
\vdots & \vdots & \vdots \\
$J= \frac{k}{2} - 2$ even / $J=\frac{k-1}{2} -1$ odd & factorizing $J=\frac{k}{2}-2$ & factorizing $J=\frac{k-1}{2} -1$\\
\hline
$J= \frac{k}{2} -1$ even / $J=\frac{k-1}{2}$ odd & analogue $BF$ & analogue $BF$ \\
\hline
$J= \frac{k}{2}$ & ``mixed'' & $-$ \\
\hline
\end{tabular}
\caption{Table of the fixed points obtained from the initial intertwiners labelled by the maximal spin $J$. For odd $k$ there is no model for $J=\frac{k}{2}$ since the initial model is only defined for even $J$. \label{tab:fixedpoints}}
\end{center}
\end{table}

As we have commented in the introduction (see also \cite{bdetal13}) the intertwiner channels are the relevant degrees of freedom in spin net models, such that it is sufficient to know the excited intertwiner channels (and the associated singular values) to identify the phase / fixed point. Thus we will not state the final fixed point tensor but focus on the description of the excited intertwiner channels. Since for all discovered fixed points the excited intertwiner channels are just excited once, they can be characterized by a matrix of singular values, where rows and columns are labelled by the representation label of the `left' and the `right' copy: A non--vanishing entry $m_{jj'}$ of this matrix indicates a contributing intertwiner channel $(j,j')$. For the fixed points the non--vanishing singular values are all equal to one, which is intwined with the triangulation independence of the corresponding three--valent models, obtained from splitting the four--valent vertices into three--valent ones \cite{refining}. The following fixed points appear as a result of the coarse graining flow:
\begin{itemize}
\item {\it Degenerate:} \\
On the degenerate fixed point, which coincides with the initial model with $J=0$, only the trivial representation $j=0$ is allowed.
\item {\it Analogue BF:} \\
In the topological (analogue) $BF$ theory all `diagonal' intertwiner channels allowed by the quantum group are excited, i.e. channels with $j' = j$. The matrix of singular values is just the identity matrix. 
\item {\it Factorizing with spin $J$:}\\
In the factorizing model all intertwiner channels $(j,j'), j,j'\leq J$ are excited with the same singular value. In particular $j' \neq j$ is allowed. As a matrix of singular values, this model has a quadratic upper left block matrix of size $J\times J$ with all entries $1$. The other entries vanish.
\item {\it Mixed:} \\
For $k$ even and the  fixed point intertwiners  with maximal spin $J=\frac{k}{2}$, i.e. the maximal spin of the quantum group, the system flows to a peculiar fixed point. It is described by exciting all intertwiner channels $(j,j')$ for which the sum $j+j'$ is even. E.g. for $k=4$ one finds the following matrix of singular values:
\begin{equation}
\left( 
\begin{matrix}
\,1\, & \,0\, & \,1\, \\
\,0 \, & \,1\, & \,0\, \\
\,1\, & \,0\, & \,1\,
\end{matrix}
\right) \quad .
\end{equation} 
Note that the difference in the even/odd behaviour originates from the fixed point intertwiner weights $a(j_1,j_2,j_5)=a'(j_1,j_2,j_5)$, that vanish if the sum $j_1+j_2+j_5$ is odd, see \cite{bw1}. This initial condition, which involves the left copy and the right copy of the model separately, results under the coarse graining flow in a non--trivial coupling condition between the left and right copy, namely that $j_1+j'_1$ has to be even and $j_2+j'_2$ has to be even.

This is a fixed points of mixed type as also occurring for the $k=8$ example in section \ref{even}. Here the `left and right copy' of the model do not decouple completely as is the case for the factorizing fixed points. 
\end{itemize}

The results are remarkable in two ways: Alone the mere existence of the additional fixed points (with respect to the degenerate and the analogue $BF$ phase) is positive evidence that spin net models possess interesting phases. More importantly we have specified initial data in form of the fixed point intertwiners for which these models actually  flow to these fixed points, thus we can state that these phases exist indeed.

At this point we would like to add a comment on the accuracy applied in our simulations. For most of our simulations we used a particular approximation: For each intertwiner channel we picked its largest singular value, namely $\mu(j,j')=1$ for all $j$ and $j'$. This truncation is suggested by the investigations in \cite{bdetal13} which put forward the intertwiner degrees of freedom as relevant degrees of freedom.  We will name this truncation `one--singular--value per intertwiner channel algorithm'.  Note that for $k=12$, for which $j_{max}=6$ this already results in a necessary bond dimension of $\chi=49=(6+1)^2$ (not even taken into account the large number of magnetic indices, that have been absorbed into the recoupling symbols appearing in the flow equation (\ref{flow})). Thus such a truncation might be unavoidable if considering more complicated structure groups, i.e. $SU(2)_k \times SU(2)_k$.

By comparison, the full algorithm takes all singular values in all intertwiner channels and picks the largest $\chi$ ones, where $\chi$ is the  fixed bond-dimension. Thus it may happen that some intertwiner channels obtain multiplicity three, whereas others do not appear at all.

In what we have presented so far, the results of both algorithms\footnote{We can confirm this up to bond dimension $\chi=22$.} are consistent, which indicates that already the `one--singular--value per intertwiner channel' algorithm is a very well suited approximation to study  the intertwiner dynamics. (Additionally it avoids the appearance of so--called non--isolated CDL fixed points \cite{guwen}, which slow down the algorithm very much and are argued to not contain essential information \cite{sherbrook}.) 
This confirms the  `one--singular--value per intertwiner channel' truncation, and the conjecture that the intertwiner degrees of freedom are the relevant ones, i.e.\ determine the large scale behaviour of the system.

As a next step it is worthwhile to study phase diagrams for these models and phase transitions between these new phases. To do so one can linearly combine the fixed point intertwiners to give new initial data and vary their respective coefficients to tune the system towards a phase transition. 

\subsection{Superposition of intertwiners}

As already pointed out in \eqref{i8} in section \ref{sec:reisenberger}, one can additionally consider a linear combination of  fixed point intertwiners as initial data. Here we will consider linear combinations of models based on the CDL fixed point intertwiners, defined in section \ref{maximalJ}, whose amplitudes are characterized by a maximal representation $J\leq j_{max}$ excited:
\begin{equation}
{t}(\{j\},\{m\},\{n\}) = \sum_{J} \alpha_J \left( \sum_{j_5,j'_5}
\begin{tikzpicture}[baseline,scale=0.75]
\draw (-0.5,-1) -- (0,-0.5) -- (0,0.5) -- (-0.5,1)
      (0.5,-1) -- (0,-0.5)
      (0,0.5) -- (0.5,1)
      (-0.5,-1.25) node {$j_3$}
      (0.5,-1.25) node {$j_4$}
      (0.5,1.25) node {$j_1$}
      (-0.5,1.25) node {$j_2$}
      (0.25,0) node {$j_5$}
      (-0.3,-0.45) node {$J$}
      (-0.3,0.45) node {$J$};
\draw[fill] (0,-0.5) circle [radius=0.1]
            (0,0.5) circle [radius=0.1];
\end{tikzpicture}
\; \otimes \;
\begin{tikzpicture}[baseline,scale=0.75]
\draw (-0.5,0.75) -- (-0.25,1) arc(180:0:0.5) -- (0.75,-1) arc(0:-180:0.5) -- (-0.5,-0.75)
      (-0.25,1) -- (0,0.75)
      (-0.25,-1) -- (0,-0.75)
      (-0,0.5) node {$j_1$}
      (-0.5,0.5) node {$j_2$}
      (-0,-0.5) node {$j_4$}
      (-0.5,-0.5) node {$j_3$}
      (1,0) node {$j'_5$}
      (-0.55,1.1) node {$J$}
      (-0.55,-1.1) node {$J$};
\draw[fill] (-0.25,1.) circle [radius=0.1]
            (-0.25,-1.) circle [radius=0.1];
\end{tikzpicture}
\right) \quad ,
\end{equation}
where we have used the notation
\begin{equation}
\begin{tikzpicture} [baseline,scale=0.75]
\draw (0,-0.75) node {$j_1$}
      (0,-0.5) -- (0.5,0) -- (1,-0.5)
      (1,-0.75) node {$j_2$}
      (0.5,0) -- (0.5,0.5) 
      (0.5,0.75) node {$j_3$}
       (0.2,0) node {$J$};
\draw[fill] (0.5,0) circle [radius=0.1];
\end{tikzpicture}
:= a^J_{CDL}(j_1,j_2,j_3) \; {}_q\mathcal{C}_{m_1\,m_2\,m_3}^{j_1 \, j_2 \, j_3} \quad , \quad 
\begin{tikzpicture}[baseline,scale=0.75]
\draw (0,-0.55) -- (0,0) -- (-0.5,0.5)
      (0,0) -- (0.5,0.5)
      (0,-0.8) node {$j_3$}
      (-0.5,0.75) node {$j_2$}
      (0.5,0.75) node {$j_1$}
         (-0.3,-0.05) node {$J$};
\draw[fill] (0,0) circle [radius=0.1];
\end{tikzpicture}
\; = \; a'^J_{CDL}(j_1,j_2,j_3) \;{}_{\bar{q}} \mathcal{C}_{m_1\, m_2 \, m_3}^{j_1 \, j_2 \, j_3}  \quad .
\end{equation}
Also these initial tensors satisfy Reisenberger's principle, at least with respect to the 2--2 move recoupling $(j_5,j_5')\rightarrow (j_6,j_6')$. 
As already mentioned before, these intertwiners are not necessarily orthonormal, hence they only satisfy the projector property for particular choices of the parameters $\alpha_J$. Nevertheless to plot the phase diagram we will choose the parameter $\alpha_J$ rather freely, we only require that $\sum_J \alpha_J =1$.

To obtain  the phase diagrams we used the `one-singular-value per intertwiner channel' algorithm, described above. Although it is very well suited for our models, as we argued above, it is not as accurate as the full algorithm with equal or larger bond dimension. This means, e.g. that possible phase transition lines might change in position and shape if we would employ the more accurate version, however, it will not change the qualitative behaviour of the phase diagram, i.e. the phases will also exist in the full algorithm.

\subsubsection{Phase diagram for $k=5$}

\begin{figure}[h!]
\begin{center}
\includegraphics[scale=0.7]{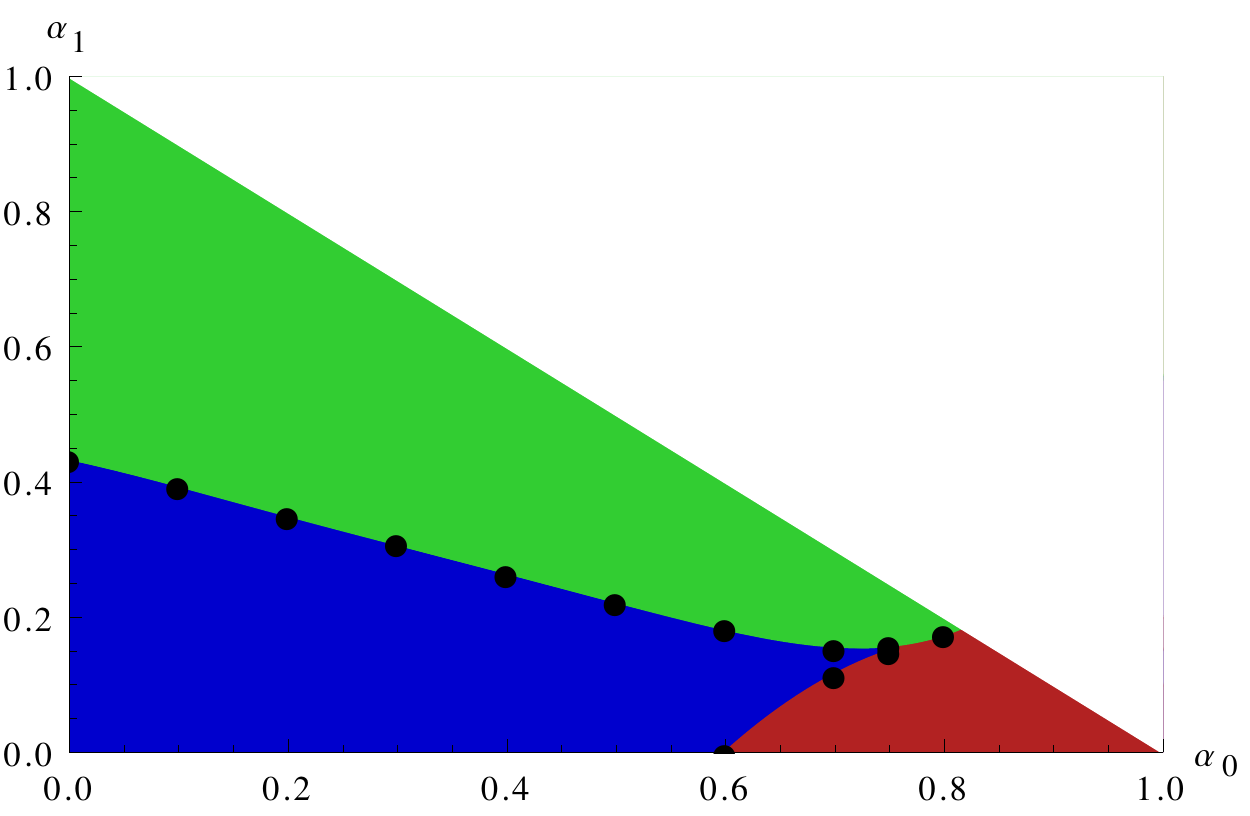}
\caption{The phase diagram for $k=5$. The red area (in the right corner of the diagram) shows the set of parameters for which the model flows to the degenerate fixed point, which is also located in the bottom right corner. For parameteres in the green area (upper corner), the system flows to the factorizing model with maximal spin $J=1$ and the blue area (bottom left corner) covers the models that flow to the analogue $BF$ fixed point. Note that neither the factorizing nor the $BF$ fixed point itself can be described by the chosen parametrization.\label{fig:phase_14}}
\end{center}
\end{figure}

For $k=5$, we have three  fixed point intertwiners that we can linearly combine: $J=0$, which is identical to the degenerate fixed point, $J=1$, which flows to the factorizing model with maximal spin $J=1$ and $J=2$, which flows to the analogue $BF$ fixed point. This gives us two parameters, $\alpha_0$ and $\alpha_1$, to tune, $\alpha_2$ is fixed by the requirement that $\sum_J \alpha_J=1$. The phase diagram is shown in figure \ref{fig:phase_14}.

From the phase diagram we can readily see that we find an extended phase for each fixed point associated to an (initial) fixed point intertwiner. This is an important difference to the results in \cite{bdetal13}, where also a factorizing fixed point has been found, however no choice of initial parameters has led to a flow (for high accuracy) to this fixed point and thus it has no associated phase. The second important observation is the relative size of the phases: The two dominating phases are the factorizing and the analogue $BF$ one whereas the degenerate one is the smallest\footnote{As mentioned above the shape of the phases is likely to change for higher accuracy.}. This is an indication that the chosen fixed point intertwiners define initial models `far away' from the degenerate fixed point. Since the degenerate phase is not of our major interest, we will thus neglect it in the next phase diagram.

\subsubsection{Phase diagram for $k=8$}

\begin{figure}[h!]
\begin{center}
\includegraphics[scale=0.6]{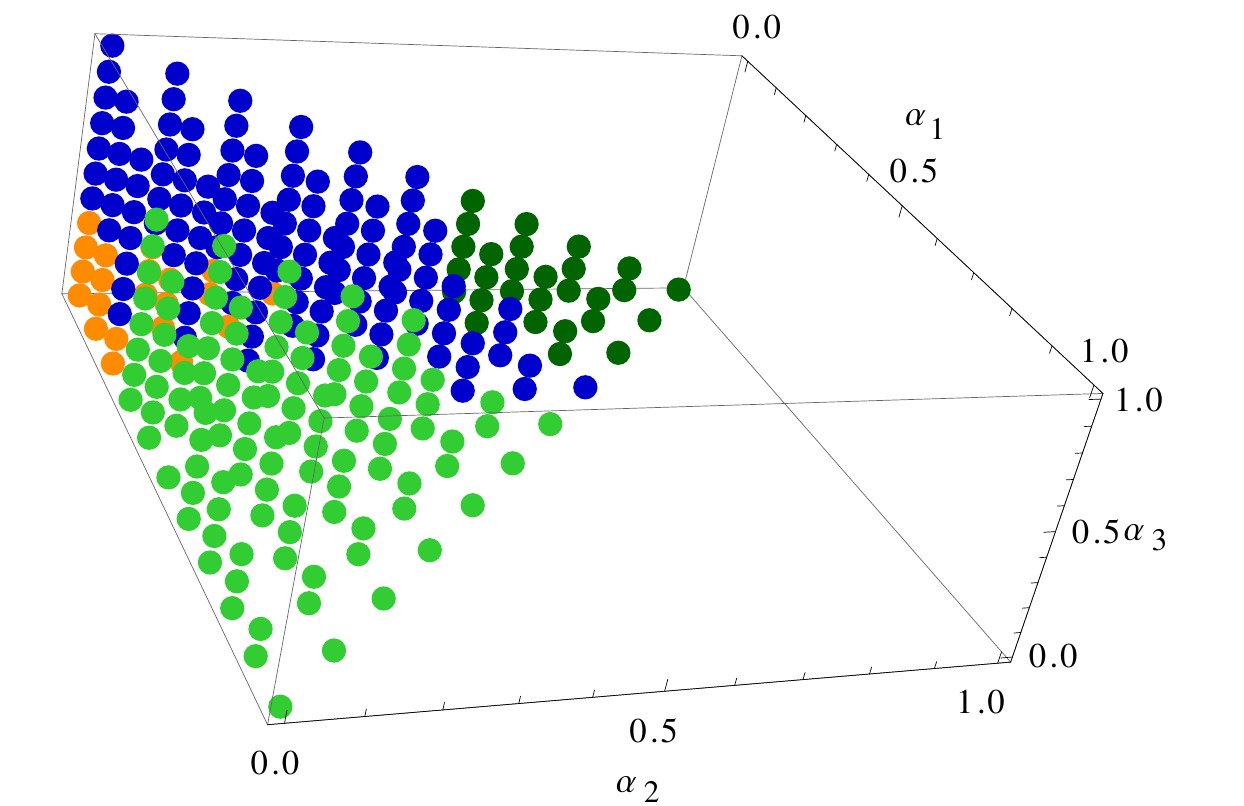}
\caption{Phase diagram for $k=8$ with $\alpha_0=0$. The coloured dots indicate to which fixed point the respective initial models flow to: The green dots show the factorizing models, lighter green for $J=1$ (area that starts at the vertex $(\alpha_1,\alpha_2,\alpha_3)=(1,0,0)$), darker for $J=2$ (area that starts at the vertex (0,1,0)). Analogue $BF$ theory is blue (area that starts at the vertex (0,0,1)). The so-called `mixed' fixed point is orange (area that starts at the vertex (0,0,0)).\label{fig:full_phase20}}
\end{center}
\end{figure}

\begin{figure}[h!]
\begin{center}
\includegraphics[scale=0.7]{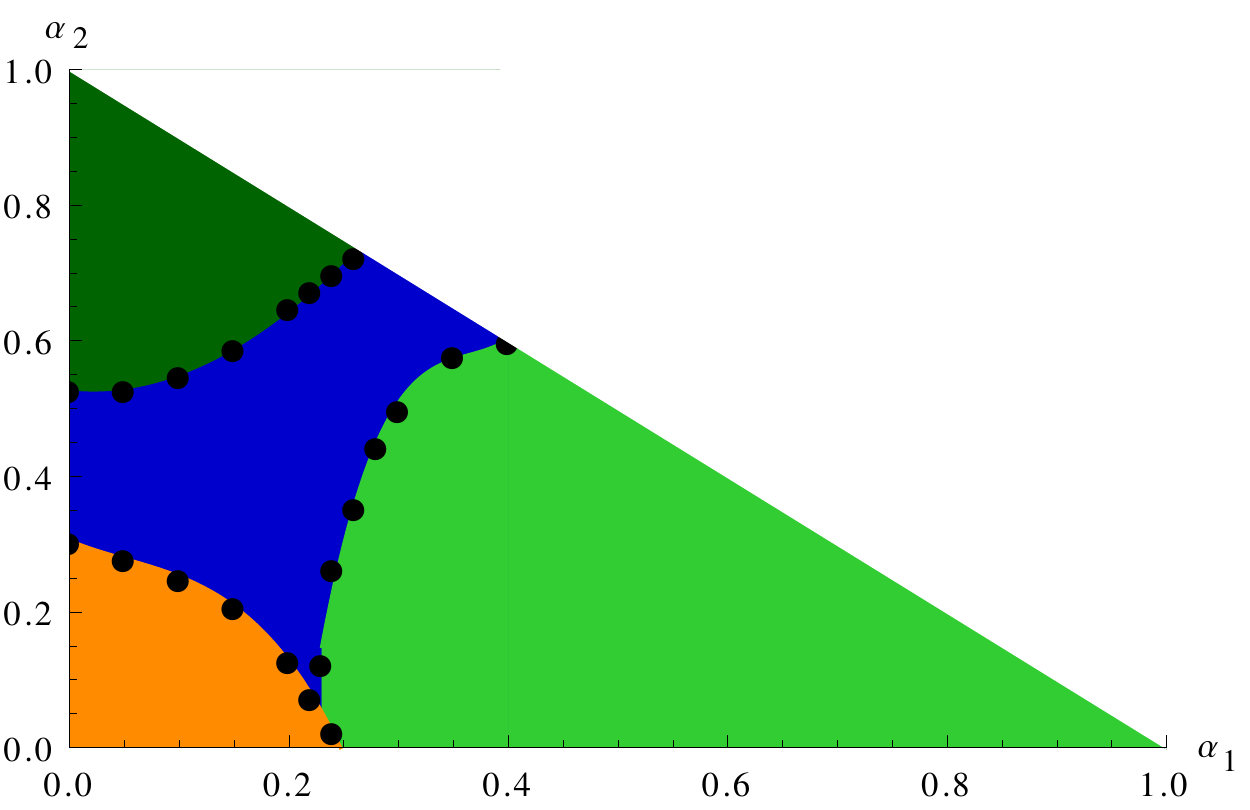}
\caption{Slice of the phase diagram for $k=8$ with $\alpha_0=\alpha_3=0$. The colouring is the same as in the previous diagram, namely the right corner corresponds to models flowing to the factorizing fixed point with $J=1$, the upper left corner corresponds to models that flow to the factorizing fixed point with $J=2$, the models at the bottom left corner flow to the `mixed' fixed point, and the phase in between these three phases corresponds to analogue BF.\label{fig:slice_phase20}}
\end{center}
\end{figure}

For the quantum group with $k=8$, we discuss the linear combintation of four  fixed point intertwiners, each labelled with a maximal (even) spin $1 \leq J \leq 4$, where we neglect $J=0$ as argued above. Together with the requirement that $\sum_{J} \alpha_J=1$, we have three free parameters. In figure \ref{fig:full_phase20} we show the full parameter space, with a raster of coloured points indicating the fixed point they flow to. In figure \ref{fig:slice_phase20} we show the interesting slice, where $\alpha_3=0$.

As in the previous diagram, we find extended phases for all fixed points, here the two factorizing fixed points for $J=1$ and $J=2$, a phase for analogue $BF$ theory and one for the `mixed' fixed point. Again, the two dominating phases are analogue $BF$ theory and the factorizing fixed point with $J=1$. Of particular interest is the special slice that we picked in figure \ref{fig:slice_phase20} because of the following two observations: First this slice shows clearly that the analogue $BF$ fixed point is very attractive, since in this slice its associated fixed point intertwiner is not excited, $\alpha_3=0$. Even if we stay on the line given by $\alpha_1 + \alpha_2 =1$, i.e. the diagonal boundary in figure \ref{fig:slice_phase20}, the system flows to $BF$ for an intermediate region between the two phases and spoils a direct phase transition between the two factorizing phases. Second we want to emphasize the `mixed' fixed point in the bottom left corner because it is a highly non-trivial fixed point and might allow for a direct phase transition to one of the factorizing models\footnote{It might happen that the system flows to analogue $BF$ theory for an intermediate region if the accuracy of the algorithm is increased.}.

\section{Spin foam interpretation of spin nets and fixed points}
\label{interpretation}

As explained shortly in section \ref{sec:reisenberger} spin nets can be understood as dimensional reductions of spin foams. Here we will make this picture more concrete, which will enable us to provide an interpretation of the coarse graining flow and the fixed points in terms of spin foams.

Let us consider a 3D spin foam defined on a cubical regular lattice. We will actually just need  to consider `one slice of cubes'.  In (\ref{i3}) we gave the partition function of a general spin foam as
\ba\label{sf1}
Z=\sum_{j_f} \prod_f \tilde \omega_f(j_f)   \,\prod_e ({\mathcal P^e})_{m_{f_1} m_{f_2} m_{f_3} m_{f_4}}^{n_{f_1} n_{f_2}  n_{f_3} n_{f_4}} \q 
\ea
where we already specialized to the case of four--valent edges. To obtain a spin net model we choose the edge intertwiners anisotropically. That is for the vertical edges we take the original ${\mathcal P}^e$ of the spin foam model. However for the horizontal edges we replace the intertwining operators with ${({\mathcal P}_{hor})}_{m_1m_2}^{n_1n_2}= \delta_{n_1,n_2} \delta^{m_1,m_2}$. Here we consider  just a two--valent intertwining operator -- this can be understood as imposing $j=0$ on all horizontal faces. Thus the magnetic indices for these horizontal faces can be omitted.  Furthermore ${{\mathcal P}_{hor}}$ is chosen such that the contributions from different horizontal rows of cubes factorize. Then, as anticipated before, we just need to consider one row of cubes.

\begin{figure}
\begin{center}
\includegraphics[width=0.4\textwidth]{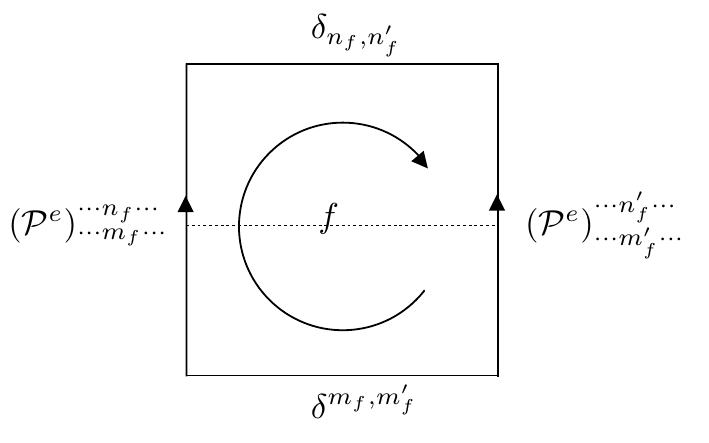}
\caption{ To each vertical face of the considered row of cubes, we attach  ${\mathcal P}^e$ to the vertical edges and $\delta$'s to the horizontal edges. These $\delta$'s contract the vertical projectors ${\mathcal P}^e$. \label{sf-face}}
\end{center}
\end{figure}

The contraction of indices among the vertical edge operators ${\mathcal P^e}$ proceeds as indicated in Figure \ref{sf-face}. That is given a face  $f$ the $\delta$'s from the horizontal edges contract the $n$ and $m$ indices of this face from the two vertical edge operators ${\mathcal P^e}$. Hence we can replace the face by an edge with indices $j,m,n$ connecting now vertices, which represent the former edges.

Thus the spin net model is given by
\ba
Z_{net}=\sum_{j_e} \prod_e \tilde \omega_e(j_e)   \,\prod_v ({\mathcal P^v})_{m_{e_1} m_{e_2} m_{e_3} m_{e_4}}^{n_{e_1} n_{e_2}  n_{e_3} n_{e_4}} \q ,
\ea
where the edge weights $\tilde \omega_e$ can be absorbed into the vertex weights $({\mathcal P^v})$. 

\begin{figure}
\begin{center}
\includegraphics[width=0.25\textwidth]{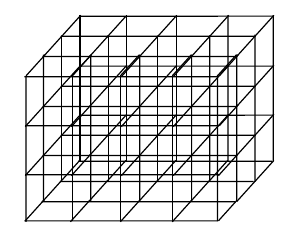}
\hspace{1cm}
\includegraphics[width=0.25\textwidth]{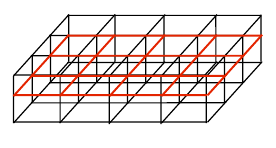}
\hspace{1.2cm}
\includegraphics[width=0.1\textwidth]{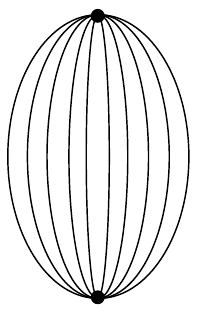}
\caption{Given a spinfoam model defined in the cubic lattice, for the corresponding spin net model we just need to consider one row of cubes. The spin net model is then defined on the 1-dimensional reduction  of the row of cubes, namely the square lattice, here distinguished by thick red lines. We can understand this reduction as making all the horizontal edges of the spinfoam belonging to the same slice to collapse in a single vertex. For the slice of cubes we then get two vertices linked by all the vertical edges, like the super melon spinfoam of the right. \label{cubes}}
\end{center}
\end{figure}

In this sense we can understand a spin net as a spin foam with just two vertices and a very large number of faces, all of which have only two edges. All edges and faces share the same two vertices, which leads to a (super) melon in the sense of \cite{melons}, see Figure \ref{cubes}.  The fact that there are only two vertices explains why the local gauge symmetry of spin foams is converted into a global symmetry for the spin net. This is given by a left and right action of the group (on the $m$ and on the $n$ indices) corresponding to the two vertices.

Such melons play a crucial role in the investigation of coloured tensor models \cite{melons}. Note however that the renormalization flow considered here is quite different from the one usually considered (which rather integrates out complete melons). In this work a (super) melon with many edges is coarse grained into another melon with fewer (effective) edges. For the spin net this corresponds to blocking a number of vertices into effective vertices.

We can therefore interpret spin net coarse graining as a highly anisotropic way of coarse graining (anisotropic) spin foams.  (In this case the analogue $BF$ phase is indeed the $BF$ phase for spin foams.) It focusses on the change of coupling between two spin foam vertices with the coarse graining flow.
Following this interpretation, we see that the models that flow to  factorizing fixed points lead in the spin foam interpretation to a factorization of vertex amplitudes. That is the contribution of the two vertices decouples into two parts associated to the upper and lower vertex of the super melon. That is gluing two spin foam vertices with many edges and blocking these into effective edges, we actually end up with two decoupled vertices, that is the melon is cut into two halves.

The decoupling of vertices is certainly a possible scenario of spin foams, of course not a very attractive one. It will be interesting to see whether this also occurs for Barrett--Crane like models with $\text{SU}(2)_k\times \text{SU}(2)_k$ structure group. For these models concerns have been voiced, that neighbouring tetrahedra are not sufficiently glued with each other, see the discussion in \cite{eprl,ryan08,alexandrov}. 

Of course we cannot be certain whether the corresponding spin foam models will really flow to the corresponding factorizing fixed points, as we are considering a very anisotropic situation. Here an investigation at least into 3D spin foams seems to be in reach employing a coarse graining algorithm proposed in \cite{eckert}, which has been already tested for Abelian models \cite{sebprivate}. Alternatively it may be possible to devise a coarse graining scheme in which the coarse graining in equatorial direction of the melons, and integrating out entire melons, alternate. 

From the spin foam point of view the `mixed fixed points' seems to be particular interesting and encouraging, as here the vertices do not decouple. It has to be investigated whether these (spin net) fixed points give also triangulation invariant spin foam models.

 Such mixed fixed points seem to result from a subtle interplay between the two different sources of the spin foam models: On the one hand spin foams can be understood as generalized lattice gauge theories. However, as noticed in \cite{bdetal13}, the phases of standard lattice gauge theory seem to be dominating. The $BF$ phase can be understood as leading to a maximal gluing between the  `two spin foam vertices' in the spin net model. Simplicity constraints indeed weaken this gluing (alone by restricting the sum over intertwiner degrees of freedom) and allow to flow out of the standard lattice gauge theory phases. However if this effect is too strong the two spin foam vertices (in the spin net model) decouple. In this sense the mixed fixed point represents a balance between these two effects. 

We think that it will be very worthwhile to push and test such an interpretation, as it could reveal the effect that the imposition of simplicity constraints has on the large scale limit of spin foams and in this way possibly resolve  the question whether these have to imposed more strongly or not, see for instance \cite{eprl,ryan08,alexandrov}.

\section{Discussion}
\label{discussion}

In this work we have taken several important steps towards a full understanding of the continuum limit of spin foam models. We in particular introduced and defined models based on the structure group $SU(2)_k$ that can encode the dynamics of the full gravitational models, but are still feasible to investigate numerically. Note that apart from certain technical subtleties (e.g. the definition of the duals) for the quantum group coarse graining, that we resolved, this nevertheless requires very  efficient numerical algorithms\footnote{Even the one--singular value algorithm leads to a bond dimension of e.g. $\chi=49$ for $k=12$. The tensors we are employing have five such indices, which run from $1$ to $49$. Thus already for  saving these tensors one needs to employ an efficient scheme, not speaking off contracting indices of several such tensors. A technique we developed for this work are `super indices' which allow to save and deal only with entries not--vanishing  as given by the coupling rules. This leads already to a huge reduction of components from $(49)^5$ to $(2359)^2$.}.  For this the symmetry protected tensor network algorithm developed here and in \cite{bdetal13} is absolutely crucial.

We considered mainly spin nets, as dimensional reductions of spin foams, in this work. However, we point out that the spin net coarse graining flow is equivalent to a coarse graining flow of spin foam melons into spin foam melons. It particular focusses on the coupling of (two) spin foam vertices with each other -- which is of course crucial to understand the macroscopic behaviour of spin foams. Indeed it allows us to describe the properties of the fixed points found via coarse graining in terms of this coupling. Note that this question cannot even be considered if one performs coarse graining only by $5-1$ Pachner moves \cite{bahretal12,unpublished}.

The coarse graining of the spin net models is encoded in a flow equation (\ref{flow}), which describes the behaviour of the intertwiner degrees of freedom. We believe that spin foam coarse graining (including non--melonic spin foams of course) will lead to quite similar coarse graining equations. Thus one might speculate whether the phases we find are already the ones one will encounter for spin foam models.

The coarse graining flow equation describes that the effective intertwiner degrees of freedom are obtained from the old intertwiner degrees of freedom basically by two operations: the first is a  contraction of the tensors encoding the intertwiners, the second is a convolution with recoupling symbols. This convolution leads to a non--trivial flow of intertwiner degrees of freedom, i.e. intertwiner channels which are not excited initially, can become excited by the coarse graining flow. This is again not the case, if considering coarse graining via $5-1$ Pachner moves only \cite{unpublished}.

Here again two effects seem to compete with each other: without simplicity constraints, i.e. models described by the Haar intertwiner where we have a restriction to $j_5=j_5'$, the recoupling symbols appearing in the flow equation, contract with each other (approximately) to a Kronecker Delta. In this case one can indeed approximate the flow by a Migdal Kadanoff recursion \cite{migdal,kadanoff}, see also  \cite{eckert}, where recoupling symbols do not appear. The flow is (approximately) restricted to the phase space where the spins of left and right copy agree, i.e. $j_5=j_5'$ throughout the coarse graining. In the case of $BF$ fixed points it leads to a gluing/coupling of the (two) spin foam vertices. This however only allows to reach the degenerate fixed point or a $BF$ theory, either of the full structure group or of a normal subgroup (in the case that one actually deals with a proper group). 

Simplicity constraints, which lead to components with $j_5 \neq j_5'$ in the tensor $\hat t_1$,  do change this picture in an essential way: now the convolution with the recoupling symbols appearing in the flow equation (\ref{flow}) becomes crucial and leads to a large set of additional fixed points. Indeed an open issue is the understanding of all possible fixed points, which might be possible with techniques developed in \cite{bw1,bw2} which allows already the understanding of all factorizing fixed points.

As we have seen for a number of examples the flow can now lead to factorizing fixed points, where the copy labelled by $j$ indices and the copy labelled by $j'$ indices decouple. In terms of (melonic) spin foams  this means a decoupling of the two spin foam vertices involved. For reasons discussed below it would be interesting to see whether such a decoupling appears in the Barrett Crane model \cite{bc} with a quantum group \cite{igor}.

However we have also examples which flow to so--called mixed fixed points, in which both a coupling between the two spin foam vertices persists and nevertheless simplicity constraints persists. It will therefore be very interesting to investigate these fixed points further and in particular study whether these can be lifted to fixed points for (non--melonic) spin foam models.  Furthermore a full classification of these fixed points, along the lines of \cite{bw1,bw2}, would reveal further possible phases for spin nets and spin foams.
  
Both cases constitute non--trivial possible phases for spin nets and for spin foams (at least the factorizing models lead to phases for spin foams as all vertices decouple). As discussed in section \ref{interpretation} a decoupling of spin foam vertices might indeed be related to concerns that tetrahedra are not properly glued to each other, e.g.\ in the Barrett Crane model \cite{eprl,ryan08,alexandrov}. It would therefore help and potentially resolve the discussions on the correct way to impose simplicity constraints, to study coarse graining for other geometric configurations. Starting from a coarse graining of melons, it might be possible to alternate the coarse graining considered here -- in equatorial direction of the melon -- with a coarse graining in vertical direction. This might already lead to a less anisotropic coarse graining, which takes also the blocking of spin foam vertices  (and not only edges as happens for the melonic flow considered here) into account.

Apart from the identification of possible phases for spin foams we can draw another important lesson from the work here, in particular compared to \cite{bdetal13}. This is related to the introduction of a new parametrization of the space of models, based on Reisenberger's construction principle and fixed point intertwiners \cite{bw1,bw2}. We have argued that this parametrization makes it much easier to satisfy the projector property, which guarantees invariance under edge subdivisions in spin foams.

Indeed the picture that this parametrization gives is quite different from the one based on $E$--functions in \cite{bdetal13}, in which the projector property is quite complicated to implement \cite{warsaw}. The parametrization in \cite{bdetal13} requires a fine tuning between the degenerate and $BF$ phase, to flow to a non--trivial (actually factorizing) model. This is not the case for the fixed point intertwiner based parametrization, which leads to a rich spectrum of non--trivial fixed points, of both factorizing and non--factorizing type. We conjecture that this is due to the projector property, which itself can be interpreted as a very weak imposition of diffeomorphism symmetry \cite{bojowaldperez}. The fine tuning in \cite{bdetal13} can thus be understood as satisfying this weak requirement of diffeomorphism invariance, as already argued in \cite{bdetal13}. The results here therefore stresses the importance of this concept \cite{bahrdittrich,ditt}. 

Interestingly the phase diagrams, obtained by considering linear combinations of fixed point intertwiners, support this conclusion even more.  Non--trivial fixed points appear in particular around the fixed point intertwiners itself. In the intermediate regions the models flow typically to the $BF$ fixed point. Although there seem to occur phase transition between non--trivial phases, e.g. in figure \ref{fig:slice_phase20}, this might disappear if a higher accuracy is employed: It might happen, that all non--trivial phases are separated by models flowing to the $BF$ fixed point. We leave the falsification or verification of this picture for future work.

Apart from the open questions mentioned so far there are numerous other directions for further work, some of which we outline here:
\begin{itemize}
\item Given the results we have found here, i.e. that fixed point intertwiner models flow to interesting fixed points, it would be of high interest to classify  such fixed points for intertwiner models, for $SU(2)_k \times SU(2)_k$, as is performed in \cite{bw1,bw2} for $SU(2)_k$. The Barrett--Crane model constitutes one such fixed point, the question is, whether a Barbero--Immirzi parameter, as appearing in the EPRL, FK and  BO models\footnote{i.e. Engle-Pereira-Rovelli-Livine, Freidel-Krasnov and Baratin-Oriti models.} \cite{eprl} can be accommodated by such a fixed point.
\item Further approximations and therefore simplifications of the flow equation (\ref{flow}) will be useful to allow the treatment of $SU(2)_k\times SU(2)_k$ and similar structure groups.
\item Coarse graining involving non--melonic spin foams should be developed along the lines of spin net coarse graining. As is discussed in \cite{refining} the tensor network algorithm might be adaptable to the case of spin foams in a more straightforward way then envisaged in \cite{eckert}, which presented a tensor network formulation of spin foams.
\item It would be interesting to investigate in further detail the phase transitions. Indeed one would expect that an interacting theory, such as 4D gravity, can only occur there. Here the first question is, whether the models along the phase transition carry `only' conformal symmetry, or lead to a fully (cylindrically) consistent continuum limit as outlined in \cite{bd12b,refining}. As is argued in \cite{refining} such a cylindrically consistent continuum limit is deeply entangled with the restoration of diffeomorphism symmetry, which  is typically broken by the discretization itself    \cite{bahrdittrich}. From a more tensor network renormalization point of view \cite{guwen,kadanoff} discusses the investigation of fixed point tensors at critical fixed points and the extraction of critical exponents. 
\end{itemize} 

In summary, we hope to have convinced the reader, that the tensor network coarse graining methods \cite{levin,guwen} and the many refinements thereof developed here and in \cite{eckert,bdetal13}, together with the conceptual understanding of how to  construct the continuum limit in \cite{bd12b,refining},  put the understanding of the possible phases of spin foam models  within the very near future.

\section*{Acknowledgements}

We are deeply thankful to Wojciech Kaminski for lots of discussions and introducing us to diagrammatical calculus. BD thanks furthermore Wojciech Kaminski for collaboration on \cite{bw1,bw2}. We also wish to thank Joseph Ben Geloun, Valentin Bonzom, Oliver Buerschaper, Lukasz Cincio, Ma{\"i}t{\'e} Dupuis, Laurent Freidel, Florian Girelli, Jeff Hnybida, Aldo Riello and Guifre Vidal for valuable discussions.
 S.St. gratefully acknowledges support by the DAAD
(German Academic Exchange Service) and would like to thank Perimeter Institute for an Isaac Newton
Chair Graduate Research Scholarship. This research was supported in part by Perimeter Institute for
Theoretical Physics. Research at Perimeter Institute is supported by the Government of Canada through
Industry Canada and by the Province of Ontario through the Ministry of Research and Innovation.

\appendix

\section{Additional diagrammatic calculations}
\label{app}

\subsection{Normalization of Haar projector}
\label{app-Haar}

The Haar projector $\mathcal{P}$ reads
\begin{equation}
\mathcal{P}_{\left(\{m\},\{m'\}\right)}(j_1,j_2,j_3,j_4) := \sum_{j_5} c(j_5) 
\begin{tikzpicture}[baseline,scale=0.75]
\draw (-0.5,-1) -- (0,-0.5) -- (0,0.5) -- (-0.5,1)
      (0.5,-1) -- (0,-0.5)
      (0,0.5) -- (0.5,1)
      (-0.5,-1.25) node {$j_3$}
      (0.5,-1.25) node {$j_4$}
      (0.5,1.25) node {$j_1$}
      (-0.5,1.25) node {$j_2$}
      (0.25,0) node {$j_5$};
\end{tikzpicture}
\; \otimes \;
\begin{tikzpicture}[baseline,scale=0.75]
\draw (-0.5,0.75) -- (-0.25,1) arc(180:0:0.5) -- (0.75,-1) arc(0:-180:0.5) -- (-0.5,-0.75)
      (-0.25,1) -- (0,0.75)
      (-0.25,-1) -- (0,-0.75)
      (-0,0.5) node {$j_1$}
      (-0.5,0.5) node {$j_2$}
      (-0,-0.5) node {$j_4$}
      (-0.5,-0.5) node {$j_3$}
      (1,0) node {$j_5$};
\end{tikzpicture}
\quad ,
\end{equation}
where $c(j_5)$ is the normalization constant that we want to determine. For $\mathcal{P}$ to be a projector, it has to satisfy the relation $\mathcal{P}^2 = \mathcal{P}$, namely
\begin{equation}
\mathcal{P} \cdot \mathcal{P} = \sum_{j_5,j'_5} c(j_5) \, c(j'_5) 
\begin{tikzpicture}[baseline,scale=0.75]
\draw (-0.5,-1) -- (0,-0.5) -- (0,0.5) -- (-0.5,1)
      (0.5,-1) -- (0,-0.5)
      (0,0.5) -- (0.5,1)
      (-0.5,-1.25) node {$j_3$}
      (0.5,-1.25) node {$j_4$}
      (0.5,1.25) node {$j_1$}
      (-0.5,1.25) node {$j_2$}
      (0.25,0) node {$j'_5$};
\end{tikzpicture}
\; \otimes \; \underbrace{
\begin{tikzpicture}[baseline,scale=0.8]
\draw (-0.25,-0.75) -- (0,-0.5) -- (0,0.5) -- (-0.25,0.75) arc(0:180:0.25) -- (-1,0.5) -- (-1,-0.5)
      (0.25,-0.75) -- (0,-0.5)
      (0,0.5) -- (0.25,0.75) arc(0:180:0.75) -- (-1,0.5)
      (-0.25,-0.75) arc(0:-180:0.25) -- (-1,-0.5)
      (0.25,-0.75) arc(0:-180:0.75) -- (-1,-0.5)
      (-1.25,0) node {$j_5$}
      (0.25,0) node {$j'_5$}
      (-0.5,0.75) node {$j_1$}
      (-0.5,-0.75) node {$j_3$}
      (-0.5,1.75) node {$j_2$}
      (-0.5,-1.75) node {$j_4$};
\end{tikzpicture}
}_{(-1)^{j_1+j_2+j_3+j_4} \left(d_{j_5}\right)^{-1} \delta_{j_5 j'_5}}\; \otimes \;
\begin{tikzpicture}[baseline,scale=0.75]
\draw (-0.5,0.75) -- (-0.25,1) arc(180:0:0.5) -- (0.75,-1) arc(0:-180:0.5) -- (-0.5,-0.75)
      (-0.25,1) -- (0,0.75)
      (-0.25,-1) -- (0,-0.75)
      (-0,0.5) node {$j_1$}
      (-0.5,0.5) node {$j_2$}
      (-0,-0.5) node {$j_4$}
      (-0.5,-0.5) node {$j_3$}
      (1,0) node {$j_5$};
\end{tikzpicture}
\; \overset{!}{=} \mathcal{P}\quad .
\end{equation}
We therefore deduce that
\begin{equation}
c(j_5) = (-1)^{j_1+j_2+j_3+j_4}d_{j_5} \quad,
\end{equation}
as we wanted to proof.

\subsection{Relation between the two different recoupling schemes}
\label{app-recouplings}

We are looking for the relation
\begin{equation}
\begin{tikzpicture}[baseline,scale=0.75]
\draw (-1,0.5) -- (-0.5,0) -- (0,0) -- (0.5,0.5)
      (-0.5,0) -- (-1,-0.5)
      (0,0) -- (0.5,-0.5)
      (-1,-0.75) node {$j_3$}
      (-1,0.75) node {$j_2$}
      (0.5,0.75) node {$j_1$}
      (0.5,-0.75) node {$j_4$}
      (-0.25,0.25) node {$j_6$};
\end{tikzpicture}
\; = \; \sum_{j_5} c(j_5,j_6) 
\begin{tikzpicture}[baseline,scale=0.75]
\draw (-0.5,-0.75) -- (0,-0.25) -- (0,0.25) -- (-0.5,0.75)
      (0.5,-0.75) -- (0,-0.25)
      (0,0.25) -- (0.5,0.75)
      (-0.75,-0.75) node {$j_3$}
      (0.75,-0.75) node {$j_4$}
      (0.75,0.75) node {$j_1$}
      (-0.75,0.75) node {$j_2$}
      (0.25,0) node {$j_5$};
\end{tikzpicture}
\quad .
\end{equation}
In order to find the coefficient $ c(j_5,j_6) $ we contract the above expression with the diagram \eqref{dualize}:
\begin{align}
\begin{tikzpicture}[baseline,scale=0.8]
\draw (-1,0.5) -- (-0.5,0) -- (0,0) -- (0.5,0.5) -- (-0.25,1) arc(180:0:0.75) -- (1.25,-1) arc(0:-180:0.75) -- (0.5,-0.5)
      (-0.5,0) -- (-1,-0.5) -- (-0.25,-1)
      (0,0) -- (0.5,-0.5)
      (-1,0.5) -- (-0.25,1)
      (-1,-0.75) node {$j_3$}
      (-1,0.75) node {$j_2$}
      (0.5,0.75) node {$j_1$}
      (0.5,-0.75) node {$j_4$}
      (-0.25,0.25) node {$j_6$}
      (1,0) node {$j'_5$};
\end{tikzpicture}
\; & = \; \sum_{j_5} c(j_5,j_6) 
\begin{tikzpicture}[baseline,scale=0.75]
\draw (-0.5,-0.75) -- (0,-0.25) -- (0,0.25) -- (-0.5,0.75) -- (0,1.25) arc(180:0:0.75) -- (1.5,-1.25) arc(0:-180:0.75) -- (0.5,-0.75)
      (0.5,-0.75) -- (0,-0.25)
      (0.5,0.75) -- (0,1.25)
      (-0.5,-0.75) -- (0,-1.25)
      (0,0.25) -- (0.5,0.75)
      (-0.75,-0.75) node {$j_3$}
      (0.75,-0.75) node {$j_4$}
      (0.75,0.75) node {$j_1$}
      (-0.75,0.75) node {$j_2$}
      (0.25,0) node {$j_5$}
      (1.25,0) node {$j'_5$};
\end{tikzpicture}
\end{align}
The evaluation of both sides of this equation gives 
\begin{align}
 (-1)^{j_1+j_2+j_3+j_4} \left( d_{j_5}d_{j_6} \right)^{-\frac{1}{2}}
\left[
\begin{matrix}
\, j_1 \, & \, j_2 \, & \, j'_5 \, \\
\, j_3 \, & \, j_4 \, & \, j_6 \,
\end{matrix}
\right]
& = (-1)^{j_1 + j_2 + j_3 + j_4} \sum_{j_5} c(j_5,j_6) \left(d_{j_5} \right)^{-1} \delta_{j_5,j'_5} \q,
\end{align}
from which we deduce
\begin{align}
\begin{tikzpicture}[baseline,scale=0.75]
\draw (-1,0.5) -- (-0.5,0) -- (0,0) -- (0.5,0.5)
      (-0.5,0) -- (-1,-0.5)
      (0,0) -- (0.5,-0.5)
      (-1,-0.75) node {$j_3$}
      (-1,0.75) node {$j_2$}
      (0.5,0.75) node {$j_1$}
      (0.5,-0.75) node {$j_4$}
      (-0.25,0.25) node {$j_6$};
\end{tikzpicture}
\; & = \; \sum_{j_5}  \sqrt{\frac{d_{j_5}}{d_{j_6}}} 
\left[
\begin{matrix}
\, j_1 \, & \, j_2 \, & \, j'_5 \, \\
\, j_3 \, & \, j_4 \, & \, j_6 \,
\end{matrix}
\right] 
\begin{tikzpicture}[baseline,scale=0.75]
\draw (-0.5,-0.75) -- (0,-0.25) -- (0,0.25) -- (-0.5,0.75)
      (0.5,-0.75) -- (0,-0.25)
      (0,0.25) -- (0.5,0.75)
      (-0.75,-0.75) node {$j_3$}
      (0.75,-0.75) node {$j_4$}
      (0.75,0.75) node {$j_1$}
      (-0.75,0.75) node {$j_2$}
      (0.25,0) node {$j_5$};
\end{tikzpicture}
\quad .
\end{align}

\subsection{Splitting of the $9j$-symbol into two $6j$-symbols}
\label{app9j}

Let us prove the equality $\eqref{9j}$. 
By using the identities

\begin{equation} \label{eq:identity1}
\begin{tikzpicture}[baseline,scale=0.75]
\draw (0,-1) -- (0,1) 
(1,-1) -- (1,1) 
         (-0.25,0) node {$j_2$}
      (1.25,0) node {$j_4$};
\end{tikzpicture}
\; = \;
\begin{tikzpicture}[baseline,scale=0.75]
\draw (-0.5,-1)-- (0,-0.5) -- (0,0.5) --(-0.5,1)
	(0,0.5) --(0.5,1)
	(0.5,-1)-- (0,-0.5)
      (0.3,0) node {$j_9$}
      (-0.75,1) node {$j_2$}
      (0.75,1) node {$j_4$}
       (-0.75,-1) node {$j_2$}
      (0.75,-1) node {$j_4$};
\end{tikzpicture}
\; (-1)^{j_2+j_4-j_9}{d_{j_9}} \;
\qquad ,\qquad
\begin{tikzpicture}[baseline,scale=0.75]
\draw [pattern=north east lines] (0,0) circle[radius = 0.5];
\draw (0,0.5) -- (0,1.5) 
      (0,-0.5) -- (0,-1.5)
      (0.25,-1) node {$j_9$}
      (0.25,1) node {$j_9$};
\end{tikzpicture}
\; = \;
\begin{tikzpicture}[baseline,scale=0.75]
\draw [pattern=north east lines] (0,0) circle[radius = 0.5];
\draw (0,0.5) -- (0,1) arc(180:0:0.5) -- (1,-1) arc(0:-180:0.5)
      (0,-0.5) -- (0,-1)
      (1.25,0) node {$j_9$};
\end{tikzpicture}
\; \frac{(-1)^{2j_9}}{d_{j_9}} \;
\begin{tikzpicture}[baseline,scale=0.75]
\draw (0,-1) -- (0,1)
      (0.3,0) node {$j_9$};
\end{tikzpicture} \quad ,
\end{equation}
we can manipulate the $9j$-symbol in the following fashion,
\begin{align}
&
\begin{tikzpicture}[baseline,scale=0.8]
\draw (-1,-1) -- (-0.5,-0.5) -- (-0.5,0.5) -- (-1,1) -- (0,1.5) arc(180:0:0.75) -- (1.5,-1.5) arc(0:-180:0.75) -- (-1,-1)
      (0,-1.5) -- (1,-1) -- (0.5,-0.5) -- (0.5,0.5) -- (1,1) -- (0,1.5)
      (-0.5,-0.5) -- (0.5,-0.5)
      (-0.5,0.5) -- (0.5,0.5)
      (0,-0.75) node {$j_3$}
      (-0.75,0) node {$j_2$}
      (0.75,0) node {$j_4$}
      (0,0.75) node {$j_1$}
      (-1.25,-1.) node {$j_7$}
      (1.25,-1.) node {$j_8$}
      (1.25,1.) node {$j_5$}
      (-1.25,1.) node {$j_6$}
      (1.75,0) node {$j_9$};
\end{tikzpicture}
\; = \;
(-1)^{j_2 + j_4 - j_9} d_{j_9}
\begin{tikzpicture}[baseline,scale=0.9]
\draw (-0.5,-0.75) -- (0,-0.25) -- (0,0.25) -- (-0.5,0.75) -- (0,1.25) arc(180:0:0.75) -- (1.5,-1.25) arc(0:-180:0.75) -- (0.5,-0.75)
      (0.5,-0.75) -- (0,-0.25)
      (0.5,0.75) -- (0,1.25)
      (-0.5,-0.75) -- (0,-1.25)
      (0,0.25) -- (0.5,0.75)
      (-0.5,0.75) -- (0.5,0.75)
      (-0.5,-0.75) -- (0.5,-0.75)
       (0,-0.5) node {$j_3$}
      (0,0.55) node {$j_1$}
      (-0.6,1) node {$j_6$}
      (0.6,1) node {$j_5$}
      (0.6,0.5) node {$j_4$}
      (-0.6,0.5) node {$j_2$}
      (0.6,-0.4) node {$j_4$}
      (-0.6,-0.4) node {$j_2$}
      (-0.6,-1) node {$j_7$}
      (0.6,-1) node {$j_8$}
      (0.25,0) node {$j_9$}
      (1.75,0) node {$j_9$};
\end{tikzpicture} 
\; = \;
(-1)^{j_2+j_4-j_9}
\begin{tikzpicture}[baseline,scale=0.9]
\draw (0,1.75) arc(180:0:0.5) -- (1.,0.75) arc(0:-180:0.5)
      (0,-0.75)  arc(180:0:0.5) -- (1,-1.75) arc(0:-180:0.5)
      (0,0.75)--(0.5,1.25)
        (0,1.75)--(0.5,1.25)
        (0,1.75)--(-0.5,1.25)
           (0,0.75)--(-0.5,1.25)
           (-0.5,1.25)--(0.5,1.25)
            (0,-0.75)--(0.5,-1.25)
        (0,-1.75)--(0.5,-1.25)
        (0,-1.75)--(-0.5,-1.25)
           (0,-0.75)--(-0.5,-1.25)
           (-0.5,-1.25)--(0.5,-1.25)
      (0.,-1.) node {$j_3$}
      (0.,1.1) node {$j_1$}
      (-0.6,1.6) node {$j_6$}
      (0.6,1.6) node {$j_5$}
      (0.6,0.75) node {$j_4$}
      (-0.6,0.75) node {$j_2$}
      (0.6,-0.75) node {$j_4$}
      (-0.6,-0.75) node {$j_2$}
      (-0.6,-1.6) node {$j_7$}
      (0.6,-1.6) node {$j_8$}
      (1.25,-1.25) node {$j_9$}
      (1.25,1.25) node {$j_9$};
\end{tikzpicture}
\q .
\end{align}
The diagrams of the right are the $6j$-symbols defined in Eq. \eqref{6j-def}, and then we get
\begin{align}
\begin{tikzpicture}[baseline,scale=0.8]
\draw (-1,-1) -- (-0.5,-0.5) -- (-0.5,0.5) -- (-1,1) -- (0,1.5) arc(180:0:0.75) -- (1.5,-1.5) arc(0:-180:0.75) -- (-1,-1)
      (0,-1.5) -- (1,-1) -- (0.5,-0.5) -- (0.5,0.5) -- (1,1) -- (0,1.5)
      (-0.5,-0.5) -- (0.5,-0.5)
      (-0.5,0.5) -- (0.5,0.5)
      (0,-0.75) node {$j_3$}
      (-0.75,0) node {$j_2$}
      (0.75,0) node {$j_4$}
      (0,0.75) node {$j_1$}
      (-1.25,-1.) node {$j_7$}
      (1.25,-1.) node {$j_8$}
      (1.25,1.) node {$j_5$}
      (-1.25,1.) node {$j_6$}
      (1.75,0) node {$j_9$};
\end{tikzpicture}
&= (-1)^{j_2 + j_4 + j_9} 
\left\{
\begin{matrix}
\, j_2 \, & \, j_4 \,  & \, j_9 \, \\
\, j_5 \, & \, j_6 \, & \, j_1 \,
\end{matrix}
\right\} 
\left\{
\begin{matrix}
\, j_2 \, & \, j_4 \,  & \, j_9 \, \\
\, j_8 \, & \, j_7 \, & \, j_3 \,
\end{matrix}
\right\} \nonumber\\
&=
\frac{(-1)^{j_2 + j_4 + j_9} (-1)^{j_5+j_6+j_7+j_8} }{d_{j_9} \sqrt{d_{j_1} d_{j_3}}}
\left[
\begin{matrix}
\, j_2 \, & \, j_4 \,  & \, j_9 \, \\
\, j_5 \, & \, j_6 \, & \, j_1 \,
\end{matrix}
\right] 
\left[
\begin{matrix}
\, j_2 \, & \, j_4 \,  & \, j_9 \, \\
\, j_8 \, & \, j_7 \, & \, j_3 \,
\end{matrix}
\right] \quad ,
\end{align}
as we wanted to proof.

~\\


\begin{thebibliography}{99}\small
\parskip -2pt


\bibitem{reisenberger}
M.~P.~Reisenberger,
  ``On relativistic spin network vertices,''
  J.\ Math.\ Phys.\  {\bf 40} (1999) 2046
  [gr-qc/9809067].
    
    \bibitem{bw1}
  B.~Dittrich and W.~Kaminski,
  ``Topological lattice field theories from intertwiner dynamics,''
  arXiv:1311.1798 [gr-qc].

\bibitem{alexreview}
 A.~Perez,
  ``The Spin Foam Approach to Quantum Gravity,''
  Living Rev.\ Rel.\  {\bf 16} (2013) 3
  [arXiv:1205.2019 [gr-qc]].
 
   \bibitem{cbook}
  C. Rovelli, ``Quantum Gravity'' (Cambridge University Press,
Cambridge 2004).

  \bibitem{eprl}
   J.~Engle, E.~Livine, R.~Pereira and C.~Rovelli,
  ``LQG vertex with finite Immirzi parameter,''
  Nucl.\ Phys.\ B {\bf 799} (2008) 136
  [arXiv:0711.0146 [gr-qc]].
E.~R.~Livine and S.~Speziale,
  		``A new spinfoam vertex for quantum gravity,''
  		Phys.\ Rev.\  D {\bf 76}, 084028 (2007),
  		[arXiv:0705.0674 [gr-qc]].
 L.~Freidel and K.~Krasnov,
  ``A New Spin Foam Model for 4d Gravity,''
  Class.\ Quant.\ Grav.\  {\bf 25} (2008) 125018
  [arXiv:0708.1595 [gr-qc]].
    A.~Baratin and D.~Oriti,
  ``Quantum simplicial geometry in the group field theory formalism: reconsidering the Barrett-Crane model,''
  New J.\ Phys.\  {\bf 13} (2011) 125011
  [arXiv:1108.1178 [gr-qc]].



\bibitem{riello}
L.~Freidel and D.~Louapre,
  ``Diffeomorphisms and spin foam models,''
  Nucl.\ Phys.\  B {\bf 662} (2003) 279
  [arXiv:gr-qc/0212001].
 C.~Perini, C.~Rovelli and S.~Speziale,
  ``Self-energy and vertex radiative corrections in LQG,''
  Phys.\ Lett.\ B {\bf 682}, 78 (2009)
  [arXiv:0810.1714 [gr-qc]].
  V.~Bonzom and M.~Smerlak,
  ``Bubble divergences from cellular cohomology,''
  Lett.\ Math.\ Phys.\  {\bf 93}, 295 (2010)
  [arXiv:1004.5196 [gr-qc]].
A.~Riello,
  ``Self-Energy of the Lorentzian EPRL-FK Spin Foam Model of Quantum Gravity,''
  arXiv:1302.1781 [gr-qc].
V.~Bonzom and B.~Dittrich,
  ``Bubble divergences and gauge symmetries in spin foams,''
  arXiv:1304.6632 [gr-qc].

\bibitem{fotini}
 F.~Markopoulou,
  ``An algebraic approach to coarse graining,''
  arXiv:hep-th/0006199.
%
%
%
R.~Oeckl,
  ``Renormalization of discrete models without background,''
  Nucl.\ Phys.\  B {\bf 657} (2003) 107
  [arXiv:gr-qc/0212047].
%
%
J.~A.~Zapata, ``Loop quantisation from a lattice gauge theory perspective," Class.\ Quant.\ Grav.\ {\bf 21} (2004) L115-L122, [arXiv:gr-qc/0401109].


\bibitem{bd12b}
 B.~Dittrich,
  ``From the discrete to the continuous: Towards a cylindrically consistent dynamics,''
  New J.\ Phys.\  {\bf 14} (2012) 123004
  [arXiv:1205.6127 [gr-qc]].
  
  
\bibitem{refining}
B.~Dittrich and S.~Steinhaus,
  ``Time evolution as refining, coarse graining and entangling,''
  arXiv:1311.7565 [gr-qc].

\bibitem{bahrdittrich}
  B.~Dittrich,
  ``Diffeomorphism symmetry in quantum gravity models,'' Adv.\ Sci.\ Lett.\ 2 (2009) 151
  [arXiv:0810.3594 [gr-qc]].
 B.~Bahr and B.~Dittrich,
  ``(Broken) Gauge Symmetries and Constraints in Regge Calculus,''
  Class.\ Quant.\ Grav.\  {\bf 26} (2009) 225011
  [arXiv:0905.1670 [gr-qc]].
 B.~Bahr and B.~Dittrich, 
 ``Improven and Perfect Actions in Discrete Gravity,''
  Phys.\ Rev.\  D {\bf 80} (2009) 124030
  [arXiv:0907.4323 [gr-qc]].
    B.~Bahr and B.~Dittrich,
 ``Breaking and restoring of diffeomorphism symmetry in discrete gravity,'' 
   AIP Conference Proceedings Volume 1196,
ed.\ by J.~Kowalski-Glikman et.\ al.\ ,pp. 10-17
  arXiv:0909.5688 [gr-qc].
    B.~Bahr, B.~Dittrich and S.~Steinhaus,
  ``Perfect discretization of reparametrization invariant path integrals,''
  Phys.\ Rev.\ D {\bf 83} (2011) 105026
  [arXiv:1101.4775 [gr-qc]].
 B.~Dittrich,
  ``How to construct diffeomorphism symmetry on the lattice,''
  PoS QGQGS {\bf 2011} (2011) 012
  [arXiv:1201.3840 [gr-qc]].

 \bibitem{sffinite}
 B.~Bahr, B.~Dittrich and J.~P.~Ryan,
  ``Spin foam models with finite groups,''
  J.\ Grav.\  {\bf 2013} (2013) 549824
  [arXiv:1103.6264 [gr-qc]].
  
  
  
 \bibitem{eckert}
 B.~Dittrich, F.~C.~Eckert and M.~Martin-Benito,
  ``Coarse graining methods for spin net and spin foam models,''
  New J.\ Phys.\  {\bf 14} (2012) 035008
  [arXiv:1109.4927 [gr-qc]].
 B.~Dittrich and F.~C.~Eckert,
  ``Towards computational insights into the large-scale structure of spin foams,''
  J.\ Phys.\ Conf.\ Ser.\  {\bf 360} (2012) 012004
  [arXiv:1111.0967 [gr-qc]].
  
  \bibitem{bahretal12} 
 B.~Bahr, B.~Dittrich, F.~Hellmann and W.~Kaminski,
  ``Holonomy Spin Foam Models: Definition and Coarse Graining,''
  Phys.\ Rev.\ D {\bf 87} (2013) 044048
  [arXiv:1208.3388 [gr-qc]].
   B.~Dittrich, F.~Hellmann and W.~Kaminski,
  ``Holonomy Spin Foam Models: Boundary Hilbert spaces and Time Evolution Operators,''
  Class.\ Quant.\ Grav.\  {\bf 30} (2013) 085005
  [arXiv:1209.4539 [gr-qc]].
  
  \bibitem{he}
 B.~Bahr, B.~Dittrich and S.~He,
  ``Coarse graining free theories with gauge symmetries: the linearized case,''
  New J.\ Phys.\  {\bf 13} (2011) 045009
  [arXiv:1011.3667 [gr-qc]].

  
  \bibitem{fat}
P.~A.~Morse,
  ``Approximate diffeomorphism invariance in near flat simplicial geometries,''
Class.\ Quant.\ Grav.\ {\bf 9} (1992) 2489. 



\bibitem{unpublished}
B. Bahr, B. Dittrich, unpublished, 2012.

  \bibitem{levin}
M.~Levin, C.~P.~Nave, ``Tensor renormalization group approach to 2D classical lattice models," Phys.\ Rev.\ Lett.\  {\bf 99} (2007) 120601, [arXiv:cond-mat/0611687 [cond-mat.stat-mech]].

\bibitem{guwen}
  Z.~-C.~Gu and X.~-G.~Wen,
  ``Tensor-Entanglement-Filtering Renormalization Approach and Symmetry Protected Topological Order,''
  Phys.\ Rev.\ B {\bf 80} (2009) 155131
  [arXiv:0903.1069 [cond-mat.str-el]].
  

\bibitem{noui}
 L.~Smolin,
  ``Linking topological quantum field theory and nonperturbative quantum gravity,''
  J.\ Math.\ Phys.\  {\bf 36} (1995) 6417
  [gr-qc/9505028].
 S.~Major and L.~Smolin,
  ``Quantum deformation of quantum gravity,''
  Nucl.\ Phys.\ B {\bf 473} (1996) 267
  [gr-qc/9512020].
   R.~Borissov, S.~Major and L.~Smolin,
  ``The Geometry of quantum spin networks,''
  Class.\ Quant.\ Grav.\  {\bf 13} (1996) 3183
  [gr-qc/9512043].
 K.~Noui and P.~Roche,
  ``Cosmological deformation of Lorentzian spin foam models,''
  Class.\ Quant.\ Grav.\  {\bf 20} (2003) 3175
  [gr-qc/0211109].
    W.~J.~Fairbairn and C.~Meusburger,
  ``Quantum deformation of two four-dimensional spin foam models,''
  J.\ Math.\ Phys.\  {\bf 53} (2012) 022501
  [arXiv:1012.4784 [gr-qc]].
  M.~Han,
  ``4-dimensional Spin-foam Model with Quantum Lorentz Group,''
  J.\ Math.\ Phys.\  {\bf 52} (2011) 072501
  [arXiv:1012.4216 [gr-qc]].
M.~Dupuis and F.~Girelli,
  ``Quantum hyperbolic geometry in loop quantum gravity with cosmological constant,''
  Phys.\ Rev.\ D {\bf 87} (2013) 121502
  [arXiv:1307.5461 [gr-qc]].
  M.~Dupuis and F.~Girelli,
``Observables in Loop Quantum Gravity with a cosmological constant,''
  arXiv:1311.6841 [gr-qc].


\bibitem{jeff}
J.~Hnybida, private communication

  
 \bibitem{bdetal13}
B.~Dittrich, M.~Mart{\'i}n-Benito and E.~Schnetter,
  ``Coarse graining of spin net models: dynamics of intertwiners,''
  New J.\ Phys.\  {\bf 15} (2013) 103004
  [arXiv:1306.2987 [gr-qc]].
  
   
  
  
\bibitem{bw2}
B.~Dittrich, W.~Kaminski, ``Ground states in fusion categories" in preparation.

\bibitem{warsaw}
W.~Kaminski, M.~Kisielowski, J.~Lewandowski,
``The EPRL intertwiners and corrected partition function,"
 Class.\ Quant.\ Grav.\  {\bf 27} (2010) 165020 [arXiv:0912.0540 [gr-qc]].


\bibitem{warsaw0}
W.~Kaminski, M.~Kisielowski and J.~Lewandowski,
  ``Spin-Foams for All Loop Quantum Gravity,''
  Class.\ Quant.\ Grav.\  {\bf 27} (2010) 095006
   [Erratum-ibid.\  {\bf 29} (2012) 049502]
  [arXiv:0909.0939 [gr-qc]].
  
B.~Bahr, F.~Hellmann, W.~Kaminski, M.~Kisielowski and J.~Lewandowski,
  ``Operator Spin Foam Models,''
  Class.\ Quant.\ Grav.\  {\bf 28} (2011) 105003
  [arXiv:1010.4787 [gr-qc]].
  
   
\bibitem{bc}
 J.~W.~Barrett and L.~Crane,
  ``Relativistic spin networks and quantum gravity,''
  J.\ Math.\ Phys.\  {\bf 39}, 3296 (1998)
  [gr-qc/9709028].
  
  \bibitem{bojowaldperez}
 M.~Bojowald and A.~Perez,
  ``Spin foam quantization and anomalies,''
  Gen.\ Rel.\ Grav.\  {\bf 42} (2010) 877
  [gr-qc/0303026].
  B.~Bahr
 ``On knottings in the physical Hilbert space of LQG as given by the EPRL model,''
  Class.\ Quant.\ Grav.\  {\bf 28} (2011) 045002
  [arXiv:1006.0700 [gr-qc]].
  B.~Dittrich and S.~Steinhaus,
  ``Path integral measure and triangulation independence in discrete gravity,''
  Phys.\ Rev.\ D {\bf 85}, 044032 (2012)
  [arXiv:1110.6866 [gr-qc]].
  
 \bibitem{wenclass} 
 X.~Chen, Z.-C.~Gu, X.-G.~Wen, ``Classification of gapped symmetric phases in 1D spin systems,"  
Phys. Rev. {\bf B 83} (2011) 035107 [arXiv: 1008.3745 [cond-mat.str-el]].

  
 \bibitem{schuch} 
 N.~Schuch, I.~Cirac, D.~Perez-Garcia,
``PEPS as ground states: Degeneracy and topology," Annals of Physics {\bf 325} (2010) 2153.
N.~Schuch, D.~Perez-Garcia, I.~Cirac, 
``Classifyinf quantum phases using MPS and PEPS," Phys. Rev. {\bf B 84} (2011) 165139 
[arXiv:1010.3732 [cond-mat.str-el]]  
  
    \bibitem{biedenharn}
L.~C.~Biedenharn, M.~A.~Lohe, ``Quantum Group Symmetries and q-Tensor  Algebras," (World Scientific, Singapore 1995)


  \bibitem{yellowbook}
  J.~S.~Carter, D.~E.~Flath, M.~Saito, ``The Classical and Quantum $6j$--symbols,"  (Princeton University Press, Princeton 1995)
  

    \bibitem{levinslides}
  M.~Levin, https://www.ipam.ucla.edu /publications/tqc2007/tqc2007 6595.ppt (2007).
  

\bibitem{sherbrook}
A. J. Ferris,
``The area law and real-space renormalization,''
Phys. Rev. B {\bf 87} (2013), 125139
[arXiv:1301.2608 [cond-mat.str-el]].
  
  
  \bibitem{vidal}
G.~Vidal,
  ``Entanglement Renormalization,''
  Phys.\ Rev.\ Lett.\  {\bf 99} (2007) 22,  220405
  [cond-mat/0512165].
  
      \bibitem{vidalsymm}
R. N. C. Pfeifer, P. Corboz, O. Buerschaper, M. Aguado, M. Troyer, G. Vidal ``Simulation of anyons with tensor network algorithms,"
Phys. Rev. B 82 (2010) 115126  [arXiv:1006.3532 [cond-mat.str-el]].
  S.~Singh, G.~Vidal,`` Tensor network state and algorithms in the presence of a global SU(2) symmetry,"
  Phys.\ Rev.\ B {\bf 86} (2012) 195114
  [arXiv:1208.3919[cond-mat.str-el]].
  
  
  \bibitem{ponzanoregge}
  G.~Ponzano and T.~Regge, 
  ``Semiclassical limit of Racah coefficients,"
in Spectroscopic and group theoretical methods in physics, ed. F. Bloch,
North-Holland Publ. Co., Amsterdam (1968) 1-58.
  
  \bibitem{turaev}
 V.~G.~Turaev and O.~Y.~Viro,
  ``State sum invariants of 3 manifolds and quantum 6j symbols,''
  Topology {\bf 31} (1992) 865.
  
    
  \bibitem{melons}
  V.~Bonzom, R.~Gurau, A.~Riello, and V.~Rivasseau,
  ``Critical behavior of colored tensor models in the large N limit,''
  Nucl. Phys. B {\bf 853} (2011) 174
  [arXiv:1105.3122].   A.~Baratin, S.~Carrozza, D.~Oriti, J.~P.~Ryan and M.~Smerlak,
  ``Melonic phase transition in group field theory,''
  arXiv:1307.5026 [hep-th].
  
  
  
\bibitem{ryan08}
B.~Dittrich and J. P.~Ryan,
``Phase space descriptions for simplicial 4d geometries,"
Class. \ Quant. \ Grav. {\bf 28} (2011) 065006 [arXiv:0807.2806 [gr-qc]].
B.~Dittrich and J. P.~Ryan,
``Simplicity in simplicial phase space,"
arXiv:1006.4295 [gr-qc].
B.~Dittrich and J. P.~Ryan,
 ``On the role of the Barbero-Immirzi parameter in discrete quantum gravity,"
 arXiv:1209.4892 [gr-qc].
  

 
   \bibitem{alexandrov}
  S.~Alexandrov, M. Geiller, and K.~Noui
  ``Spin Foams and Canonical Quantization,"
 SIGMA {\bf 8} (2012) 055 [arXiv:1112.1961 [gr-qc]].
  M. Geiller and K.~Noui
``Testing the imposition of the Spin Foam Simplicity Constraints'"
Class. \ Quant. \ Grav. {\bf 29} (2012) 135008 [arXiv:1112.1965 [gr-qc]].
 
  \bibitem{sebprivate}
 B.~Dittrich, S.~Steinhaus, w.i.p.  

\bibitem{migdal}
A.A. Migdal,  
Sov. Phys. JETP {\bf 42} (1975) 413.
L P.  Kadanoff
Annals Phys. {\bf 100} (1976) 359.

\bibitem{kadanoff}
E.~Efrati, Z.~Wang, A.~Kolan, and L. P.~Kadanoff
``Real Space Renormalization in Statistical Mechanics,"
 arXiv:1301.6323 [cond-mat].



\bibitem{igor}
I. Khavkine and J. D. Christensen, 
``q-Deformed spin foam models of quantum gravity," 
Class. Quant. Grav. {\bf 24} (2007) 3271 [arXiv:0704.0278 [gr-qc]].



   \bibitem{ditt}
  C.~Rovelli,
  ``Discretizing parametrized systems: the magic of Ditt-invariance,''
  arXiv:1107.2310 [hep-lat].
  
  




\end{thebibliography}
\end{document}